\documentclass[preprint,authoryear,11pt,3p,times,review]{elsarticle}

\usepackage{amssymb}
\usepackage{amsthm}

\usepackage{color}
\usepackage[breaklinks]{hyperref}
\hypersetup{colorlinks=true,linkcolor=blue,citecolor=blue,filecolor=blue,urlcolor=blue}
\renewcommand{\v}[1]{\ensuremath{\mathbf{#1}}}

\usepackage{etoolbox}
\usepackage{breqn}
\usepackage{multirow}
\usepackage{threeparttable}
\usepackage{arydshln}
\usepackage{adjustbox}
\usepackage[skip=2pt,font=small]{caption}
\usepackage{scrextend}
\deffootnotemark{\textsuperscript{(\thefootnotemark)}}

\usepackage{lineno}
\BeforeBeginEnvironment{dmath}{\begin{nolinenumbers}}
\AfterEndEnvironment{dmath}{\end{nolinenumbers}}

\journal{Planetary \& Space Science}

\begin{document}

\begin{frontmatter}



\title{Exploring terrestrial lightning parameterisations for exoplanets and brown dwarfs}


\author[a1,a2,a3,a5]{G. Hodos\'an\corref{ref1}}
\author[a1,a2,a4]{Ch. Helling}
\author[a1]{and I. Vorgul}

\cortext[ref1]{E-mail: hodosan.gabriella@gmail.com}

\address[a1]{Centre for Exoplanet Science, University of St Andrews, St Andrews KY16 9SS, UK}
\address[a2]{SUPA, School of Physics and Astronomy, University of St Andrews, St Andrews KY16 9SS, UK}
\address[a3]{Instituto de Astrof\'isica de Andaluc\'ia, (IAA-CSIC), Glorieta de la Astronom\'ia s/n, 18008, Granada, Spain}
\address[a4]{SRON Netherlands Institute for Space Research, Sorbonnelaan 2, 3584 CA Utrecht, NL}
\address[a5]{RAL Space, STFC Rutherford Appleton Laboratory, Didcot, Oxfordshire OX11 0QX, UK}

\begin{abstract}
Observations and models suggest that the conditions to develop lightning may be present in cloud-forming extrasolar planetary and brown dwarf atmospheres. Whether lightning on these objects is similar to or very different from what is known from the Solar System awaits answering as lightning from extrasolar objects has not been detected yet. We explore terrestrial lightning parameterisations to compare the energy radiated and the total radio power emitted from lightning discharges for Earth, Jupiter, Saturn, extrasolar giant gas planets and brown dwarfs.
We find that lightning on hot, giant gas planets and brown dwarfs may have energies of the order of $10^{11}$--$10^{17}$ J, which is two to eight orders of magnitude larger than the average total energy of Earth lightning ($10^9$ J), and up to five orders of magnitude more energetic than lightning on Jupiter or Saturn ($10^{12}$ J), affirming the stark difference between these atmospheres. Lightning on exoplanets and brown dwarfs may be more energetic and release more radio power {\bf} than what has been observed from the Solar System. Such energies would increase the probability of detecting lightning-related radio emission from an extrasolar body.
\end{abstract}


\end{frontmatter}

% \begin{linenumbers}

%%%%%%%%%%%%%%%%%%%%%%%%%%%%%%%%%%%%%%%%%%%%%%%%%%

%%%%%%%%%%%%%%%%% BODY OF PAPER %%%%%%%%%%%%%%%%%%

%__________________________________________________________________
%__________________________________________________________________
\section{Introduction} \label{sec:int}

Lightning on Earth has been studied for hundreds of years \citep[e.g.][]{rakov2003, yair2008, siingh2015, helling2016}. Its role within the terrestrial global electric circuit \citep{wilson1921} and its importance in pre-biotic chemistry  \citep{miller1953, miller1959, cleaves2008, rimmer2016, 2019RSPTA.37780398H} is subject of ongoing research.

Lightning signatures span the whole electromagnetic spectrum, from very low frequency (VLF) radio emission (from a few Hz to few hundreds of MHz) to very high energy X-rays \citep[][Table 1]{bailey2014}. Lightning releases about 1\% of its energy in both optical \citep[][p. 334]{borucki1987, hill1979, krider1968} and radio \citep{volland1982, volland1984, farrell2007} frequencies, however radio emission can be more prominent than optical, as the radio background produces less noise than the optical background, e.g. due to the host star \citep{hodosan2016b}. 
  
Lightning radio emission has been observed not only in Earth thunderclouds, but in volcanic plumes on Earth \citep[e.g.][]{mather2006, james2008}, as well as in the atmospheres of other Solar System planets. Saturn Electrostatic Discharges (SEDs) were observed by \textit{Voyager 1} during its close approach in 1980 \citep{warwick1981}, by the RPWS (Radio and Plasma Wave Science) instrument of the \textit{Cassini} spacecraft between 2004 and 2011 \citep{fischer2006, fischer2007, fischer2011b}. Ground-based observations of SEDs were reported by \citet{zakharenko2012}, who conducted their observations with the UTR-2 (Ukrainian T-shaped Radio telescope) and compared their results with simultaneous \textit{Cassini} measurements. Sferics\footnote{\textit{Sferics} (or atmospherics), in general, are the emission in the low-frequency (LF) range with a power density peak between 4 and 12 kHz on Earth \citep{volland1984} produced by lightning discharges. Since only radio emission in the higher frequency range can penetrate through the ionosphere, high frequency (HF) radio emission caused by lightning on other planets are also called sferics \citep{desch2002}. This type of emission is the most probable to be observed coming from other planetary bodies, since it is the only type of lightning radio emission capable of escaping the ionosphere or the magnetosphere of the object.} were detected inside Jupiter's atmosphere by the Galileo probe in 1996 \citep{rinnert1998} and whistlers\footnote{Electromagnetic (EM) waves propagating along magnetic field lines and emitting in the very low-frequency (VLF) range. We can distinguish two main types. The majority of the EM signals traverse the ionosphere into the magnetosphere and after reaching the opposite hemisphere they travers the ionosphere again to the downward direction \citep{rakov2003}. The minority of whistlers propagate through the ionosphere into the magnetosphere and there dissipate \citep{rakov2003}.} as well in the planet's magnetosphere $\sim$20 years earlier by the \textit{Voyager 1} plasma wave instrument \citep{gurnett1979}. The outer two giant planets, Neptune and Uranus also showed radio emission, which were attributed to lightning activity \citep{zarka1986}. 

Even though, there is no direct electromagnetic detection of lightning on an extrasolar object yet \citep{2020MNRAS.495.3881H}, studies have shown that both exoplanets and brown dwarfs host environments with the necessary ingredients (i.e. charged particles, seed electrons, charge separation) for lightning to initiate. Both observations \citep[e.g.][]{kreidberg2014, sing2009, sing2013, sing2015} and kinetic cloud models \citep[e.g.][]{helling2008, helling2011a, helling2011b} showed that clouds form in extrasolar atmospheres. Thundercloud formation involves convection, which is one of the main processes on Earth to separate two oppositely charged regions from each other. On extrasolar planets and brown dwarfs, gravitational settling was suggested to be a mechanism for large-scale charge separation \citep{helling2013a}. The result of charge separation is the build-up of an electric potential and, therefore, the electric field that is necessary for the initiation of lightning discharges \citep{rakov2003,aplin2013,helling2013a,helling2016}. \citet{helling2013a} and \citet{bailey2014} suggested that the processes building up the electric field necessary for lightning are able to produce lightning discharges in extrasolar atmospheres. \citet{zarka2012} concluded that emission $10^5$ times stronger than radio emission observed on Jupiter or Saturn from a distance of $10$ pc, with a bandwidth of 1--10 MHz (integration time: 10--60 min)  would be detectable for exoplanets. This conclusion is based on scaling up the same radio emission observed from Jupiter and Saturn. Recently, similar excursuses were undertaken to explore the observability of exoplanet radio emission with LOFAR.
 
The current paper explores questions related to lightning properties in extrasolar planetary and brown dwarf atmospheres, such as: How does our expectation of lightning radiation on exoplanets compare to what is known from the Solar System? Could lightning be more energetic? What would be the lightning energy deposited into the atmosphere of the extrasolar body? To address these questions, we utilise lightning parameterisations that were originally developed for Earth, and we explore the parameter space that may affect the energy dissipated from lightning discharges and the power radiated at radio frequencies. 

The paper utilised the general dipole model of lightning developed for Earth (Sect.~\ref{sec:model}) that we explore for extrasolar atmospheres. In Sect.~\ref{sec:param}, our modelling approach and the required input parameters are discussed. Section~\ref{sec:val} presents our results for Earth, Jupiter and Saturn, provides a comparison to literature results, and evaluates the effect of parameter uncertainties. Section~\ref{sec:resdis} presents our results for exoplanet and brown dwarf atmospheres, including an assessment of parameter uncertainties. Section~\ref{sec:conc} concludes this paper.

%__________________________________________________________________
%__________________________________________________________________
\section[]{Model Description} \label{sec:model}

We adopt a modelling ansatz that utilises lightning model parameterisations that were developed and tested for Earth lightning \citep[e.g.][]{bruce1941, rakov2003}, and further explored for Solar System lightning \citep[e.g.][]{farrell2007,lammer2001,yair2012}. We will utilise this ansatz in order to explore lightning properties in exoplanetary and brown dwarf atmospheres. The aim is to determine the total radiation energy released from a single lightning flash, and explore the properties of the emitted power spectrum. The radiation energy will determine the power emitted at certain frequencies and the radio flux observable from lightning discharges. Once the energy of a single lightning discharge is determined, the total lightning energy affecting the atmosphere \citep{hodosan2016a} can be estimated and potential effects on the atmosphere chemistry studied \citep{bailey2014,2017MNRAS.470..187A}.

Lightning discharges are complex phenomena and different aspects are simulated by different models \citep[e.g.][]{gordillo2010,ebert2010}. The lightning channel itself is often tortuous, built up of several segments and branches \citep{levine1978}. \citet{moss2006} used a Monte Carlo model to obtain properties of lightning streamers and found that 10$E_k$\footnote{$E_k$ is the conventional breakdown threshold field \citep{helling2013a}.} fields, which are produced at streamer tips, can accelerate part of the low energy electrons emitted from the streamers to energies high enough to function as seed electrons for a thermal runaway electron avalanche. \citet{babich2015} presented a model of the local electric field increase in front of a lightning stepped leader, and showed that the front electrons are capable of initiating relativistic runaway avalanches. \cite{2019Icar..333..294K} modelled the streamer propagation in gas representing Titan's atmosphere. \citet{gordillo2010} modelled the conductivity in a sprite\footnote{Sprites are red transient luminous events occurring after an intense positive cloud-to-ground discharge above the thunderstorm in the altitude range $< 40-90$ km \citep{rousseldupre2008}.} streamer channel and found that the changes caused by sprites in the atmospheric conductivity last for several minutes after the discharge. 

When modelling extraterrestrial lightning, parameterisations tested for Earth lightning are applied \citep{farrell1999,farrell2007,lammer2001,bailey2014,yair2012}. \citet{farrell2007} showed that changing one parameter, the duration of the discharge, Saturnian lightning energies could well be in the super-bolt ($<10^{11}$--$10^{12}$ J), rather than average Earth-like energy range. Although the optical detection of Saturnian lightning \citep{dyudina2010} has confirmed its very high optical energy-release, the work of  \citet{farrell2007} show the importance of parameter studies when observations do not constrain physical properties well. Such is the case with exoplanetary lightning.

In this paper, we consider a lightning return stroke model\footnote{A return stroke is the most luminous part of a lightning flash, which occurs when the ionized channel connecting two oppositely charged regions becomes discharged.}, based on a simple dipole radiation model, not taking into account channel tortuosity, branching \citep[e.g.][]{bruce1941, rakov2003}, nor any microphysics in the lightning channel. Its advantage is the low number of parameters and specification of the channel current in order to achieve an agreement between the electromagnetic field predicted by the model and observed at distances ($r$) up to several hundreds of km. Such return stroke models use a relatively simple relation between the current in the lightning channel at any time and height, $I(z,t)$, and the current in the channel base, $I(0,t)$ \citep{rakov2003}. To estimate the radiation energy dissipated from a single lightning discharge, the following properties are calculated:\\
1) Electric current, $i(t)$: Ideally, it is derived from the number of charges, $Q$, that accumulate in the current channel, but is often used as parameter as $Q$ is largely unknown (Sect. \ref{sec:current}).\\
2) Electric field, $E(t)$, from the dipole moment of the the lightning channel with current $i(t)$ (Sect. \ref{sec:efield}).\\
3) Frequency and power spectra, $E(f)$ and $P(f)$, of the electric field, $E(t)$ (Sect. \ref{sec:freqsp}). The power spectrum possesses properties important for characterizing lightning radio emission: $f_0$ is the peak frequency, the frequency at which the largest amount of power is released; and $n$ is the negative slope of the power spectrum at high frequencies ($f > f_0$), which carries information on the amount of power released at these frequencies. \\
4) Radiated power at different frequencies, $P'(f)$, and radiated discharge energy, $W_{\rm rad}$ (Sect. \ref{sec:disen}).

\smallskip
\noindent
The return stroke model was developed to model cloud-ground lightning (CG) on Earth, in comparison to intra-cloud (IC) lightning. As of yet, it is not immediately clear which of these two, CG or IC, may be best used to model lightning in atmospheres of extrasolar, ultra-cool objects like exoplanets and brown dwarfs. Exoplanets are very diverse, ranging from rocky to gaseous planets. Most of the known exoplanets are in fact rocky planets \citep{2020arXiv200514671B}. Giant gas planets and brown dwarfs have rather warm and high-pressure ($>$100bar) inner atmospheres which turn these gases into conducting plasmas. We utilise the ideas of the CG models in our following work because the CG model is more widely applied in the Earth and Solar System community, hence, a wider range of parameters has been studied, and also because it appears more suitable to the large diversity of exoplanetary atmosphere environments.

%__________________________________________________________________
\subsection{Lightning as radiating dipole} \label{sec:dipole}
 
A common way to model lightning discharges in the Solar System is to assume that lightning radiates as a dipole \citep[e.g.][]{bruce1941}.  The parameters that determine the dipole radiation are the length of the dipole (or characteristic length of charge separation), $h$; the charges that run through it, $Q(t)$; the characteristic time of the duration of the discharge, $\tau$; and finally the velocity with which the discharge event occurs, $\v v_0$. In case of return stroke modelling, the velocity, $\v v_0$ will be the velocity of the return stroke itself. The current, $i(t)$, in the discharge channel is determined by the charges, $Q(t)$, accumulating there:

\begin{equation} \label{eq:charge}
Q(t) = \int_0^t{i(t') dt'}.
\end{equation}
We note that the charges that initiate a lightning discharge may be orders of magnitude more abundant compared to the charges forming the lightning channel current $i(t)$. The charges $Q(t)$ also determine the electric dipole moment $M(t)$, 
\begin{equation} \label{eq:mom}
M(t) = h Q(t).
\end{equation}
Once the extension of the discharge, $h$, and the velocity, ${\v v_0}$, of the event is known, the duration (time scale) of the discharge, $\tau$, can be calculated,

\begin{equation} \label{eq:tau1}
\tau = \frac{h}{\v v_0}. 
\end{equation}

\noindent For large source-observer distances of $r \geq 50$ km, the discharge time scale $\tau$ can be estimated according to \citet{volland1984} as

\begin{equation} \label{eq:tau}
\tau = \frac{2 \pi}{\sqrt{\alpha \beta}}, 
\end{equation}

\noindent where $\alpha$ and $\beta$ are frequency-type constants. $\alpha^{-1}$ represents the overall duration of the return stroke, while $\beta^{-1}$ represents the rise time of the current wave \citep[$\alpha < \beta$;][]{dubrovin2014}. Because the radio emission of lightning is the result of the acceleration of electrons, the duration of the lightning discharge{\bf} (and consequently the extension of the main lightning channel, with no branches) will determine the frequency (and therefore, the wavelength) where the radiated power reaches its peak ($f_0$). The discharge duration $\tau$ is linked to the peak frequency \citep{zarka2004},

\begin{equation} \label{eq:taufr}
f_0 = \frac{1}{\tau}.
\end{equation}

%__________________________________________________________________
\subsection{Current wave function} \label{sec:current}

%Table - Bi-exp. params
\begin{table*}
 \caption{Parameters of bi-exponential and Heidler current functions (Figs.~\ref{fig:compi}, \ref{fig:ibh}). The peak current, $i_0$=30 kA in all cases. \textbf{Top:} Bi-exponential current function parameters  (Eq. \ref{eq:1}) as given in the literature for Earth and Jupiter. The last column relates the value-pairs with the plotted current functions and associated electric fields, frequency and power spectra in  Figs.~\ref{fig:compi}, \ref{fig:ebh}, and \ref{fig:fpbh} (left). \textbf{Middle:} Heidler current function parameters (Eq. \ref{eq:3}); $m=2$ in all cases. The last column relates the value-pairs with the plotted current functions and associated electric fields, frequency and power spectra in Figs. \ref{fig:compi}, \ref{fig:ebh}, and \ref{fig:fpbh} (right). \textbf{Bottom:} Parameters for which both current functions are the same (Fig. \ref{fig:ibh}).}
 \vspace{0.3cm}
  \resizebox{\textwidth}{!}{
  \begin{tabular}{@{}llllll@{}}	
	\hline
	\multicolumn{6}{c}{\it Bi-exponential function} \\
	$\alpha$ [$1/s$] & $\beta$ [$1/s$] & \multicolumn{2}{l}{Reference} & Planet & Figs \ref{fig:compi}, \ref{fig:ebh}, and \ref{fig:fpbh}; left \\ 
	\hline
	$4.4 \times 10^4$ & $4.6 \times 10^5$ & \multicolumn{2}{l}{\citet{bruce1941}} & Earth & $[1]$ \\
	$2 \times 10^4$ & $2 \times 10^5$ & \multicolumn{2}{l}{\citet{levine1978}} & Earth & $[2]$ \\
	$1.5 \times 10^3$ & $1.75 \times 10^3$ & \multicolumn{2}{l}{\citet{farrell1999}} & Jupiter & $[3]$ \\
	\hline \hline
	\multicolumn{6}{c}{\it Heidler function}	\\
	$\eta$ & $\tau_1$ [$\mu$s] & $\tau_2$ [$\mu$s] & \multicolumn{2}{l}{Comments} & Figs \ref{fig:compi}, \ref{fig:ebh}, and \ref{fig:fpbh}; right \\ 
	\hline
	0.73 & 0.3 & 0.6 & \multicolumn{2}{l}{\citet[][table 1]{diendorfer1990}} & $[1]$ \\ 
	0.37 & 0.3 & 0.6 & \multicolumn{2}{l}{$\eta$ calculated by Eq.~\ref{eq:3}} & $[2]$ \\
	0.92 & 0.454 & 143.0 & \multicolumn{2}{l}{\vtop{\hbox{\strut $\eta$ calculated by Eq.~\ref{eq:3};}\hbox{\strut $\tau_1$ and $\tau_2$ are values for subsequent stroke;}\hbox{\strut \citet[][table 1]{heidler2002}}}} & $[3]$ \\
	\hline \hline
	\multicolumn{6}{c}{\it Comparison of the two current functions as in Fig. \ref{fig:ibh}} \\
	$\alpha$ [$1/s$] & $\beta$ [$1/s$] & $\eta$ & $\tau_1$ [$\mu$s] & $\tau_2$ [$\mu$s] & Comment \\
	\hline
	$6.97 \times 10^{3}$ & $2.2 \times 10^6$ & 1 & 0.454 & 143.0 & \vtop{\hbox{\strut $\alpha$ and $\beta$ are expressed}\hbox{\strut by $\tau_1$ and $\tau_2$ (Eq. \ref{eq:ab}).}\hbox{\strut $\tau_1$ and $\tau_2$: \citet[][table 1]{heidler2002}}} \\
	\hline
  \label{table:3}
  \end{tabular}
	}
\end{table*}
%Table 

\begin{figure*}
	\resizebox{\textwidth}{!}{
  \includegraphics[]{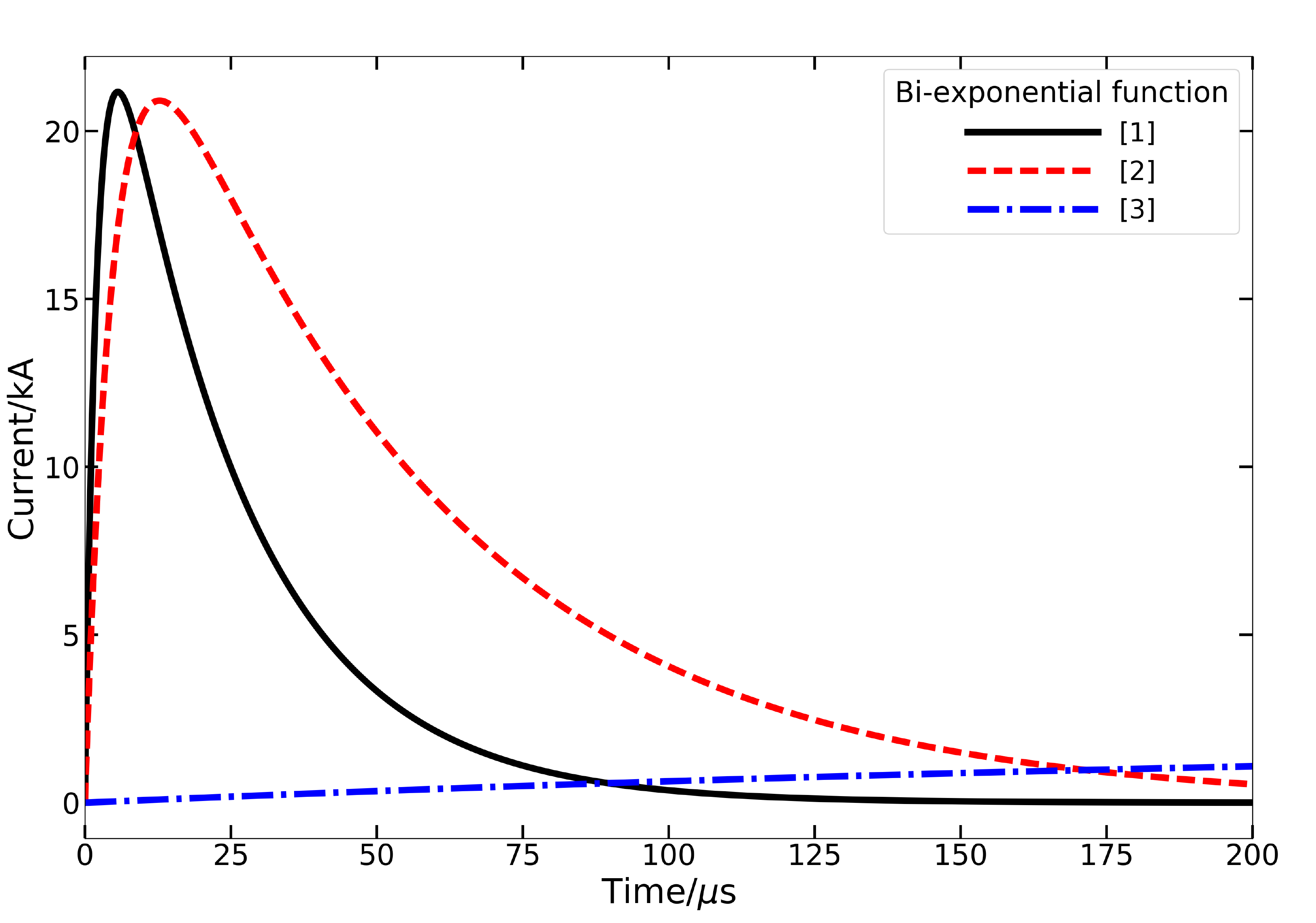}
  \includegraphics[]{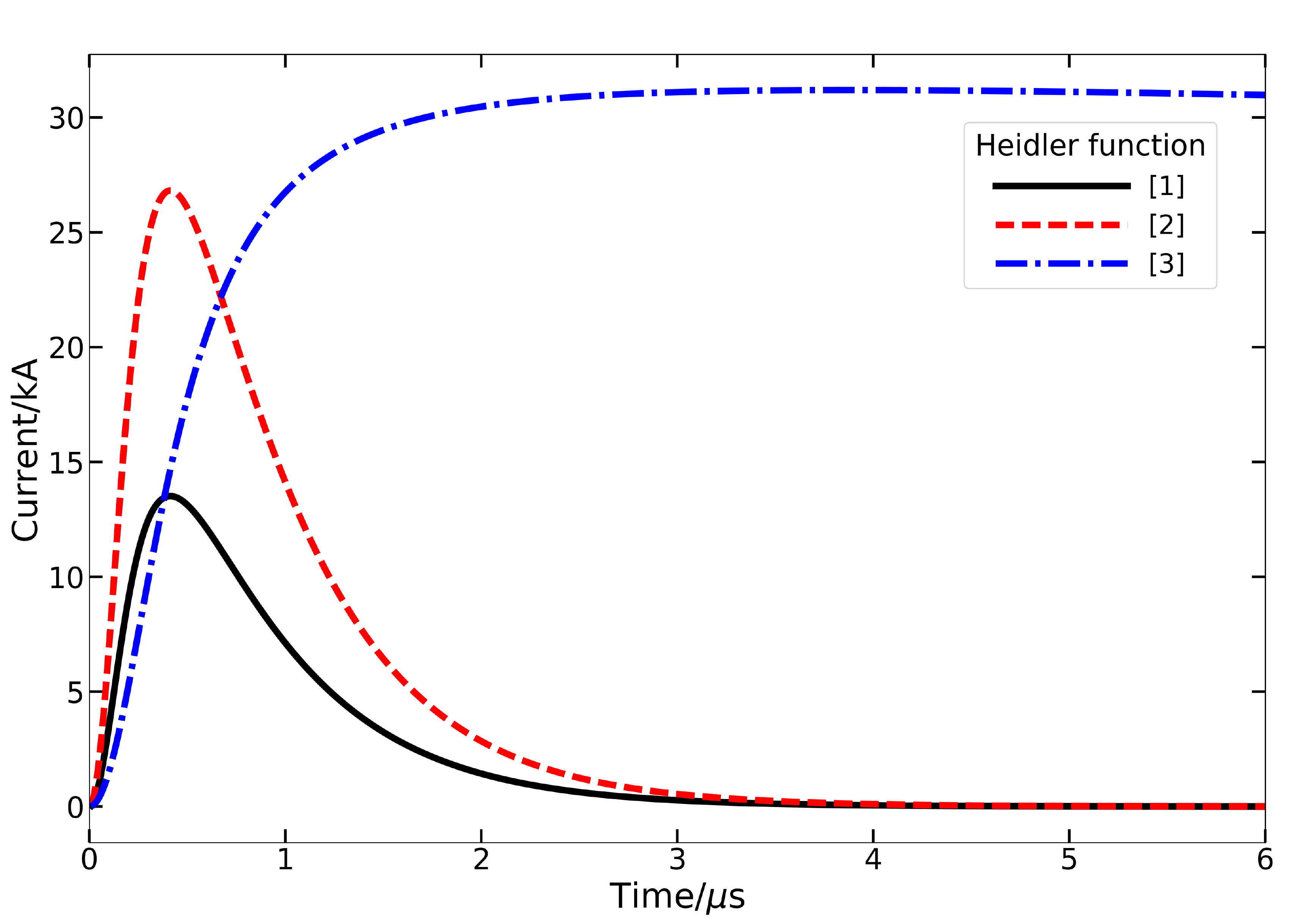}
	}
  \caption{Current wave functions. \textbf{Left:} Bi-exponential  (Eq. \ref{eq:1}), for  $\alpha$ and $\beta$  cases [1], [2], and [3] (top of Table \ref{table:3}). The larger $\alpha$ the shorter the discharge event, while the larger $\beta$ the quicker the current reaching its peak. \textbf{Right:} Heidler (Eq. \ref{eq:3}), for $\eta$, $\tau_1$, and $\tau_2$ cases  [1], [2], and [3] (middle of Table \ref{table:3}). The peak current, $i_0$ = 30 kA in all cases. Case [3] of the bi-exponential function resembles Jupiter, while case [3] of the Heidler function resembles a subsequent stroke of Earth lightning.}
  \label{fig:compi}
\end{figure*}

\begin{figure}
  \centering
  \includegraphics[width=0.7\columnwidth]{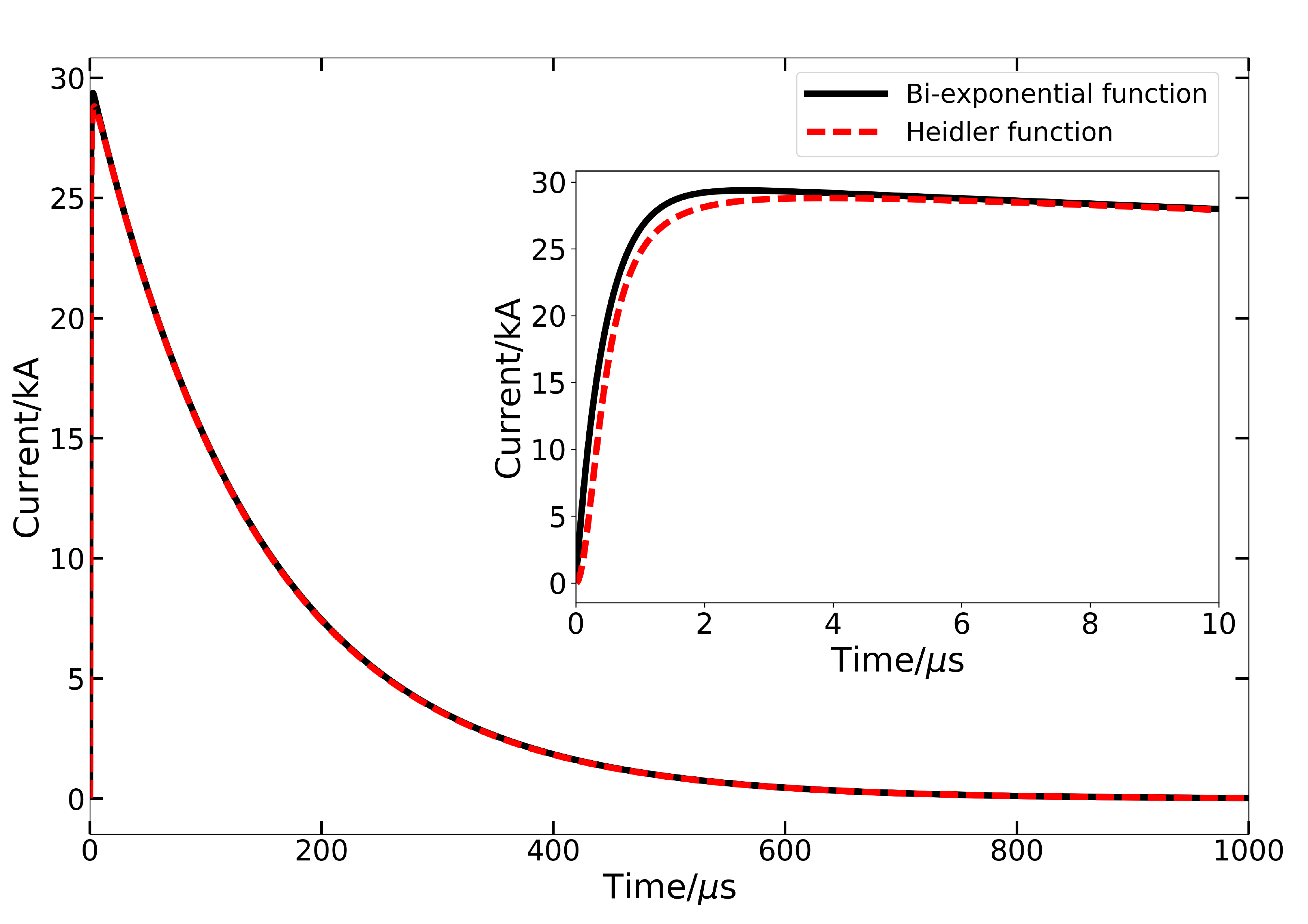}
  \caption{Comparison of the bi-exponential (black, solid line, Eq. \ref{eq:1}) and the Heidler (red, dashed line, Eq. \ref{eq:3}) current functions for  parameters listed in the bottom part of Table \ref{table:3}. Both current functions represent the same current when $\alpha$ and $\beta$ are expressed by $\tau_1$ and $\tau_2$  as in Eq. \ref{eq:ab} and $\eta= 1$.}
  \label{fig:ibh}
\end{figure}

The current at the channel base (i.e., $z = 0$) produced by electrons moving from one charged region to another, has been modelled by various current functions in the literature. The most used ones are the double- (or bi-) exponential function (Eq. \ref{eq:1}) introduced by \citet{bruce1941}, and the Heidler function \citep[Eq. \ref{eq:3};][]{heidler1985}. A combination of multiple Heidler functions or bi-exponential and Heidler functions tend to reproduce observed current shape better \citep[][Section 4.6.4]{rakov2003}. We explore how the two current functions affect the resulting electric field, its frequency and power spectra. We investigate their sensitivity against the parameters and utilise those derived from measurements for Solar System planets.

The bi-exponential current function is expressed as
\begin{equation} \label{eq:1}
i(z=0,t) = i(t) = i_0(e^{-\alpha t}-e^{-\beta t}),
\end{equation}

\noindent where $\alpha$ and $\beta$ are the same parameters as in Eq. \ref{eq:tau}, and $i_0$ is the current peak, the global maximum of the current function, $z=0$ represents the channel base. For simplicity we use $i(t)$ when referring to $i(0,t)$.  Table \ref{table:3} summarises  the parameters $\alpha$ and $\beta$ from the literature \citep{bruce1941, levine1978, farrell1999}. Difference among the $\alpha$ and $\beta$ parameters results from different assumptions of the current channel: \citet{levine1978} considered the tortuosity of the channel, while \citet{bruce1941} did not. \citet{farrell1999} modelled lightning on Jupiter trying to reproduce a current waveform less steep than the one for Earth lightning, resulting in lower $\alpha$ and $\beta$ parameters. \citet{levine1978} modified the bi-exponential current function slightly  by adding an intermediate current (with a current peak of 2.5 kA) to the formula, and making it continuous in $t=0$. Hence the lower $\alpha$ and $\beta$ parameters that describe the main current pulse in their work. Figure \ref{fig:compi} (left) illustrates the effect of the parameters $\alpha$ and $\beta$ on the shape of the current function: As $\alpha$ decreases, the duration of the discharge event becomes longer, while as $\beta$ decreases, the rise time of the current (the time between $t=0$ s and the peak) becomes longer.

The Heidler function \citep{heidler1985} is expressed as
\begin{equation} \label{eq:3}
i(t) = \frac{i_0}{\eta}\frac{\left(\frac{t}{\tau_1}\right)^m}{\left(\frac{t}{\tau_1}\right)^m+1}e^{-\frac{t}{\tau_2}},
\end{equation}
\noindent where $m \in \mathbb{N}$, $i_0$ [kA] is the current peak, $\eta e^{-\frac{\tau_1}{\tau_2}\left(m\frac{\tau_2}{\tau_1}\right)^{1/m}}$ is a correction factor for the current peak 
\citep[][p.~8]{paolone2001}, $\tau_1$ [s] is the time constant determining the current-rise time and $\tau_2$ [s] is the time constant determining the current-decay time \citep{diendorfer1990}.

%\begin{equation} \label{eq:eta}
%\eta = e^{-\frac{\tau_1}{\tau_2}\left(m\frac{\tau_2}{\tau_1}\right)^{1/m}}.
%\end{equation}

The Heidler function is preferred to the bi-exponential one because its time-derivative is zero at $t=0$ (unlike the bi-exponential function, which shows a discontinuity at $t=0$), which is consistent with the measured return-stroke current wave shape \citep{paolone2001, heidler2002}. Figure \ref{fig:compi} (right) shows Heidler functions with different parameter combinations. All these waveforms are for Earth lightning parameterisations. The black and red lines show values for first return strokes \citep{diendorfer1990} and the blue line represents subsequent return stroke waveforms \citep{heidler2002}.
 
Figure \ref{fig:ibh} demonstrates that the bi-exponential and the Heidler functions show a very similar functional behaviour when $\alpha$ and $\beta$ are expressed with $\tau_1$ and $\tau_2$ through:

\begin{equation} \label{eq:ab}
\alpha = \frac{1}{\tau_1+\tau_2}, \\
\beta = \frac{1}{\tau_1},
\end{equation}

\noindent with $\eta=1$ and $m=2$ in Eq. \ref{eq:3}. 

For simplicity, we will use the bi-exponential current function for our further studies. Also, our research is focused on distant objects with only far-field observations possible. There is, thus, no benefit in using the near field at the channel itself (see Sect. \ref{ssec:elcom}), as given by the waveguide model of a lightning channel in \citet{volland1981}. The results of the waveguide model show that the Bruce \& Golde aperiodic waves dominate the radiation field (damped oscillation), which is the field with the largest effects at largest distances. Hence, the bi-exponential current model is sufficient for our investigation.

%__________________________________________________________________
\subsection{Electric field} \label{sec:efield}

\begin{figure*}
	\resizebox{\textwidth}{!}{
  \includegraphics{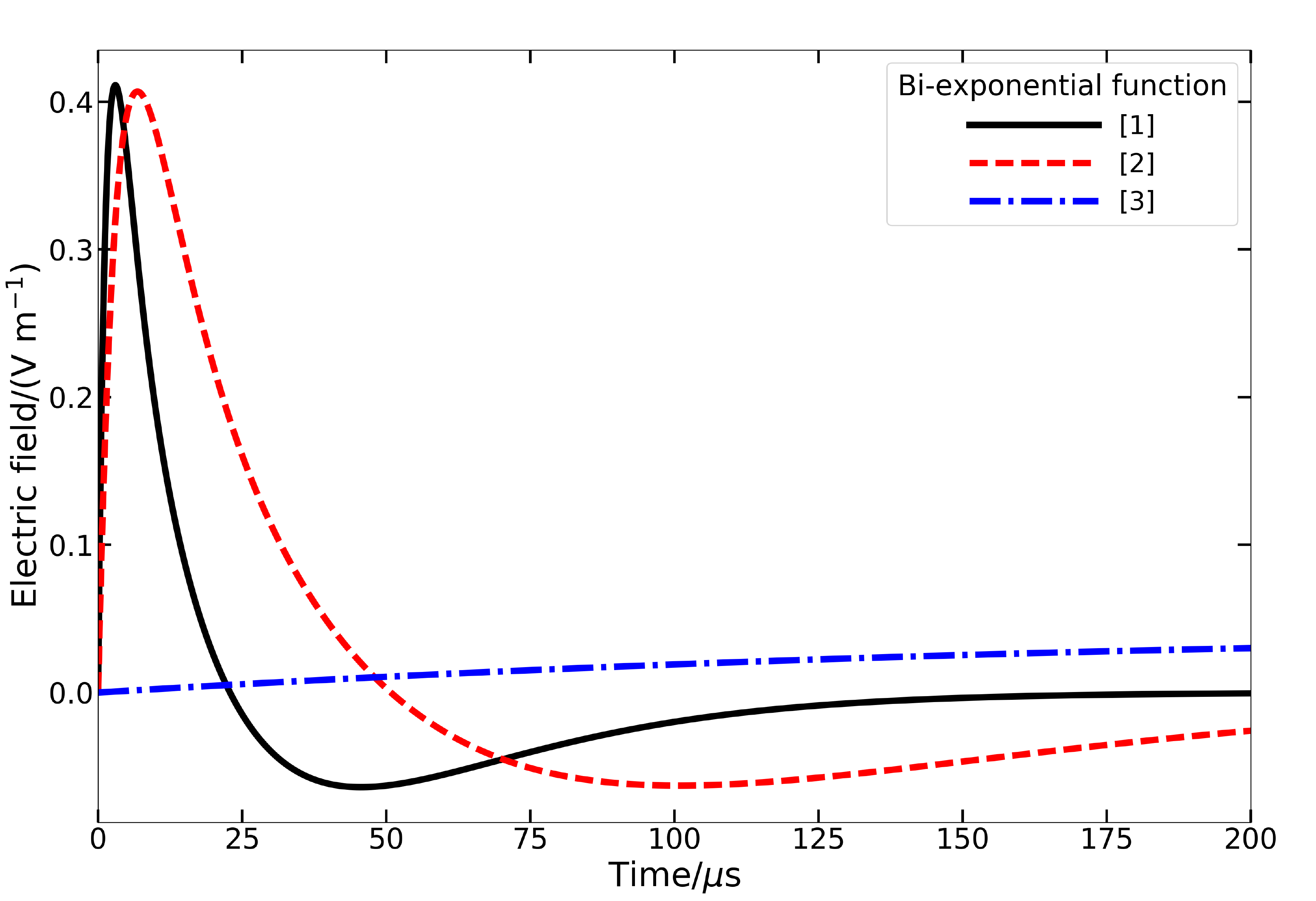}	
  \includegraphics{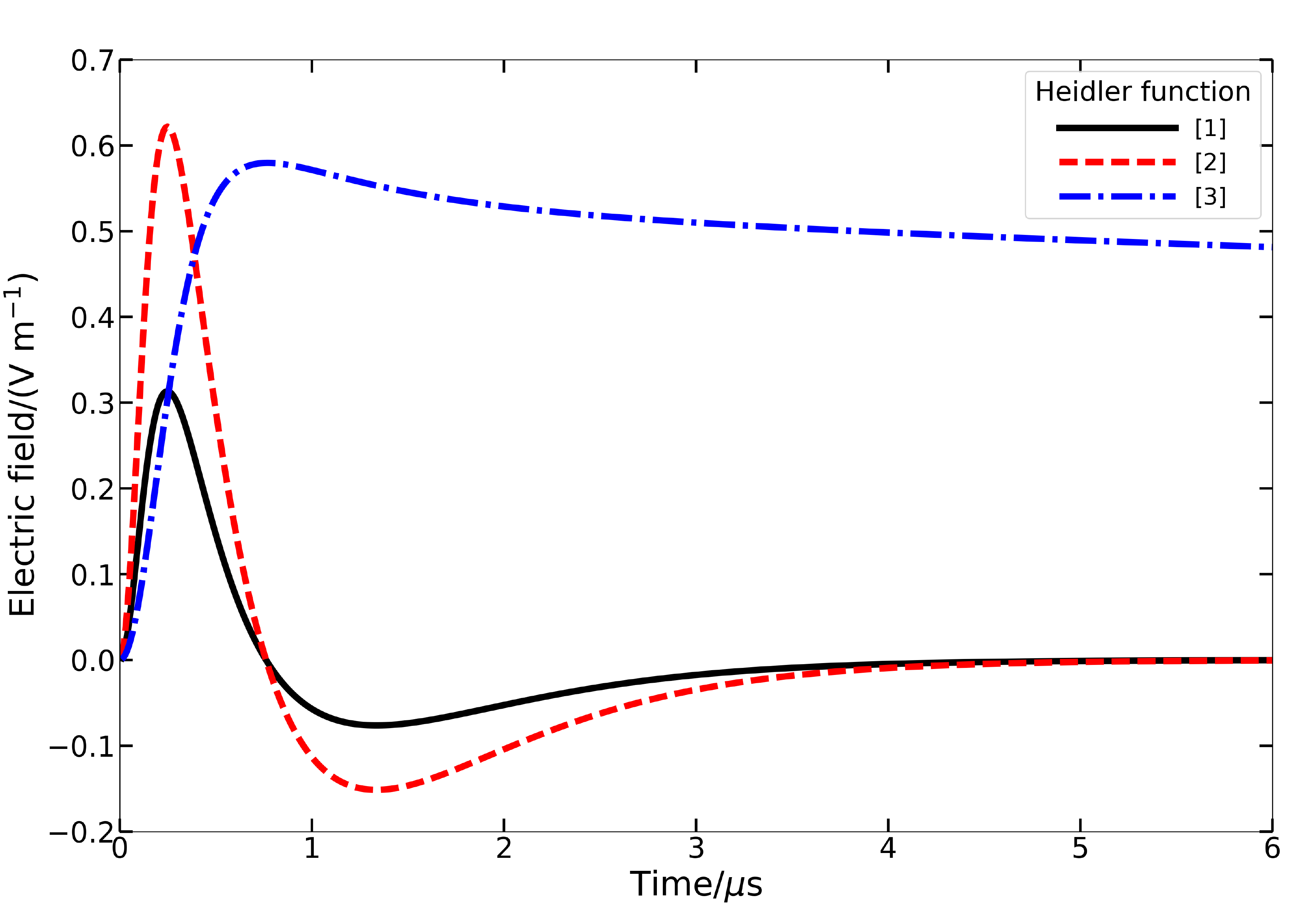}
	}
  \caption{Electric fields resulting form the bi-exponential (left, Eq. \ref{eq:10}) and the Heidler (right, combined Eq. \ref{eq:hddm}, \ref{eq:hdm} with Eq. \ref{eq:4}) current functions as in Fig. \ref{fig:compi}. The parameters are cases [1], [2], and [3]  in the top and middle of Table \ref{table:3}. The observer-source distance is $r = 1000$ km, and the current velocity is $\v v_0 = 8 \times 10^7$ m s$^{-1}$ in all cases. Case [3] of the left plot resembles Jupiter, while case [3] of right plot resembles a subsequent stroke of Earth lightning.}
  \label{fig:ebh}
\end{figure*}

%%% Electric field %%%

The electric field  resulting from the electric current (Eq. \ref{eq:1},  \ref{eq:3}) is given by the following general expression for a linear dipole model of a lightning discharge \citep{bruce1941}:
\begin{equation} \label{eq:4}
E(t) = \frac{1}{4 \pi \epsilon_0}\left[\frac{M(t)}{r^3}+\frac{1}{c r^2}\frac{dM(t)}{dt}+\frac{1}{c^2 r}\frac{d^2M(t)}{dt^2}\right],
\end{equation}

\noindent where $M(t)$ is the electric dipole moment given by Eq. \ref{eq:mom}, $\epsilon_0$ is the permittivity of the vacuum in F$/$m (Farads per meter), and all physical quantities have SI units. The first term in Eq. \ref{eq:4} is the electrostatic field, the second term is the magnetic induction field and the third term is the radiation field.

The time derivative of the dipole moment, $M(t)$,  is
\begin{equation} \label{eq:5}
\frac{dM(t)}{dt} = 2i(t)\int_0^t{\v v(t') dt'},
\end{equation}

\noindent where $i(t)$ is the current in the lightning channel at time $t$ and $\v v(t)$ is the velocity of the return stroke \citep{bruce1941}, and the factor of two is because of the consideration of an image current. \citet{bruce1941} found that this velocity decreases as the lightning stroke propagates upwards into the low-pressure atmosphere, 
\begin{equation} \label{eq:6}
\v v(t) = \v v_0 e^{-\gamma t},
\end{equation}
\noindent with  $\v v_0 \sim 8 \times 10^7$ m s$^{-1} = 0.3 c$, with $c$ the speed of light, and $\gamma \sim 3 \times 10^4$ s$^{-1}$ describes the time-dependent velocity of the discharge for Earth. The drop of the velocity observed for the upward-propagating stroke may be due to the loss of energy, in which case the stroke velocity would decrease not only upward-propagating but downward-propagating as well.

\subsubsection{Electric field from the bi-exponential current function}

First we consider the bi-exponential function (Eq. \ref{eq:1}) as the current function at the channel base ($i(0,t) = i(t)$). Two approaches can be followed when calculating the electric field: the velocity of the return stroke can be considered either constant ($\v v(t) = \v v_0$) or varying in time (Eq. \ref{eq:6}).
Combining Eqs \ref{eq:1}, \ref{eq:5} and \ref{eq:6} results in:
\begin{equation} \label{eq:7}
\frac{dM(t)}{dt} = 2\frac{i_0\v v_0}{\gamma}(e^{-\alpha t}- e^{-\beta t})(1 - e^{-\gamma t}).
\end{equation} 
For $\v v_0$=const, Eq. \ref{eq:7} simplifies to
\begin{equation} \label{eq:9}
\frac{dM(t)}{dt} = 2 i_0 \v v_0 t(e^{-\alpha t}- e^{-\beta t}),
\end{equation}
which was used by \citet{farrell1999} for deriving the electric field of Jovian lightning discharges. Combining Eqs. \ref{eq:4}, \ref{eq:9} with the derivative of Eq. \ref{eq:9}, and by neglecting the first term in Eq.~ \ref{eq:4}, results in
\begin{dmath} \label{eq:10}
E(t) =\frac{2 i_0 \v v_0}{4\pi \epsilon_0}\left[\frac{1}{c r^2}t(e^{-\alpha t}- e^{-\beta t})+\frac{1}{c^2 r}(e^{-\alpha t}- e^{-\beta t}-t\alpha e^{-\alpha t}+t\beta e^{-\beta t})\right].
\end{dmath}

\begin{figure*}
  \includegraphics[width=0.5\columnwidth]{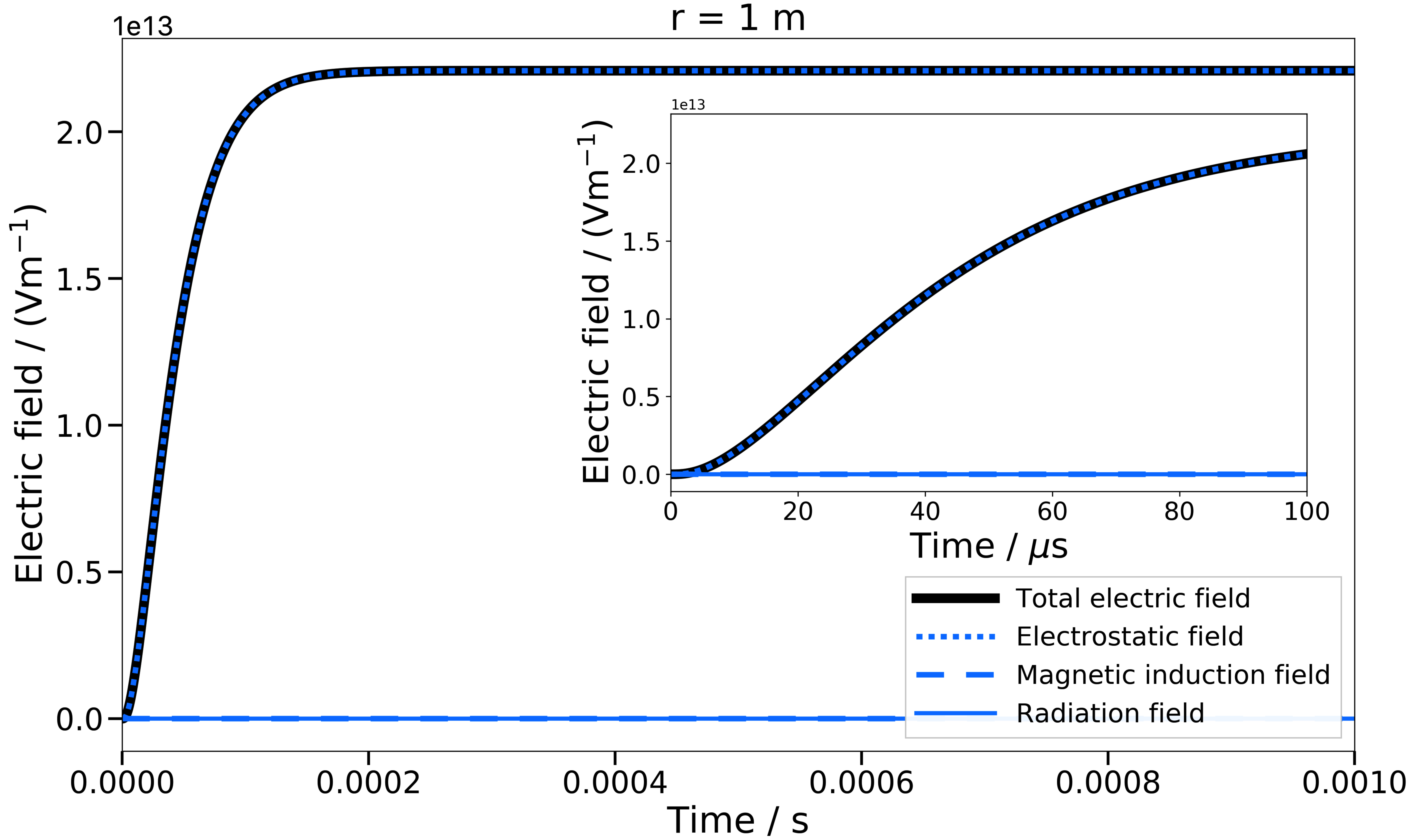}
  \includegraphics[width=0.5\columnwidth]{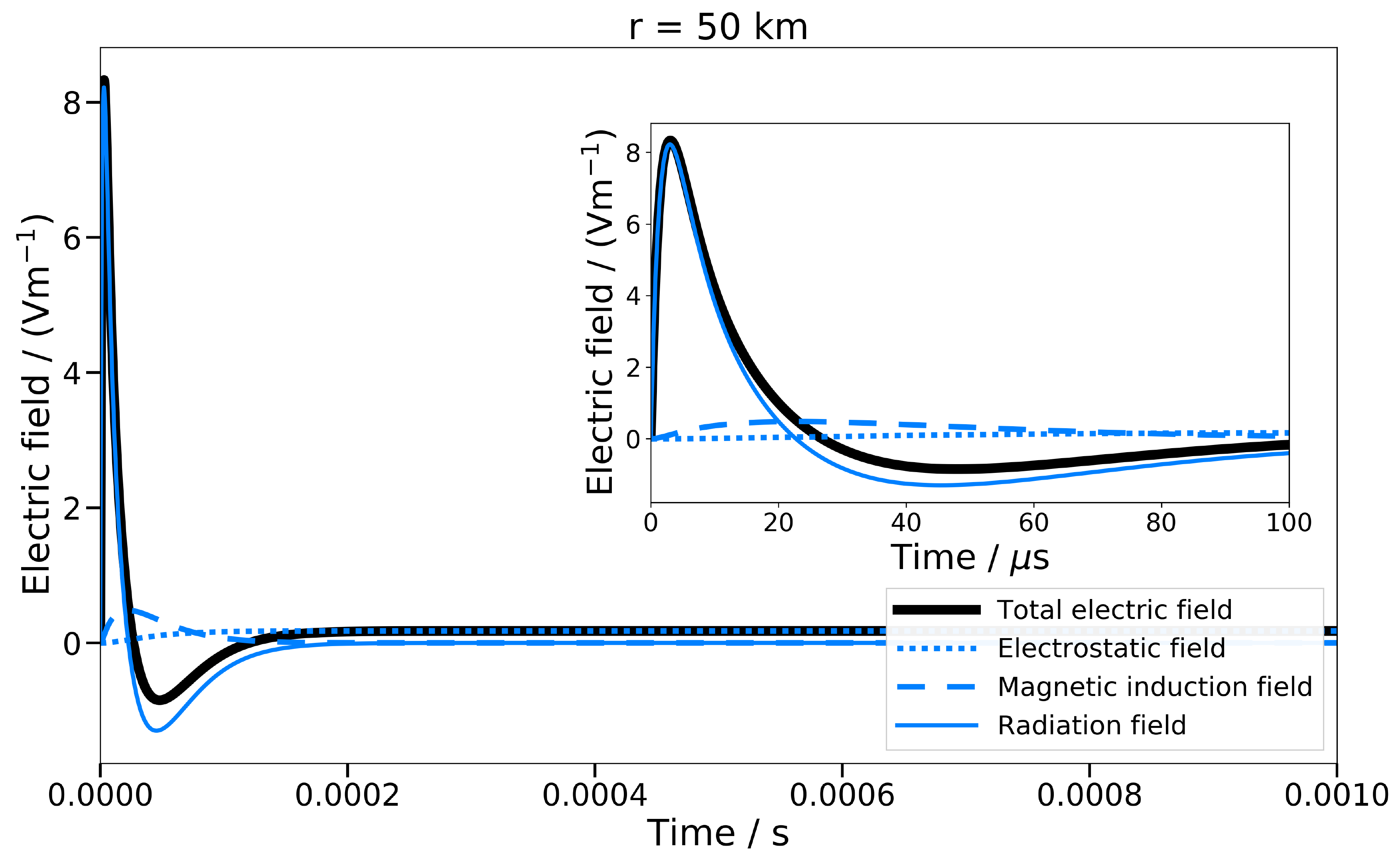} \\
  \includegraphics[width=0.5\columnwidth]{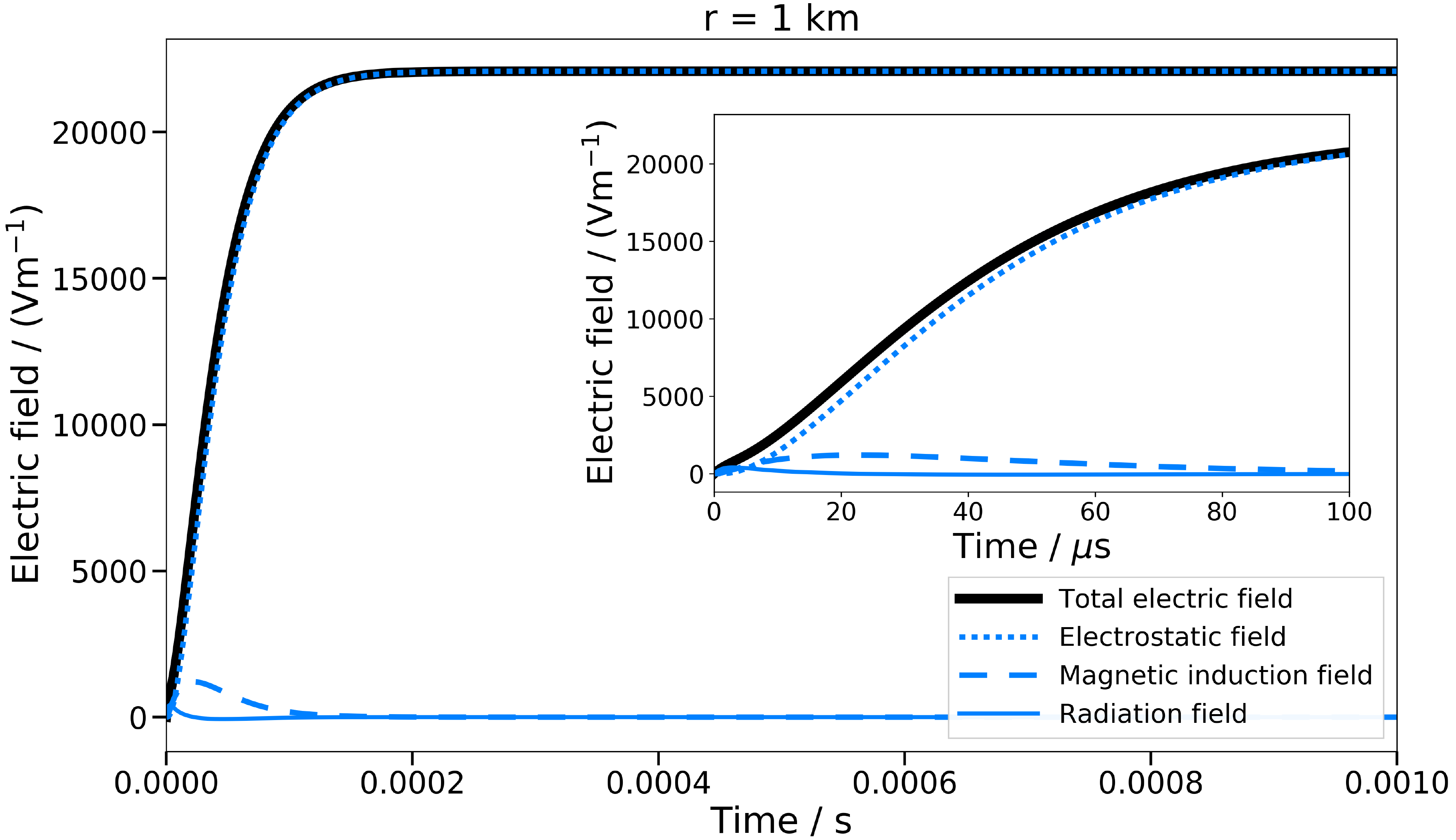}
  \includegraphics[width=0.5\columnwidth]{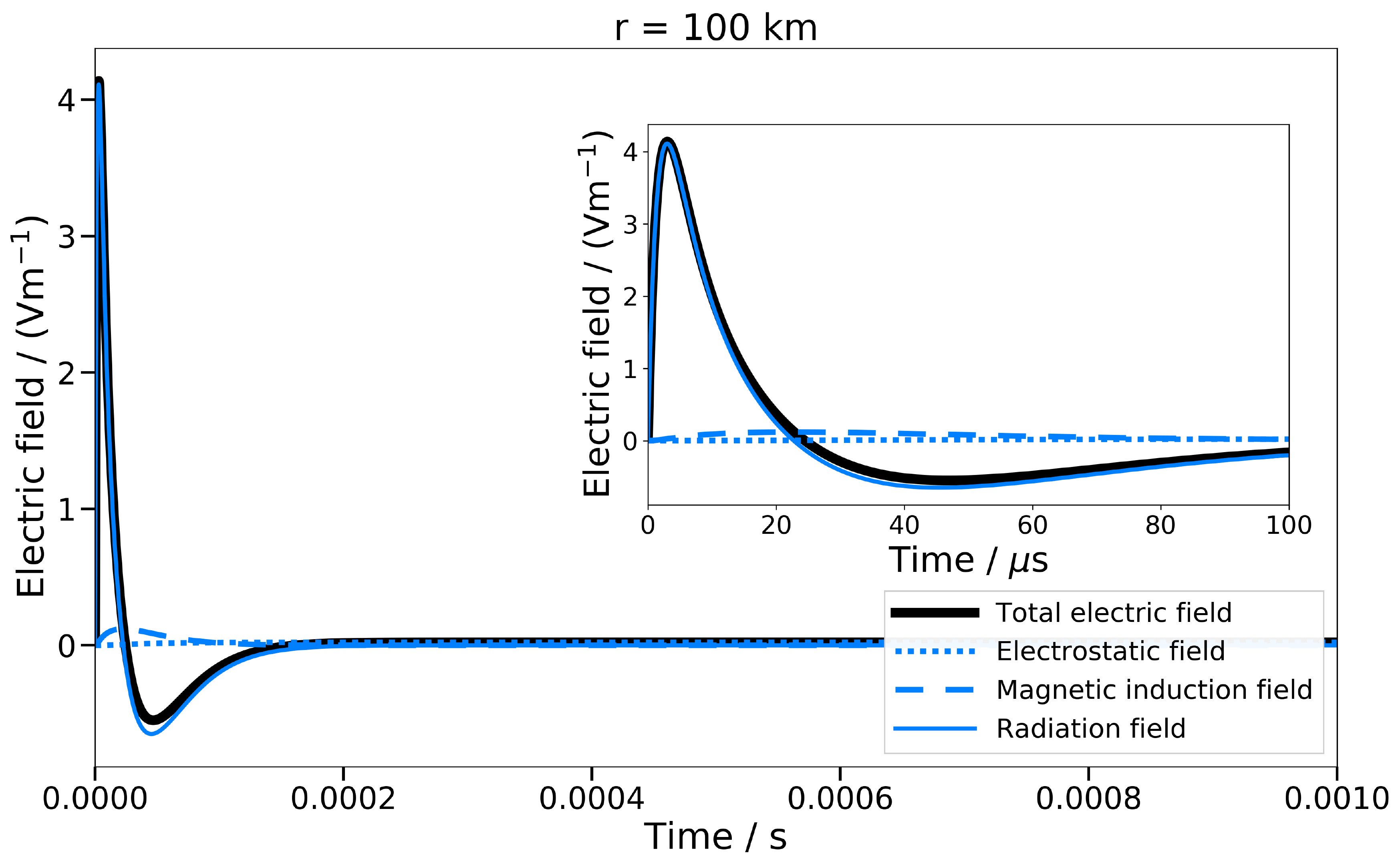} \\
  \includegraphics[width=0.5\columnwidth]{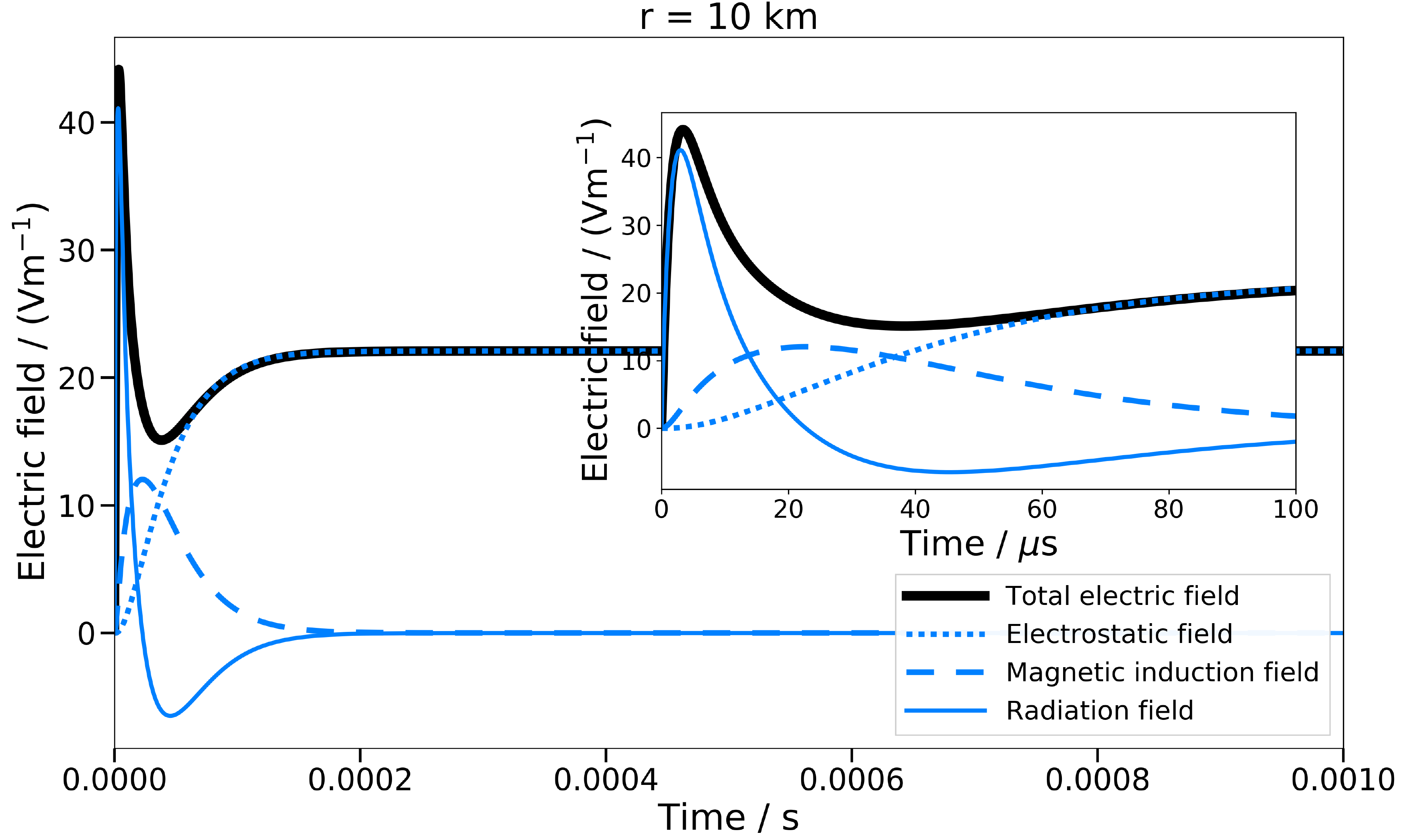}
  \includegraphics[width=0.5\columnwidth]{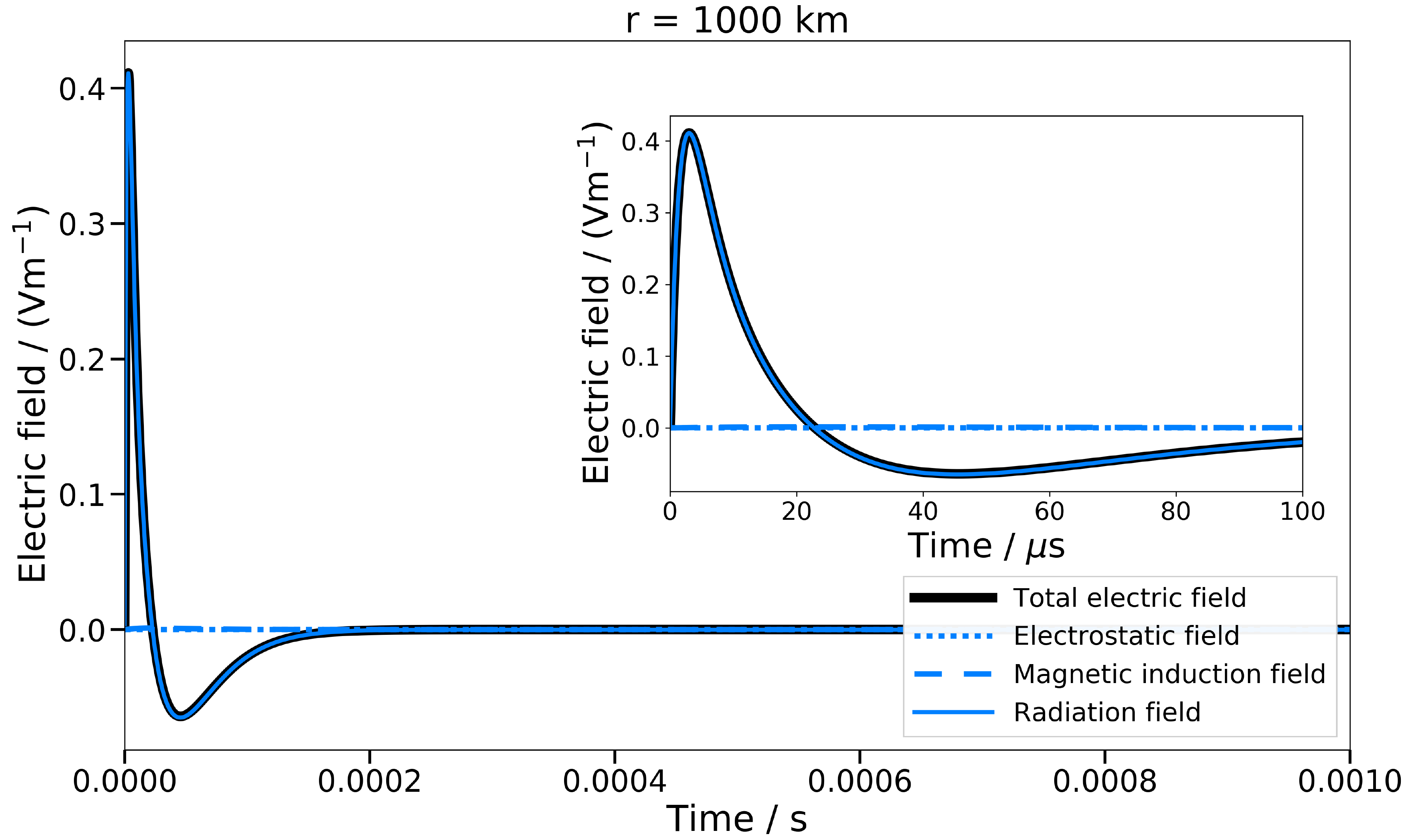}
  \caption{The three components of the electric field (Eq. \ref{eq:4}): total electric field (solid black line),  electrostatic field (dotted blue line), magnetic induction field (dashed blue line),  radiation field (solid blue line). The different panels demonstrate how the electric field components change with the lightning-observer distance, $r$ (Sect. \ref{ssec:elcom}).}
  \label{fig:ecomp}
\end{figure*}

\subsubsection{Electric field from the Heidler current function} \label{ssec:heid}

Now we consider a current given by the Heidler function (Eq. \ref{eq:3}). With $\v v_0$=const,  Eq. \ref{eq:9} changes to
\begin{dmath} \label{eq:hdm}	%hdm - heidler moment 1st derivative
\frac{dM(t)}{dt} = 2 \v v_0 t \frac{i_0}{\eta}\frac{\left(\frac{t}{\tau_1}\right)^m}{\left(\frac{t}{\tau_1}\right)^m+1}e^{-\frac{t}{\tau_2}},
\end{dmath}
and the second derivative of the dipole moment is
\begin{dmath} \label{eq:hddm}	%hddm - heidler moment 2nd derivative
\frac{d^2M(t)}{dt^2} = \frac{2 \v v_0 i_0}{\eta} \left[\frac{\left(\frac{t}{\tau_1}\right)^m}{\left(\frac{t}{\tau_1}\right)^m+1}e^{-\frac{t}{\tau_2}}\left(1-t \frac{1}{\tau_2}\right) + t e^{-\frac{t}{\tau_2}} \frac{\frac{m}{\tau_1} (\frac{t}{\tau1})^{m-1}}{((\frac{t}{\tau_1})^m + 1)^2}\right].
\end{dmath}

By combining Eqs \ref{eq:hdm} and \ref{eq:hddm} with Eq. \ref{eq:4}, we calculate the electric field at large distances from the source ($r >> h$). Figure \ref{fig:ebh} demonstrates how the electric field changes when derived from different current functions. The distance between the source and observer for all figures is $r = 1000$ km, and the propagation velocity is constant, $\v v_0 = 8.0 \times 10^7 {\rm m s}^{-1}$. Due to the large distance, $r$, the first term in Eq. \ref{eq:4} can be  neglected in Fig. \ref{fig:ebh}. To derive the electric field, SI units are used. The curves of the electric fields in Fig. \ref{fig:ebh} show the same forms relative to each other as that shown by the current functions in Fig. \ref{fig:compi}.

\subsubsection{Comparing the three electric field components} \label{ssec:elcom}

The dipolar electric field of a lightning discharge has three main components (Eq. \ref{eq:4}): electrostatic, magnetic induction and radiation fields. The three components depend on $r$, the distance between the lightning channel and the observer. Figure \ref{fig:ecomp} illustrates this dependence. All parameters are the same for all figures ($i_0=30 \ {\rm kA;} \ \v v_0=8\times10^{7} \ {\rm m s}^{-1}$), and all electric fields were calculated from the bi-exponential current function ($\alpha=4.4\times10^4 \ {\rm s}^{-1}; \ \beta=4.6\times10^5 \ {\rm s}^{-1}$) for changing distances $r$. It is clearly seen, that the electrostatic field (blue dotted line) has an effect on the total electric field (black solid line) close to the discharge event. This effect decreases as we get further away from the source and at a few tens of km it becomes negligible. In the meantime, the induction field (blue dashed line) becomes stronger and still affects the overall shape at $\sim$50 km, slightly increasing the total electric field. Kilometres away from the source, the effect of the radiation field increases. From a few tens to hundreds of km, the radiation field is the dominant component of the electric field (similar to Fig. 1 in \citet{dubrovin2014} who analysed the electric field of a Saturnian lightning discharge).

Hence, at large distances ($r>50$ km) the radiation field will be the dominant component of the electric field, with a slight contribution from the induction field. From now onwards, we only use these parts of the electric field to calculate lightning electric field frequency spectrum and radio energy of lightning for exoplanets that are parsecs away from our Solar System.

%__________________________________________________________________
\subsection{Electric field frequency and power spectra} \label{sec:freqsp}

\begin{figure*}
	\resizebox{\textwidth}{!}{
  \includegraphics{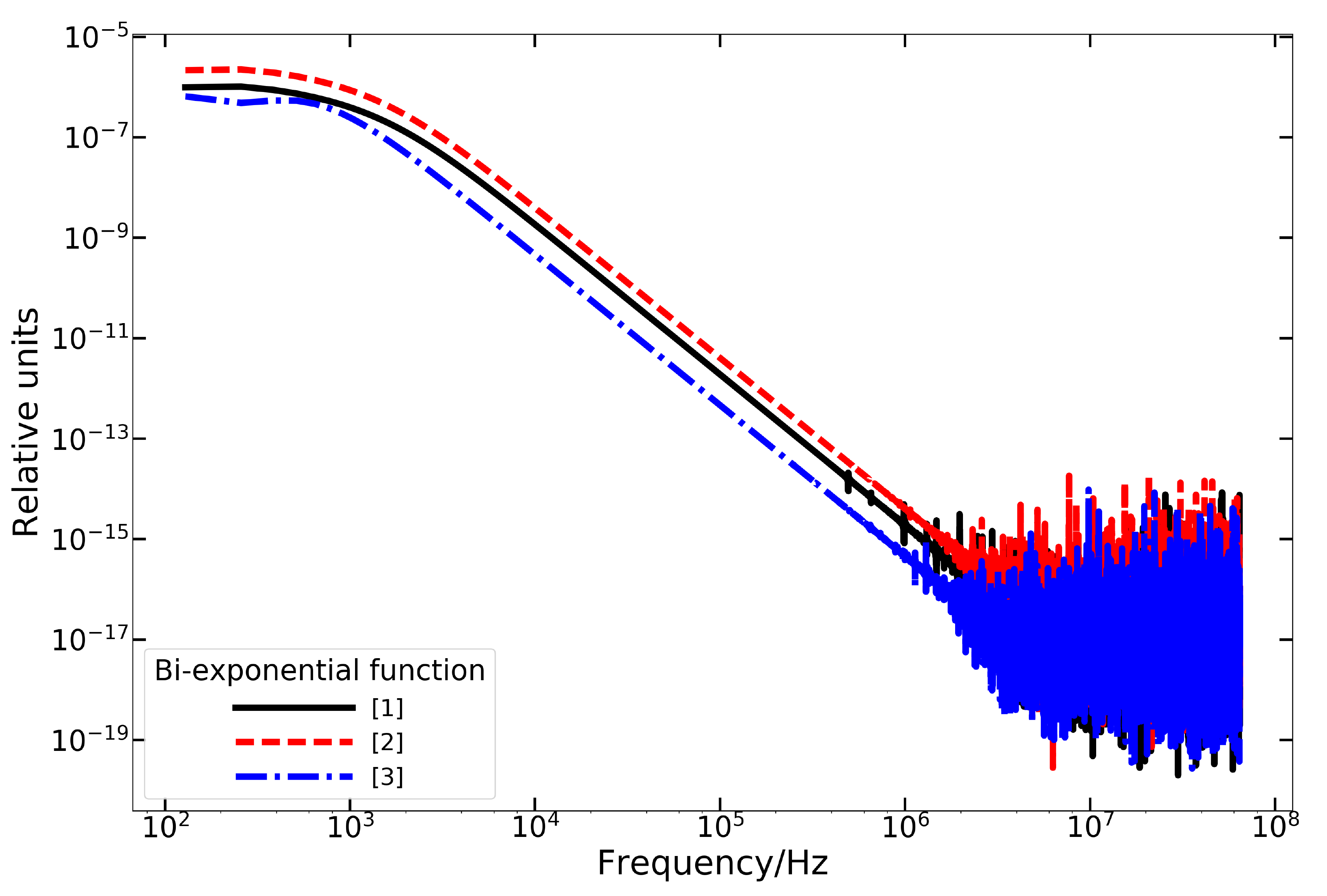}
  \includegraphics{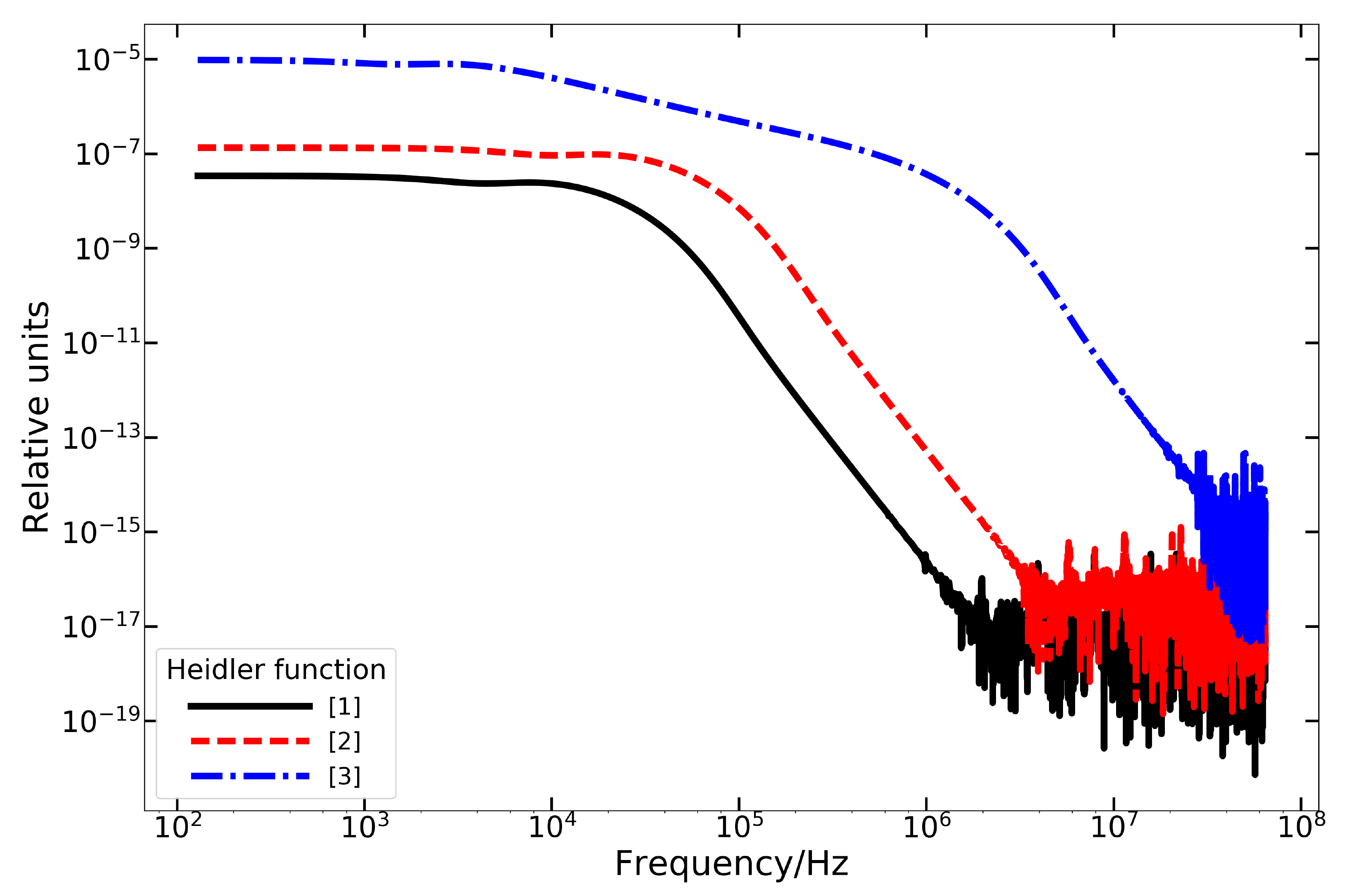}
	}
  \caption{Power spectra calculated from the electric fields of bi-exponential (left, Eq. \ref{eq:10}) and the Heidler (right, combined Eq. \ref{eq:hddm} and \ref{eq:hdm} with Eq. \ref{eq:4}) current functions represented in Fig. \ref{fig:ebh} and with cases [1], [2], and [3] given in Table \ref{table:3}, top and middle panel. The observer-source distance is $r = 1000$ km, and the current velocity is $\v v_0 = 8 \times 10^7 {\rm ms}^{-1}$ in all cases. Spectra obtained with numerical Fast Fourier Transform. Case [3] of the left plot resembles Jupiter, while case [3] of the right plot resembles a subsequent stroke of Earth lightning.}
  \label{fig:fpbh}
\end{figure*}

\begin{figure}
  \centering
  \includegraphics[width=0.7\columnwidth]{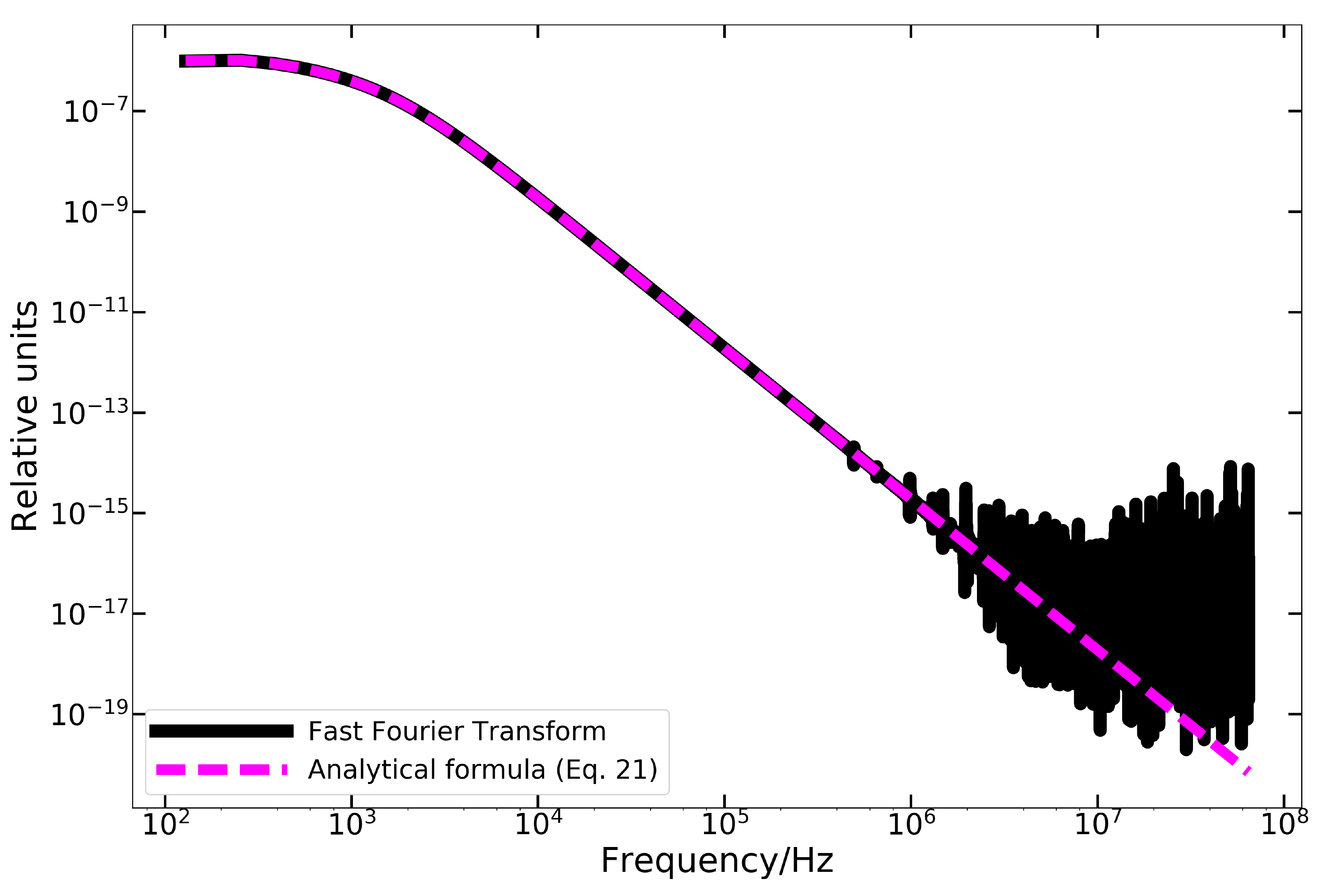}
  \caption{Numerical Fourier transform (black, solid) compared to the analytical Fourier transform (magenta, dashed) of the same electric field function. Used parameters are the same as in Fig. \ref{fig:fpbh} left-hand side panel, black curve. The numerical version (calculated by IDL's \textit{fft} function) carries a forest-like noise at very high frequencies, which is eliminated from our model by using the analytical formula (Eq. \ref{eq:pow_an}).}
  \label{fig:facom}
\end{figure}

The power spectrum of the electric field carries information on the amount of power released at certain frequencies. It quickly reaches its peak then slowly decreases with a power law. The characteristics of the power spectrum (peak frequency, $f_0$, and spectral roll-off, $\frac{dP(f)}{df}$) help predict the amount of power released by lightning in a frequency band, i.e. at a band that is used for observations. Because of the ionosphere and other limitation factors, such as the large distance between the observer and the source, which triggers the contamination of the radio signal by background and foreground sources, only part of the radio lightning spectrum can be observed, and the peak of the emission is often unknown. In our model we generate a radio electric field frequency spectrum and the power spectrum in order to determine the total lightning power released at radio frequencies and the lightning radiated energy at these frequencies. We also estimate $f_0$, and $n$, where possible, which will be important for predictions of future lightning radio observations.

The electric field frequency spectrum, $E(f)$, is the Fourier transform (FT) of the electric field, $E(t)$. Eq. \ref{eq:8}. defines the relation between the electric field in the time domain and the frequency domain,

\begin{equation} \label{eq:8}
E(f) = \int_{0}^{\infty}{e^{-i 2 \pi f t}E(t)dt},
\end{equation}
where $i = \sqrt{-1}$ and $f$ is the frequency in [Hz].

We derive the following analytical form of the FT of the electric field, $E(t)$ given in Eq. \ref{eq:10}:

{\small
\begin{dmath}\label{eq:88}
E(f) = \frac{2 i_0 \v v_0}{4\pi \epsilon_0}\left[\frac{1}{c r^2}\left(\frac{1}{(\alpha+i 2 \pi f)^2}-\frac{1}{(\beta+ i 2 \pi f)^2}\right)+\frac{1}{c^2 r}\left(\frac{\beta}{(\beta+ i 2 \pi f)^2}-\frac{\alpha}{(\alpha + i 2 \pi f)^2}-\frac{1}{\beta + i 2 \pi f}+\frac{1}{\alpha + i 2 \pi f}\right)\right].
\end{dmath}
\normalsize
}

\noindent In practice the FT of a function is usually calculated using the numerical Fast Fourier Transform (FFT) method. For basic functions such as Eq. \ref{eq:4}, it is easy to determine the analytical form of the frequency spectrum, however for more complicated ones, such as the electric field derived from the Heidler-function (Sect. \ref{ssec:heid}), it is easier to use the FFT, which we are going to use in the following calculations\footnote{IDL's inbuilt \textit{fft} function}. Generally, the FT of a function is calculated when the function contains a periodic signal. The challenge in using numerical FFT on the non-periodic electric field of a lightning discharge is, firstly, numerical integrations cannot be made to infinity, which means a part of the function has to be cut out; and secondly, by cutting the function, a sharp edge will appear, which results in a badly calculated FT. To solve this problem the Hann Window function was used which is a technique for signal processing and it smooths the edges of the curve so that the FFT could be calculated properly. Furthermore, an FFT code does not know the time step with which the data are sampled, it assumes that the sample has a step-size of 1 s. This is not true in our case, and to correct for it, we multiply the result of the FFT with our own time-step, which depends on the duration of the discharge.

The magnetic component of the electromagnetic field, $B$, radiated by lightning can be expressed by the electric field as in $B = E/c$. Because of this relation, the magnitude of the magnetic field is much smaller than that of the electric field, therefore traditionally the electric field measured from lightning is used to obtain the radiated power. By squaring the electric field one obtains the power radiated by the field. The frequency-dependent power spectrum, $P(f)$, is the FT of $E(t)^2$: 

\begin{equation} \label{eq:pow}
P(f) = \int_{0}^{\infty}{e^{-i 2 \pi f t}E(t)^2dt}.
\end{equation}

\noindent We obtain the exact form of Eq. \ref{eq:pow} with the help of WolframAlpha\footnote{https://www.wolframalpha.com/}, and with $E(t)$ as in Eq. \ref{eq:10}:

{\small
\begin{dmath} \label{eq:pow_an}
P(f) = \left(\frac{2 i_0 \v v_0}{4 \pi \epsilon_0}\right)^2 \left[\frac{1}{c^2 r^4} \left( \frac{1}{4(\alpha+i \pi f)^3} + \frac{1}{4(\beta+i \pi f)^3} - \frac{2}{(\alpha+\beta+2 i \pi f)^3}\right) + \frac{1}{c^4 r^2} \left(\frac{1}{2(\alpha+i \pi f)} + \frac{1}{2(\beta+i \pi f)} + \frac{\alpha^2}{4(\alpha+i \pi f)^3} + \frac{\beta^2}{4(\beta+i \pi f)^3} + \frac{2(\alpha+\beta)}{(\alpha+\beta+2 i \pi f)^2} - \frac{4 \alpha \beta}{(\alpha+\beta+2 i \pi f)^3} - \frac{2}{\alpha+\beta+2 i \pi f} - \frac{\alpha}{2(\alpha+i \pi f)^2} - \frac{\beta}{2(\beta+i \pi f)^2} \right) + \frac{1}{c^3 r^3} \left(\frac{4 (\alpha+\beta)}{(\alpha+\beta+2 i \pi f)^3} - \frac{4}{(\alpha+\beta+2 i \pi f)^2} - \frac{\alpha}{2(\alpha+i \pi f)^3} - \frac{\beta}{2(\beta+i \pi f)^3} + \frac{1}{2(\alpha+i \pi f)^2} + \frac{1}{2(\beta+i \pi f)^2}\right)\right].
\end{dmath}
}

\noindent
Then, the frequency-dependent power spectral density, $P'(f)$ [W Hz$^{-1}$], at the source of the emission can be expressed by:

\begin{equation} \label{eq:pow2}
P'(f) = P(f) 2 c \pi \epsilon_0 r^2.
\end{equation} 

\noindent Equation \ref{eq:pow2} is the time-averaged power, assuming sinusoidal wave functions\footnote{The time-averaged power, in general, is the intensity, $I$, times the area of a sphere, where the source radiates to: $<P> = I 4 \pi r^2$. The time average of a sinusoidal wave function is $1/2$, resulting in $I = (1/2) c \epsilon_0 E^2$, where $E$ is the radiating electric field. Hence, $P(f)$ in Eq.~\ref{eq:pow2} is multiplied by 2 instead of 4.}. The radiation term in Eq. \ref{eq:pow_an}, i.e. the term proportional to $1/r^2$, dominates at large distances, hence the power spectral density, Eq. \ref{eq:pow2}, is nearly independent of the distance. We solve the complete Eq.~\ref{eq:pow2}. To obtain the total released power in a frequency interval, we integrate $P'(f)$ over the frequency range it has been released at.

Figure \ref{fig:fpbh} shows the power spectra resulting from the electric fields shown in Fig. \ref{fig:ebh}. The distance between the source and observer for all figures is $r = 1000$ km, and the propagation velocity is  $\v v_0 = 8.0 \times 10^7 {\rm m s}^{-1}$. At very high frequencies a forest-like noise is introduced with the numerical FFT. This noise is eliminated from our model by using the analytical form of the FT (Fig. \ref{fig:facom}).

To calculate the discharge energy, we need to know how the power spectrum varies with frequency and where its peak, $f_0$, is. To obtain the slope, $n$, we fit a linear function to the part of the spectrum that is at frequencies larger than the peak frequency. The peak is obtained from the discharge duration $\tau$ (Eq. \ref{eq:taufr}). In the further sections, we use the analytical form of the electric field frequency and power spectra obtained from the bi-exponential function (Eqs. \ref{eq:88}, \ref{eq:pow_an}).

%__________________________________________________________________
\subsection{Radiated discharge energy and radiated power density} \label{sec:disen}

The radiated discharge energy, $W_{\rm rad}$, and the discharge dissipation energy, $W_d$, are calculated in order to estimate how energetic lightning discharges may be on extrasolar objects with different atmospheric conditions (T$_{\rm gas}$, p$_{\rm gas}$, chemical composition, atmospheric extension). By knowing the energy dissipated from lightning, we can estimate the changes in the local chemical composition of the atmosphere and determine whether observable signatures can be produced as a result of the production of non-equilibrium species \citep[e.g.][]{hodosan2016b,2017MNRAS.470..187A} or photometric variabilities \citep{2020MNRAS.495.3881H}. The radiated discharge energy, $W_{\rm rad}$ [J], derives from the total radiated power, $P_{\rm rad}$ [W], released during the discharge as
\begin{equation} \label{eq:nwd}	%nwd == new wd
W_{\rm rad} = P_{\rm rad} \tau,
\end{equation}

\noindent where $\tau$ [s] is the discharge duration.
The total radiated power is given by 
\begin{equation} \label{eq:ptot}
P_{\rm rad} = \int_{f_{\rm min}}^{f_{\rm max}}{P'(f') df'}.
\end{equation}

\noindent The integral boundaries are calculated when the time sample of the model is converted into frequencies. $f_{\rm min} = t_{\rm max}^{-1}$ and $f_{\rm max} = t_{\rm min}^{-1}$, where $t_{\rm max}-t_{\rm min} = \tau$. This enables us to obtain the frequency spectrum of lightning above the peak frequency, starting from $f_0$.

The total dissipation energy of lightning, $W_d$, is derived from the radiation discharge energy $W_{\rm rad}$, assuming a certain radio efficiency, $k$. $k$ represents the amount of energy radiated into the radio frequencies from the total dissipated energy, and is between 0 and 1, 
%\begin{equation} \label{eq:wd2}
$W_d = \frac{1}{k} W_{\rm rad}$.
%\end{equation}

%__________________________________________________________________
%__________________________________________________________________
\section{Approach} \label{sec:param}

\begin{figure}
  \centering
  \includegraphics[width=0.5\columnwidth]{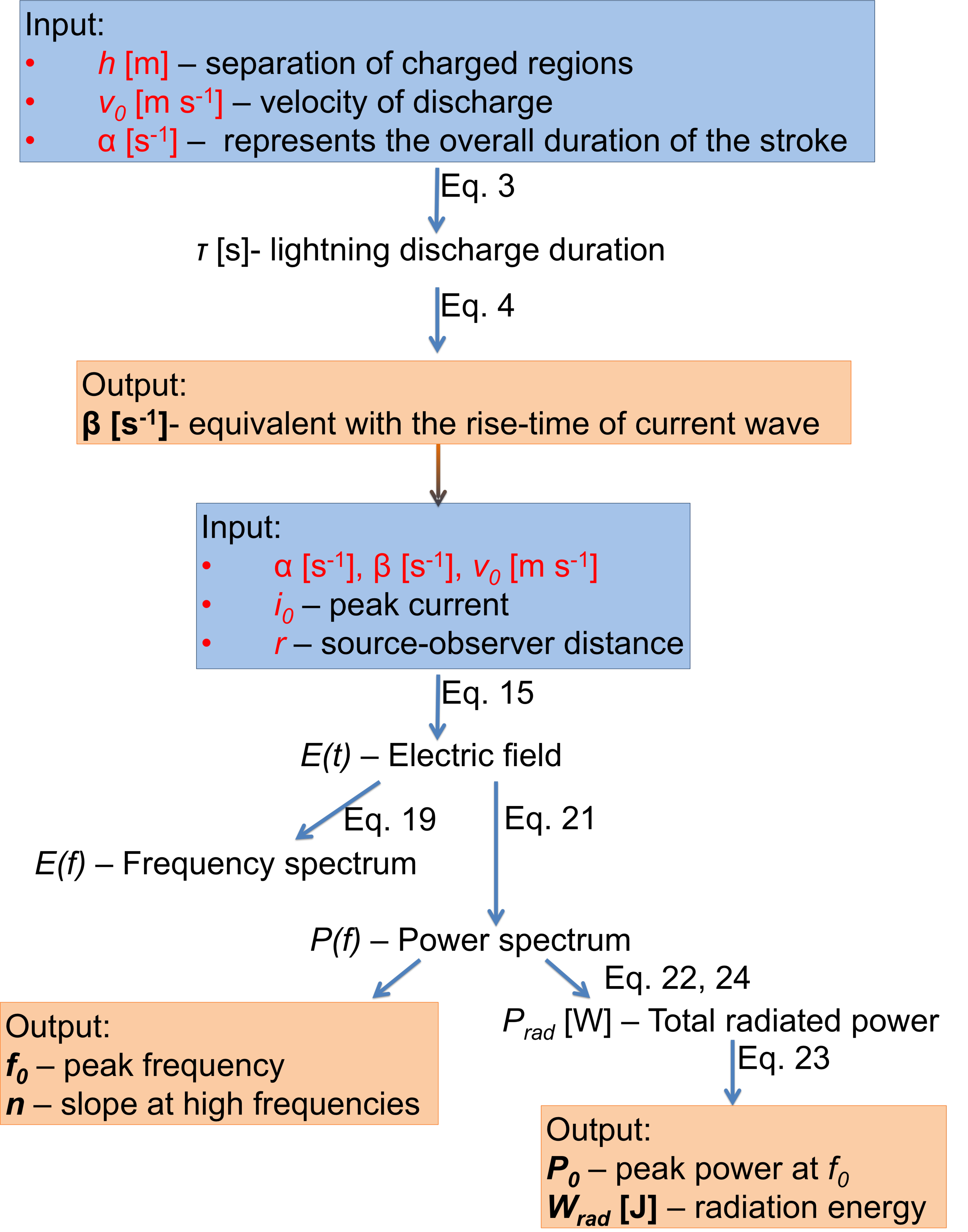}
  \caption{Flow chart for deriving lightning energy and lightning power.}
  \label{fig:chart2}
\end{figure}

Our computational procedure can be summarised as follows (Fig. \ref{fig:chart2}): From a bi-exponential current function (Eq. \ref{eq:1}), the electric field and its frequency and power spectra are calculated. Next, properties of the power emitted at high frequencies are determined: $f_0$ is the frequency at which the peak of the power is emitted; $n$ is the spectral roll-off of the power spectrum at frequencies $f>f_0$. These properties of the power spectrum are necessary when making predictions of the lightning radio power emitted at chosen observed frequencies. Finally, the lightning energy radiated into the radio frequencies is estimated. The following parameters need to be specified:\\
-- The distance between the source and the observer, $r$ (i.e. distance between the extrasolar object and Earth)\\
-- The characteristic length of the charge separation, or length of the discharge, $h$. Ideally, a fully developed streamer-lightning model would be used here. Instead, scaling laws may be used to explore possible discharge length scale in non-terrestrial environments \citep[][]{bailey2014}.\\
-- The velocity of the return stroke, $\v v_0$, was taken from \citet{bruce1941} as $\v v_0 = 0.3 {\rm c} = 8.0 \times 10^7 {\rm m s}^{-1}$ for Earth lightning. Ideally a kinetic, microscopic model would be applied to derive these velocities for the varying chemical compositions on exoplanets/brown dwarfs and their day/night sides. \\
-- The current peak value, $i_0$. We used $i_0=30$ kA as the reference value. It is the average current in a terrestrial return stroke channel.

The flow chart in Fig. \ref{fig:chart2} provides a run through the subsequently used parameters: From the two variables $h$ and $\v v_0$, the discharge duration $\tau$ is determined from Eq. \ref{eq:tau1}. Thereafter, $\alpha$, one of the parameters of the bi-exponential current function (Eq. \ref{eq:1}), is randomly picked from a Gaussian distribution, which has a mean and standard deviation that ensure that $\alpha$ is $\sim$1 orders of magnitude smaller than $\beta$. This is an empirical choice based on literature values (Table \ref{table:3}, top). The mean and standard deviation of the normal distribution are proportional to $\alpha \beta$ (Eq. \ref{eq:tau}). From $\tau$ and $\alpha$, the second  parameter of the current function, $\beta$, is calculated from Eq. \ref{eq:tau}. The next step is to determine the electric field produced by the lightning current from  Eq. \ref{eq:10} (in SI units). The electric field frequency and power spectra are calculated according to Sect.~\ref{sec:freqsp}. The power spectrum represents the distribution of radiated power in frequency space. The total power of lightning radiated at radio frequencies (Eq. \ref{eq:ptot}), which will determine the energy radiated at radio from lightning discharges (Eq. \ref{eq:nwd}), can now be calculated. The results of these calculations are the properties of the power spectrum, $f_0$ and $n$, the duration of the discharge, $\tau$, the peak current, $i_0$, the total radio power, $P_{\rm rad}$, and the radio energy of lightning, $W_{\rm rad}$. Some of the outputs, $f_0$, $n$, will be used to estimate the radiated power density, and hence the observability of lightning radio emission, at a given frequency, where observations are planned. 

In addition to exploring values for Earth, Saturn, and Jupiter in the majority of our study, we wish to link $i_0$ to some physically motivated number of charges. Therefore, we explore values for $i_0$ guided by the minimum number of charges, $Q_{\rm min}$, necessary to overcome the electrostatic breakdown field in extrasolar atmospheres according to \citet[their fig. 7]{bailey2014}. This must be considered as an upper limit for two reasons: Firstly, it does not include the idea of runaway breakdown \citep{rousseldupre2008}, therefore overestimates the critical field strength necessary to initiate a breakdown. This suggests that the obtained $Q_{\rm min}$ for each atmosphere will be an upper limit necessary for breakdown and the actual values in nature may be lower. Secondly, it may be unlikely that all charges that cause the breakdown of an electrostatic field in a thundercloud will be able to discharge through just one lightning channel. It is, however, unknown, which fraction of these charges would be linked to the lightning current for exoplanets or for non-terrestrial Solar System planets. When exploring lightning power for exoplanets / brown dwarfs, we perform tests for which we obtained $i_0$ from $Q_{\rm min}$ from the simplified  Eq. \ref{eq:charge}, $i_0 = \frac{\Delta Q}{\Delta t}$ with $\Delta Q \equiv Q_{\rm min}$ and $\Delta t \equiv \tau$.

%__________________________________________________________________
%__________________________________________________________________
\section[]{Model results for Earth, Jupiter, and Saturn} \label{sec:val}

In this section, we explore our modelling approach (Sects. \ref{sec:model}, \ref{sec:param}) for literature parameters for Earth, Jupiter and Saturn which were derived from observations. We note that these are the three best observed planets with respect to lightning in the Solar System. However, the data for Jupiter and Saturn are very limited, compared to Earth such that parameter studies will be a valuable tool to explore possible ranges of lightning energies outside Earth, and to appreciate possible uncertainties. On Earth, lightning discharges emerge with a whole energy spectrum with a maximum energy of $\sim$$10^{10}$ J, but only the high energy events are observable on Jupiter and Saturn (see Fig.~6 in \citealt{hodosan2016a}).

%__________________________________________________________________
\subsection{Earth} \label{sec:earval}

\begin{figure}
	\resizebox{\textwidth}{!}{
  \includegraphics[width=\columnwidth]{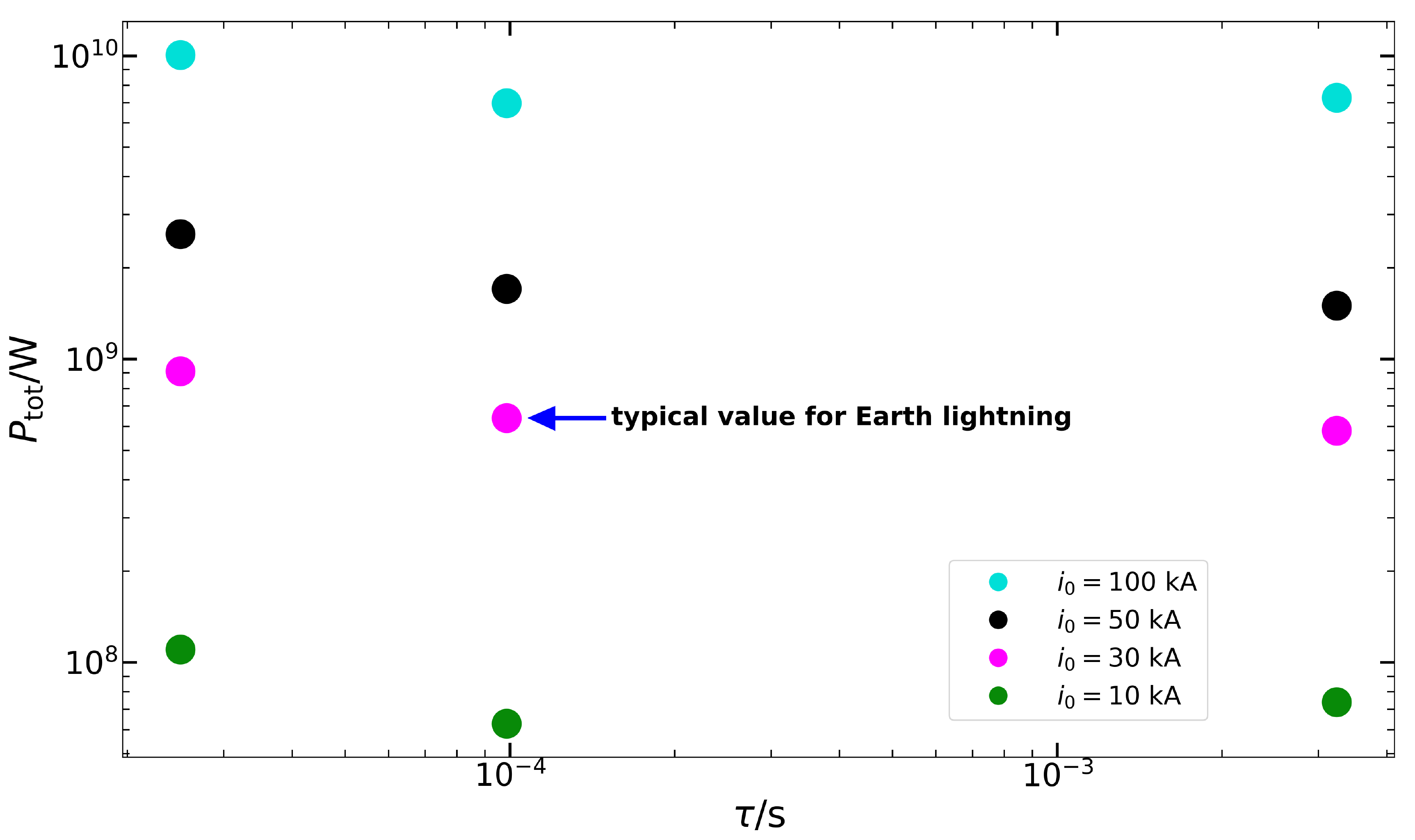}
	\includegraphics[width=\columnwidth]{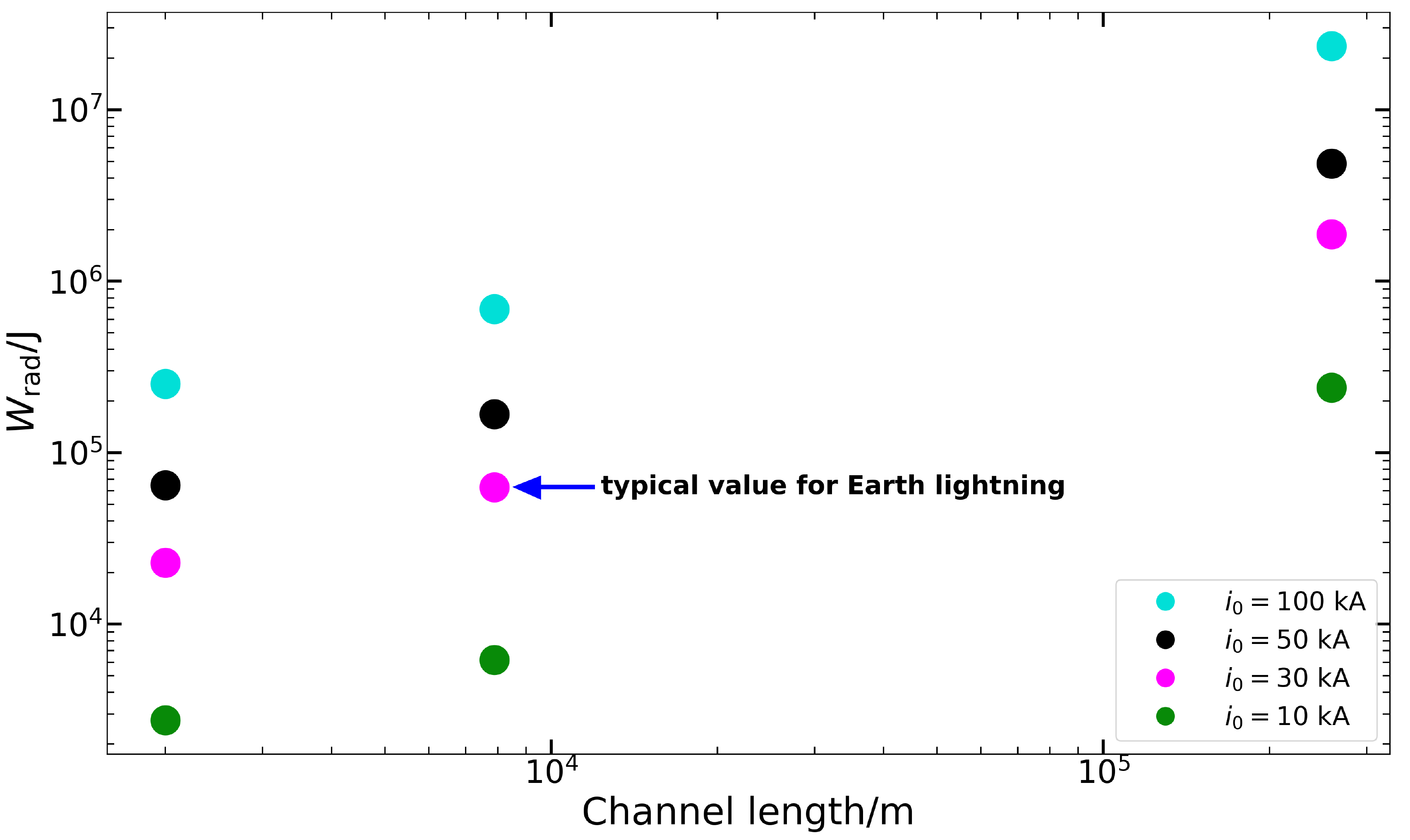}
	}
  \caption{Radio power and radiated energy for terrestrial lightning: \textbf{Left:} Total radio power of lightning, $P_{\rm rad}$ for diffferent discharge durations, $\tau$. \textbf{Right:} Radiated energy, $W_{\rm rad}$ for different discharge length, $h$. The two panels depend on each other through $h$ (Eq. \ref{eq:tau1}). The four colours indicate four different peak currents, $i_0$.  All input parameters are listed in Table \ref{table:yv}. The blue arrow points to a typical value of an Earth lightning stroke, with $\tau = 100$ $\mu$s, and $W_{\rm rad} \sim$$6 \times 10^4$ J \citep[][table 6.2]{volland1984}. Even though $P_{\rm rad}$ does not change significantly with $\tau$ and hence with $h$, $W_{\rm rad}$ will be higher if $h$ increases, and therefore $\tau$ increases (Eq. \ref{eq:nwd}).}
  \label{fig:yv}
\end{figure}

\begin{table}  
 \centering
 \small
 \caption{Earth lightning parameters as in Fig. \ref{fig:yv}. The values are the discharge lengths, $h$, and peak currents, $i_0$, observed and modelled for  Earth.  $\tau$ depends on $h$ through Eq. \ref{eq:tau1},  $\v v_0 = 0.3 {\rm c}$.}
 \vspace{0.3cm}
  \begin{tabular}{@{}lll@{}} 
	\hline
	$h$ [m] & Reference ($h$) & $\tau$ [s] \\
	\hline
	$2 \times 10^3$ & \citet{baba2007} & $2.5 \times 10^{-5}$ \\
	$7.89 \times 10^3$ & \citet[][p. 124]{rakov2003} & $10^{-4}$ \\
	$2.59 \times 10^5$ & \citet{bruning2015} & $3 \times 10^{-3}$ \\
	\hline
  \end{tabular}
	
	\vspace{0.15cm}

  \begin{tabular}{@{}ll@{}} 
	\hline
	$i_0$ [A] & Reference ($i_0$) \\
	\hline
	$10^4$ & arbitrary example \\
  $3 \times 10^4$ & \citet{farrell1999} \\
	\vtop{\hbox{\strut $5 \times 10^4$}\hbox{\strut $10^5$}} & \vtop{\hbox{\strut \citet{heidler2002}}\hbox{\strut arbitrary example}} \\
	\hline
  \end{tabular}
	\label{table:yv}
\end{table}

%Table Validation
\begin{table*}
\caption{Modelling Earth lightning based on the parameter choice in \citet{volland1984} (table 6.1, first row ("$G-R_1$")).}
\begin{adjustbox}{max width=\textwidth}
\begin{threeparttable}  
 \vspace{0.3cm}
  \begin{tabular}{@{}lllllllllll@{}} 
	\hline
	 & $i_0$ [kA] & $\alpha$ [s$^{-1}$] & $\beta$ [s$^{-1}$] & $\tau$ [s] & $h$ [m] & $|Q|$ [C] & $f_0$ [kHz] & $W_d$ [J] & $W_{\rm rad}$ [J] & $k$ \\
	\hline
	\citet[][table 6.2]{volland1984} & 30 & $2 \times 10^4$ & $2 \times 10^5$ & $9.9 \times 10^{-5}$\tnote{(1)} & 7890 & 1.35 & 10.1 & $6.97 \times 10^7$ & $1.46 \times 10^6$\tnote{(2)} & 0.021 \\
	\hdashline
	Model input & 30 & $2 \times 10^4$ & $2 \times 10^5$ & - & - & - & - & - & - & 0.021 \\
	Model output (this work) & - & - & - & $9.93 \times 10^{-5}$ & 7948 & 2.98 & 10.06 & $6.96 \times 10^6$\tnote{(3)} & $1.46 \times 10^5$ & - \\
	\hline
  \end{tabular}
  \begin{tablenotes}
	\item[(1)] $\tau$ is not listed in \citet[][table 6.2]{volland1984}. Applying Eqs \ref{eq:tau1} and \ref{eq:tau} on the values in the first row, the obtained $\tau$ is 98.5 and 99.35 $\mu$s, respectively.
	\item[(2)] $W_{\rm rad}$ obtained from $W_d$ and $k$, which are listed in \citet[][table 6.2]{volland1984}.
	\item[(3)] Calculated from $W_{\rm rad}$ (model output)  through $W_d = \frac{1}{k} W_{\rm rad}$.  
  \end{tablenotes}
\end{threeparttable}
\end{adjustbox}
	\label{table:val}
\end{table*}

Earth lightning is the most well-studied form of lightning discharges in the Solar System. The modelling of Earth lightning goes back to the first half of the $20^{\rm th}$ century \citep[e.g.][]{bruce1941, drabkina1951}, and then further develops in the 1970s--80s \citep[e.g.][and references therein]{uman1969, heidler1985, rakov2003}. Therefore, we apply measured parameters of Earth lightning discharges (Table \ref{table:yv}) as a template to test our combined modelling approach. Figure~\ref{fig:yv} (left) presents the lightning radio power $P_{\rm rad}$ for different {\bf} discharge durations, $\tau$, and for different peak currents $i_0$ (different colours). Figure~\ref{fig:yv} (right) shows how the radiated energy $W_{\rm rad}$ varies with $h$. The figure shows that the larger the peak current the more power and energy is released from a lightning stroke. It also demonstrates that the longer the channel length (i.e. the longer the discharge duration, since $\tau$ and $h$ are directly proportional; see Eq. \ref{eq:tau1}) the more energy is radiated from lightning, even though the released power rather decreases with larger $\tau$. The blue arrow in Fig.~\ref{fig:yv} points out a lightning return stroke with $\tau = 100$ $\mu$s, which corresponds to a peak frequency, $f_0 = 10$ kHz, and $i_0 = 30$ kA. The radio energy of such a stroke is $\sim$$6 \times 10^4$ J, while the radio power is $\sim$$6 \times 10^8$ W. \citet{borovsky1998} estimated the energy dissipated from a return stroke from the electrostatic energy density stored around the lightning channel. They found the dissipation energy per unit length to be $2 \times 10^2$--$10^4$ J m$^{-1}$. They note that their result is in agreement with previous studies considering hydrodynamic models for lightning channel expansion \citep[e.g][]{plooster1971}. \citet{plooster1971} found the total energy of Earth lightning to be $4 \times 10^2$--$9 \times 10^2$ J m$^{-1}$ for $i_0=20$ kA, and $1.7 \times 10^3$ J m$^{-1}$ for $i_0 = 40$ kA. \citet{borovsky1998}, however, pointed out that their results are 1-2 orders of magnitude lower than calculated by e.g. \citet{krider1968}, who estimated the total dissipation energy from optical measurements assuming an optical efficiency of 0.38 to be $2.3 \times 10^5$ J m$^{-1}$. Assuming a channel length of $h = 7890$ m, as for the data point marked by a blue arrow in Fig. \ref{fig:yv}, the dissipated energy of lightning, $W_d$, according to the above authors is between $1.6 \times 10^6$ J and $1.8 \times 10^9$ J. Applying  a radio efficiency $k=0.01$, the radiated energy in the radio band is between $1.6 \times 10^4$ J and $1.8 \times 10^7$ J. Our value of $\sim$$6 \times 10^4$ J is within this range, however, closer to the values obtained by \citet{borovsky1998} and \citet{plooster1971}.

We also  tested our modelling  approach for values from \citet{volland1984} (their table 6.2). We chose the first row of the table ("$G-R_1$"), which was derived from parameters used in \citet{bruce1941}. It represents a first return stroke of a lightning discharge. We list the parameters and the results of our model in Table \ref{table:val}. The table suggest that our results of discharge energy are approximately one order of magnitude lower than the ones obtained by \citet{volland1984}.

%__________________________________________________________________
\subsection{Saturn} \label{sec:satval}

%Table Validation - Saturn
\begin{table*} 
\caption{Saturnian lightning parameters. Two sets of tests were performed: \textbf{Case (1)} Modelling a quick discharge, $\tau_{\rm stroke} = 1$ $\mu$s, as in \citet{farrell2007}, and \textbf{Case (2)} Reproducing the measured energy of SEDs and the shape of their power spectra, with $\tau_{\rm stroke} = 100$ $\mu$s as in \citet{mylostna2013}. The extension of the discharge, $h$, was obtained from Eq. \ref{eq:tau1} with $\v v=0.3$ c. Input values are marked in  italics in the table.}	
\begin{adjustbox}{max width=\textwidth}
\begin{threeparttable}  
 \vspace{0.3cm}
  \begin{tabular}{@{}llllllllllll@{}}
	\hline
	 & $\tau_{\rm stroke}$ [$\mu$s] & $\tau_{\rm SED}$ [s] & $i_0$ [kA] & $|Q|$ [C] & $P_{\rm rad}$ [W]\tnote{(1)} & $W_{\rm rad}$ [J]\tnote{(1)} & $W_{\rm SED}$ [J] & $W_d$ [J]\tnote{(2)} & $k$ & \vtop{\hbox{\strut stroke/}\hbox{\strut SED}} & \vtop{\hbox{\strut SED/}\hbox{\strut flash}} \\
	\hline
	(1a) & \textit{1} & $\mathit{10^{-6}}$ & \textit{30} & 0.03 & $6.4 \times 10^8$ & $6.4 \times 10^2$ & $6.4 \times 10^2$ & $6.4 \times 10^5$ & \textit{0.001} & \textit{1} & \textit{1} \\
	(1b) & \textit{1} & $\mathit{10^{-6}}$ & 35000 & 35 & $10^{15}$ & $10^9$ & $10^9$ & $\mathit{10^{12}}$ & \textit{0.001} & \textit{1} & \textit{1} \\
	 & \textit{1} & \textit{0.23} & 75 & 0.075 & $4.5 \times 10^9$ & $4.5 \times 10^3$ & $10^9$ & $\mathit{10^{12}}$ & \textit{0.001} & $2.3 \times 10^5$ & \textit{1} \\
	\hdashline
	(2) & \textit{100} & \textit{0.23} & 75 & 7.5 & $4.5 \times 10^9$ & $4.5 \times 10^5$ & $10^9$ & $\mathit{10^{12}}$ & \textit{0.001} & $2.3 \times 10^3$ & \textit{1} \\
	 & \textit{100} & \textit{0.035} & 135 & 13.5 & $1.6 \times 10^{10}$ & $1.6 \times 10^6$ & $5.5 \times 10^8$ & $\mathit{1.1 \times 10^{12}}$ & \textit{0.001} & 350 & \textit{2} \\
	\hline
  \end{tabular}
  \begin{tablenotes}
	\item[(1)] Here $P_{\rm rad}$ represents stroke power, and $W_{\rm rad}$ represents the stroke energy.
	\item[(2)] $W_d$ is the total dissipation energy of a lightning flash.
  \end{tablenotes}
 \end{threeparttable}
 \end{adjustbox}
   \label{table:val3}
\end{table*}

\begin{table}
	\begin{center}
	\caption{Stroke durations, $\tau$, for testing Saturnian discharges, and the resulting parameters that depend on $\tau$ (Eqs.~\ref{eq:tau1}, \ref{eq:tau}, \ref{eq:taufr}). These values are outputs from the model as described in Sect. \ref{sec:satval} and Table \ref{table:val3}.}
 \vspace{0.3cm}
  \begin{tabular}{@{}llllll@{}}
  \hline
   & $\tau_{\rm stroke}$ [$\mu$s] & $\alpha$ [s$^{-1}$] & $\beta$ [s$^{-1}$] & $h$ [m] & $f_0$ [kHz] \\
	 \hline
  (1) & \textit{1} & $3.5 \times 10^6$ & $1.1 \times 10^7$ & 90 & 1000 \\
	(2) & \textit{100} & $3.3 \times 10^4$ & $1.2 \times 10^5$ & 9000 & 10 \\
  \hline
  \end{tabular}
	\label{table:satval_2}
	\end{center}
\end{table}

Saturn Electrostatic Discharges (SED\footnote{Broadband lightning radio emission on Saturn.}) have been observed since \textit{Voyager 1} and \textit{2} passed by the planet \citep{warwick1981,zarka1983}. Apart from Earth, the largest amount of knowledge we have about lightning radio emission is of Saturn. The measured SED spectrum shows a relatively flat part below ~10 MHz, and it becomes a bit steeper till 40 MHz (\textit{Voyager} PRA (Planetary Radio Astronomy) cut-off limit) with a slope of $f^{-1}$ to $f^{-2}$ \citep{zarka1983,zarka2004}. \citet{warwick1981} deduced the shortest time structure of SEDs to be 140 $\mu$s, while \citet{zarka1983} measured a burst duration of 30 to 450 ms from \textit{Voyager} data. \textit{Cassini} data showed a slight roll-off of $f^{-0.5}$ of the spectrum at the range of 4--16 MHz, with power spectral density of 40 to 220 W Hz$^{-1}$, and burst duration of 15 to 450 ms \citep{fischer2006}. The peak frequency of SED emission cannot be determined from the data, which means it is below the ionospheric cut-off. Assuming an Earth-like discharge, with peak frequencies around 10 kHz, to reproduce the measured power densities \citep[on average $\sim$60 W Hz$^{-1}$, ][]{zarka2004}, a very strong discharge is needed, with energies of the order of $10^{13}$ J. Therefore, \citet{farrell2007} suggested a much faster discharge, which would result in a peak frequency at $\sim$1 MHz, and would shift the whole power spectrum to higher frequencies. The result of this would be that discharges less energetic than previously estimated \citep[$\sim$$10^6$ J,][]{farrell2007} could produce the observed power density. However, this theory was excluded by the first optical detections of lightning on Saturn \citep{dyudina2010}. \citet{dyudina2010} measured an optical energy of $10^9$ J of a single lightning flash, while \citet{dyudina2013} obtained optical energies between $10^8$--$10^9$ J, both suggesting that the total energy of a lightning flash is of the order of $10^{12}$ J, assuming that 0.1\% of the total energy of lightning is radiated in the optical \citep{borucki1987}. \citet{mylostna2013} observed SEDs with the Ukrainian T-shaped Radio telescope (UTR-2), and mapped their temporal structure. They found that the finest structure observable was 100 $\mu$s short, deduced a spectral roll-off of $f^{-2}$ between 20 kHz and 200 kHz, and a peak frequency $f_0 = 17$ kHz. Their findings further support the super-bolt scenario of Saturnian lightning flashes.

We conducted two sets of calculations for Saturnian lightning (Tabel~\ref{table:val3}): Saturn case (1): Modelling a quick discharge, $\tau_{\rm stroke} = 1$ $\mu$s, as in \citet{farrell2007}, and (2) reproducing the measured total energy of a flash \citep[e.g.][]{dyudina2010,dyudina2013,mylostna2013} and the shape of SED power spectra \citep[e.g.][]{zarka1983,fischer2006}, with $\tau_{\rm stroke} = 100$ $\mu$s as in \citet{mylostna2013}. Saturn case (1a) is a test of the model with an Earth-like current peak \citep[$i_0 = 30$ kA,][]{volland1984}, while Saturn case (1b) takes into account the obtained total energy from optical measurements \citep[$W_d = 10^{12}$ J][]{dyudina2013}, and iterates the corresponding peak current (increasing $i_0$ by 5000 kA during each step). Saturn case (1b) also considers an SED built up of several strokes (second row of case (1b) in Table \ref{table:val3}). Saturn case (2) consists of various cases depending on how many strokes/SED and SED/flash are used, and each time it iterates the current peak (with 15 kA during each step) to match $W_d = 10^{12}$ J. In each case, the distance, $r$, was taken to be the distance between the \textit{Cassini} spacecraft and Saturn during the measurements, $r = 46.5$ $R_{\rm Sat}$ \citep{fischer2007}. We assume that the radio efficiency, $k$, is the same as the optical one determined for Jupiter by \citet{borucki1987}. Each time the steps described in Sect. \ref{sec:param} were followed (Fig. \ref{fig:chart2}), except the starting input parameter was $\tau$ and not $h$. Furthermore, we consider cases when the duration of an SED burst is not equal to the duration of a stroke or a flash. $\tau_{\rm SED} = 0.23$ s is the average value in \citet{fischer2006} for an SED burst duration, while $\tau_{\rm SED} = 0.035$ s is the duration \citet{dyudina2013} considered for their energy estimates. In this case, \citet{dyudina2013} also assumed that one flash has a duration of 70 ms, therefore, one flash consisting of two SEDs. The "stroke/SED" and "SED/flash" values listed in Table \ref{table:val3} are a direct result of the time considerations.

Table \ref{table:val3} lists the input parameters and the results of the tests. Case (1) shows what a quick discharge would look like on Saturn. First, we assume that nothing else is known about SEDs than what was considered in \citet{farrell2007}, we also assume that one flash consists of one SED, which consists of one stroke. Such a discharge would release $6.4 \times 10^2$ J energy in the radio band, and dissipates $6.4 \times 10^5$ energy in total. Next, we apply the total dissipation energy previously obtained and confirmed by optical measurements, $W_d = 10^{12}$ J, to the quick discharge theory, and find that, an incredibly large, 35000 kA current is necessary to produce such an energy from one flash consisting of one SED made up of one stroke. Finally, it is known that SED bursts are longer in duration than 1 $\mu$s. Here, we apply the average SED duration found in \citet{fischer2006}, $\tau_{\rm SED} = 0.23$ s. We find that 75 kA current has to run through one stroke, to produce $4.5 \times 10^3$ J radio energy, and $2.3 \times 10^5$ strokes are needed to produce $W_d = 10^{12}$ J energy of a flash consisting of one SED with the duration of $\tau_{\rm SED} = 0.23$ s.

\begin{figure}
  \centering
  \includegraphics[width=0.7\columnwidth]{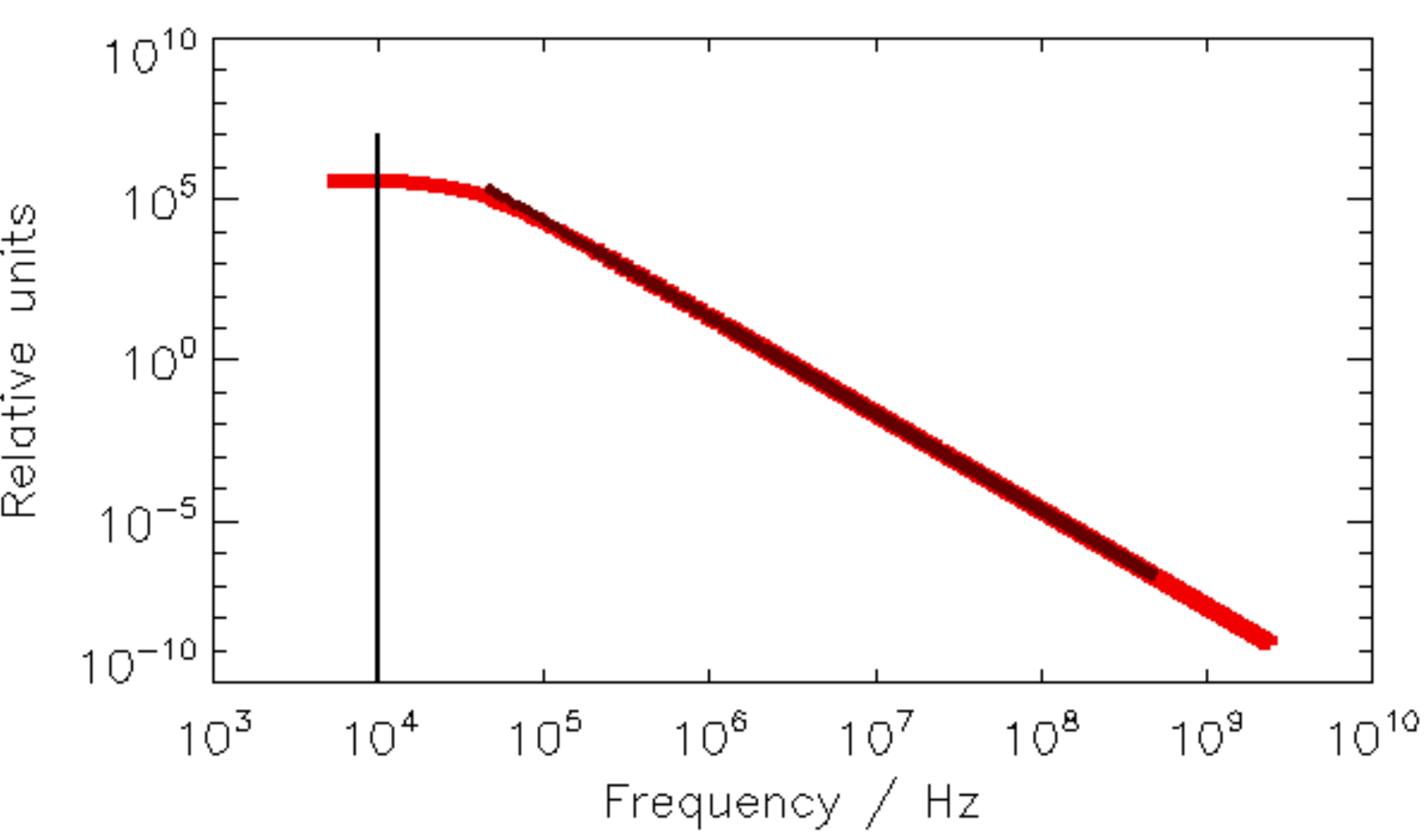} 
  \caption{Saturnian lightning electric field power spectrum (red). The fitted black line varies with frequency as $f^{-3}$, which is the slope of the power spectrum. The vertical black line indicates the peak frequency, $f_0 = 10$ kHz. The figure is based on the parameters listed in the second row of Case (2) in Tables \ref{table:val3}, \ref{table:satval_2}.}
  \label{fig:satshape}
\end{figure}

Saturn Case (2) uses more realistic values for SED and stroke duration, based on previous measurements. The stroke duration was assumed to be $\tau_{\rm stroke} = 100$ $\mu$s in each case, after \citet{mylostna2013}. First, we considered an SED of $\tau_{\rm SED} = 0.23$ s \citep[average in][]{fischer2006}, and one SED in one flash. The results suggest that 75 kA peak current would be enough to produce a dissipation energy of $10^{12}$ J, with 2300 strokes/SED with $W_{\rm rad} = 4.5 \times 10^5$ J/stroke. Second, we assume that $\tau_{\rm SED} = 0.035$ s, and one flash consists of two SED bursts. We find that 350 strokes/SED would produce enough energy, with $i_0 = 135$ kA to account for the total dissipation energy of a flash, $W_d = 10^{12}$ J. In Table \ref{table:satval_2} we further see that a quick, $\tau = 1$ $\mu$s, discharge would propagate  90 m, while a slower discharge of $\tau = 100$ $\mu$s, would result in an extension, $h = 9$ km, similar to Earth discharges. It is unknown how far Saturnian lightning can propagate.

Finally, we tested our modelling approach for the shape of the observed power spectrum. Figure \ref{fig:satshape} shows a representative curve. All other curves show the same shape, but with a shifted peak depending on the discharge duration. This suggests that our model, which is based on a simple, vertical dipole radiation model, with no branches and tortuosity of the channel, does not reproduce the shape of the power spectrum well. The resulting power spectrum shows a slope of $f^{-3}$ ($n = 3$) at high frequencies, no matter what the input parameters are. Since for Saturn, a flatter spectrum has been observed, our results may underestimate the power released at these frequencies. To obtain the same discharge energy, one has to increase the peak current if the modelled power spectrum is steeper than the actual one, and therefore, our model most probably overestimates the peak current obtained for each case in this section. We note that the modelling parameterisations used here and in the literature were developed for Earth lightning originally.

%__________________________________________________________________
\subsection{Jupiter} \label{sec:jupval}

%Table Validation - Jupiter
\begin{table*}  
	\caption{Jovian lightning parameters. \textbf{Case (i)} Reproduce values obtained from the Galileo probe by \citet{rinnert1998} with $\v v=0.1$c. \textbf{Case (ii)} The discharge duration, $\tau$, is the average of the duration interval (260$-$520 $\mu$s) given by \citet{rinnert1998}, and the radio efficiency is taken to be the same as the optical efficiency determined by \citet{borucki1987}. The difference between (ii,a) and (ii,b) is the used radio efficiency. The peak current is iterated so that the final dissipation energy would be at least $10^{12}$ J.}
 \begin{adjustbox}{max width=\textwidth}
 \begin{threeparttable}
 \setlength{\tabcolsep}{5pt} 
 \vspace{0.3cm}
  \begin{tabular}{@{}llllllllll@{}}
	\hline
   & $\tau$ [$\mu$s] & $i_0$ [kA] & $|Q|$ [C] & $M(t)$ [C m] & $h$ [m] & $P_{\rm rad}$ [W] & $W_{\rm rad}$ [J] & $W_d$ [J] & $k$ \\
	\hline
	(i), \citet{rinnert1998} & 240 & $6 \times 10^3$ & 1500 & $10^7$ & 7000 & $10^{14}$ & $2.5 \times 10^{10}$ & $10^{12}$ & 0.025 \\
	Model input & 240 & - & - & - & - & - & - & $10^{12}$ & 0.025 \\
	Model output (this work) & - & $2.5 \times 10^4$ & 5930 & $4.3 \times 10^7$ & 7190 & $1.07 \times 10^{14}$ & $2.57 \times 10^{10}$ & $1.03 \times 10^{12}$\tnote{(1)} & - \\
	\hdashline
	(ii,a), Model input & 390\tnote{(2)} & - & - & - & - & - & - & $10^{12}$ & 0.001\tnote{(3)} \\
	Model output (this work) & - & $2.8 \times 10^3$ & 1100 & $3.87 \times 10^7$ & $3.5 \times 10^4$ & $2.9 \times 10^{12}$ & $1.13 \times 10^9$ & $1.13 \times 10^{12}$\tnote{(1)} & -  \\
	(ii,b), Model input & 390\tnote{(2)} & - & - & - & - & - & - & $10^{12}$ & 0.025 \\
	Model output (this work) & - & $1.24 \times 10^4$ & 4850 & $1.7 \times 10^8$ & $3.5 \times 10^4$ & $6.6 \times 10^{13}$ & $2.57 \times 10^{10}$ & $1.03 \times 10^{12}$\tnote{(1)} & -  \\
	\hline
  \end{tabular}
  \begin{tablenotes}
	\item[(1)] Final $W_d$, the result of the last iteration of $i_0$
 	\item[(2)] Average duration of 260 $\mu$s and 520 $\mu$s from \citet{rinnert1998}.
	\item[(3)] Optical efficiency from \citet{borucki1987}. We assume that the radio efficiency is the same.
  \end{tablenotes}
 \end{threeparttable}
 \end{adjustbox}
 	\label{table:val2}
\end{table*}

%Table Validation - Jupiter
\begin{table*}
 \caption{Jovian lightning parameters. \textbf{Case (iii):} Testing  Jovian values  based on the modelling approach of \citet{farrell1999} ($\alpha = 1.5 \times 10^3$ s$^{-1}$, $\beta = 1.75 \times 10^3$ s$^{-1}$ \citealt{farrell1999}; $W_d = 10^{12}$ \citealt{rinnert1998}; $k = 0.001$). \citet{farrell1999} modelled Jovian discharges with peak frequency, $f_0 = 500$ Hz, and duration of 1 to 2 ms.}
 \vspace{0.3cm}
  \resizebox{\textwidth}{!}{
  \begin{tabular}{@{}lllllllllll@{}}
	\hline
   & $\tau$ [$\mu$s] & $i_0$ [kA] & $|Q|$ [C] & $M(t)$ [C m] & $h$ [m] & $P_{\rm rad}$ [W] & $W_{\rm rad}$ [J] & $W_d$ [J] & $k$ & $f_0$ [Hz] \\
	\hline
  (iii) & 3800 & $3.2 \times 10^3$ & $1.25 \times 10^4$ & $4.37 \times 10^9$ & $3.5 \times 10^5$ & $2.67 \times 10^{11}$ & $10^9$ & $10^{12}$ & 0.001 & 257 \\
	\hline
  \end{tabular}
	\label{table:val2_2}
	}
\end{table*} 

Jovian lightning was observed by several spacecraft both in the optical and radio bands \citep[e.g.][]{cook1979,borucki1982,borucki1992,rinnert1998,little1999,baines2007}. The \textit{Voyagers}, \textit{Galileo}, \textit{Cassini}, and \textit{New Horizons} all measured the average optical power of lightning on Jupiter to be $\sim$$10^9$ J, with values between $3.4 \times 10^8$ J \citep{baines2007} and $2.5 \times 10^{10}$ J \citep{rinnert1998}. \citet{borucki1987} determined from laboratory experiments that the optical efficiency of lightning on Jupiter is 0.001. \citet{rinnert1998} estimated both the radio energy and total dissipation energy from data gathered by the \textit{Galileo} probe during its descent into Jupiter's atmosphere. Their results suggest that the radio efficiency is 0.025, with $W_{\rm rad} = 2.5 \times 10^{10}$ J, and $W_d = 10^{12}$ J. The data of the probe provide us with valuable information on the radio spectrum of lightning on this Solar System gas giant. \citet{rinnert1998} obtained pulse durations between 266 and 522 $\mu$s, with inter-pulse gaps between 680 $\mu$s and 1 s. Such slow discharges have their peak power radiated at $\sim$500 Hz \citep{farrell1999}. \citet{rinnert1998} estimated several properties of discharges on Jupiter, which we use as comparison for our model results (see Table \ref{table:val2}). We consider three cases (Tables~\ref{table:val2},~\ref{table:val2_2}). In the Jovian case (i), we test our modelling approach against the example  in \citet{rinnert1998}, who deduced lightning parameters from \textit{Galileo} probe data. For the sake of comparison,  we use the same $\v v=0.1$c as in \citet{rinnert1998}. In the Jovian case (ii), we apply information of the duration of the discharge measured by the \textit{Galileo} probe \citep{rinnert1998}, and experimental results from \citet{borucki1987}, who estimated the optical efficiency of lightning on Jupiter. Here, we assume this efficiency is the same for radio emission. Results for the Jovian case (iii) are given in Table~\ref{table:val2_2} when we ran the model with input parameters from \citet{farrell1999}. For each case, the peak current was iterated to reach a dissipation energy that is at least $W_d = 10^{12}$ J, at $r = 1000$ km \citep[like in][]{farrell1999}.

The results of Jovian case (i) (Table~\ref{table:val2}) show that to reach the desired dissipation energy, $W_d = 10^{12}$ J, our model requires $\sim 4$ times more charges and more peak current in the channel, than what was estimated by \citet{rinnert1998}. This suggests that our model underestimates the released power, and the shape of the electric field power spectrum is flatter than what we obtain. Jupiter cases  (ii,a) and (ii,b) (Table \ref{table:val2}) illustrate the importance of radio efficiency, $k$. When $k$ is lower, a lower amount of radio energy is necessary to obtain the required total dissipation energy, which results in lower number of necessary charges and amount of peak current in the channel. Jupiter cases (iii) has a ten times slower discharge for Jupiter than cases (i) and (ii). Though the necessary peak current to obtain $W_d = 10^{12}$ J is not much higher than, e.g., in {\bf} Jupiter case (ii,a), the resulting charges and charge moment are orders of magnitude larger. This is because a ten times slower discharge will create a ten times longer discharge channel with the same velocity, resulting in very large $Q$ and $M(t)$ values.

%__________________________________________________________________
\subsection{Evaluating the limits of an Earth focused lightning approach} \label{sec:5_eval}

The combination of fundamental physics in form of a dipole model and the necessary parameterisations of current functions, discharge times and discharge lengths are the base of the lightning modelling approach that has been applied so far in this paper. This approach has originally been developed and tested for lightning in the Earth atmosphere for which large amounts of detailed data are available. We have demonstrated that this approach consequently  works best for lightning parameters that were determined for Earth, and that larger uncertainties occur when applied in comparison to the few detections available for Jupiter and Saturn. Evaluating these uncertainties is of interest as we wish to apply the same modelling approach to exoplanets and brown dwarfs, astronomical objects for which it will be even harder to conduct lightning observations in the classical Solar System sense. Using these Solar System data enables us to evaluate the limits of {\bf} Earth-focused lightning modelling approach, and hence, the uncertainties that our calculations for exoplanets will carry. We conclude that the shape of the Saturn/Jupiter power spectrum cannot be well reproduced with the present modelling approach, but we caution that the power spectrum can only be observed for a limited frequency range and that the peak is often not resolved. In exploring this modelling approach for un-tested parameter combinations that may represent extrasolar lightning, we observe the following points of caution with respect to the range of parameters that was tested for Earth, Jupiter and Saturn:

\begin{itemize}
	 \setlength\itemsep{1.0em}
		\item The tests for Earth lightning suggest that we underestimate the energy by one order of magnitude. The power spectrum has a slope of $f^{-3}$, instead of the observed $-2$ and $-4$. We suggest that the overall shape of the electric field power spectrum of lightning is relatively steeper than the observed one, and therefore, results in lower amount of calculated power and energy.
		\item The tests for Saturn suggest that Saturnian discharges are indeed super-bolt-like discharges, with peak currents around 70$-$130 kA. Our modelling approach gives an overall shape of the electric field power spectrum of lightning that is steeper than the observed one, and therefore results in a lower amount of calculated radio power. Hence, our model likely overestimates the necessary current in the channel to produce an observed discharge dissipation energy of $W_d = 10^{12}$ J.
		\item The tests for Jupiter help to constrain differences to observed Jovian and Saturnian-like discharges. The measurements of the \textit{Galileo} probe \citep{rinnert1998,lanzerotti1996} provide us with valuable information on the behaviour of lightning radio emission on Jupiter. It seems, the electric field power spectrum of these discharges \citep[from $f^{-1.5}$ to $f^{2}$,][]{farrell1999} is flatter than what is known from Earth \citep[from $f^{-2}$ to $f^{-4}$,][]{rakov2003}, but not as flat as Saturnian spectra \citep[from $\sim$$f^{-0.5}$ to $f^{-2}$,][]{fischer2006,mylostna2013}. Our results  overestimate the necessary peak current to reach $W_d = 10^{12}$ J by a factor of 4. Hence, the produced lightning power is underestimated due to a modelled power spectrum that is steeper ($f^{-3}$) than observations suggest.  
\end{itemize}

In conclusion, the modelling approach used underestimates the released power and energy by a factor of four to ten. We will consider this as source of uncertainty when we discuss our results of exoplanetary lightning modelling.

%Table h + atmospheres
\begin{table*}
 \caption{The breakdown and lightning characteristic for exoplanet and brown dwarf atmospheres from  \citet[][Figs. 7, 9]{bailey2014}. We use $\v v_0=0.3$c.}
 \vspace{0.3cm}
  \resizebox{\textwidth}{!}{
  \begin{tabular}{@{}llllll|lllll@{}}	
	\hline
	\multicolumn{6}{c|}{solar metalicity ([M/H] = 0.0)} &\multicolumn{4}{c}{sub-solar metallicity ([M/H] = -3.0)} \\
	 & T$_{\rm eff}$ [K] & $h_1$ [m] & $\tau_1$ [s] & $Q_{\rm min,1}$ [C] & $i_{0,1}$ [A] & $h_2$ [m] & $\tau_2$ [s] & $Q_{\rm min,2}$ [C] & $i_{0,2}$ [A] \\ 
	\hline
	\multirow{4}{*}{\vtop{\hbox{\strut Brown dwarf}\hbox{\strut log($g$) = 5.0}}} & 1500 & 168 & $2.1 \times 10^{-6}$ & 70 & $3.3 \times 10^7$ & 890 & $1.1 \times 10^{-5}$ & 312 & $2.8 \times 10^7$ \\
	 & 1600 & 66 & $8.2 \times 10^{-7}$ & 33 & $4.0 \times 10^7$ & 753 & $9.4 \times 10^{-6}$ & 237 & $2.5 \times 10^7$ \\	
	 & 1800 & 58 & $7.3 \times 10^{-7}$ & 22 & $3.0 \times 10^7$ & 623 & $7.8 \times 10^{-6}$ & 216 & $2.8 \times 10^7$ \\
	 & 2000 & 27 & $3.3 \times 10^{-7}$ & 12 & $3.6 \times 10^7$ & 286 & $3.6 \times 10^{-6}$ & 80 & $2.2 \times 10^7$ \\	
	\hline
  \end{tabular}
	}
	
	\vspace{0.3cm}

	\resizebox{\textwidth}{!}{
  \begin{tabular}{@{}llllll|llll@{}}	
	\hline
	\multicolumn{6}{c|}{solar metalicity ([M/H] = 0.0)} &\multicolumn{4}{c}{sub-solar metallicity ([M/H] = -3.0)} \\
	 & T$_{\rm eff}$ [K] & $h_1$ [m] & $\tau_1$ [s] & $Q_{\rm min,1}$ [C] & $i_{0,1}$ [A] & $h_2$ [m] & $\tau_2$ [s] & $Q_{\rm min,2}$ [C] & $i_{0,2}$ [A] \\ 
	\hline
	\multirow{4}{*}{\vtop{\hbox{\strut Giant gas planet}\hbox{\strut log($g$) = 3.0}}} & 1500 & 69 & $8.6 \times 10^{-7}$ & $3.8 \times 10^3$ & $4.5 \times 10^9$ & 2494 & $3.1 \times 10^{-5}$ & $1.0 \times 10^5$ & $3.2 \times 10^9$ \\
	 & 1600 & 20 & $2.5 \times 10^{-7}$ & $1.8 \times 10^3$ & $7.5 \times 10^9$ & 2370 & $3.0 \times 10^{-5}$ & $8.2 \times 10^4$ & $2.8 \times 10^9$ \\	
	 & 1800 & 19 & $2.3 \times 10^{-7}$ & $1.7 \times 10^3$ & $7.2 \times 10^9$ & 1844 & $2.3 \times 10^{-5}$ & $6.2 \times 10^4$ & $2.7 \times 10^9$ \\
	 & 2000 & 15 & $1.9 \times 10^{-7}$ & $1.1 \times 10^3$ & $5.6 \times 10^9$ & 942 & $1.2 \times 10^{-5}$ & $3.4 \times 10^4$ & $2.9 \times 10^9$ \\	
	\hline
  \end{tabular}
	\label{table:h}
	}
\end{table*}
%Table 

%__________________________________________________________________
%__________________________________________________________________
\section{Exploring  the lightning power for atmospheres of exoplanets and brown dwarfs} \label{sec:resdis}

Lightning in extrasolar atmospheres is largely unexplored. Firstly, because of the challenges to detect lightning in Solar System objects other than Earth, Jupiter and Saturn (see \citealt{2020SSRv..216...26A}), and secondly because of the immense modelling effort required to derive observable quantities from first principles. Most modelling efforts are directed to Earth lightning (see \citealt{helling2016,2018JGRD..123.7615P,2020arXiv200514588N}), and parameterisations for the lightning frequency, channel lengths and volume, and sometimes also chemical fluxes are applied to study the effect of lightning on the Earth atmosphere chemistry (e.g., \citealt{bruning2015,2019E&SS....6.2317G}). More recently, \cite{2019Icar..333..294K} addressed the inset of lightning by modelling streamers for a gas-mixture representative of Titan's atmosphere. Observations of lightning on exoplanets have been attempted in the radio with LOFAR \citep{2019A&A...624A..40T}, and through photometric variability in the optical with the Danish telescope \citep{2020MNRAS.495.3881H}.  \citet{2019A&A...624A..40T} conclude that exoplanet lightning needs to be $10^5\times$ more powerful in the radio spectral range than Jupiter at a distance of 5 pc in order to be observable by LOFAR.
 
In this paper, we use a lightning dipole model, including current wave functions, which has been developed for lightning on Earth. We, therefore, tested our modelling  approach in Sect. \ref{sec:val} for known parameter combination from Earth, and for data from Jupiter, and Saturn. These tests suggest that the model underestimates the released power and energy by a factor of four to ten. This is because the shape of the electric field power spectrum does not change during our modelling approach, as we do not include the effects of channel tortuosity and branching. Further uncertainties will be linked to the values of the peak current, $i_0$, the discharge extension, $h$, and associated time scales, $\tau$, which are treated as parameters and not derived from first principles nor are they derived consistently in the literature. Changing $\tau$, the discharge duration, the resulting power spectrum shifts, and its peak closely follows the expression in Eq. \ref{eq:taufr}. We study the effects of these input parameters on the energy release and radio power output of lightning discharges. Figure~\ref{fig:yv} demonstrates that the larger the peak current the more power and energy are released from a lightning stroke, and that the slower the discharge the more energy is radiated from lightning, even though the power released rather slightly decreases with larger $\tau$.

\begin{figure*}
	\resizebox{\textwidth}{!}{
  \includegraphics[width=\columnwidth, trim=0cm 0cm 0cm 0cm]{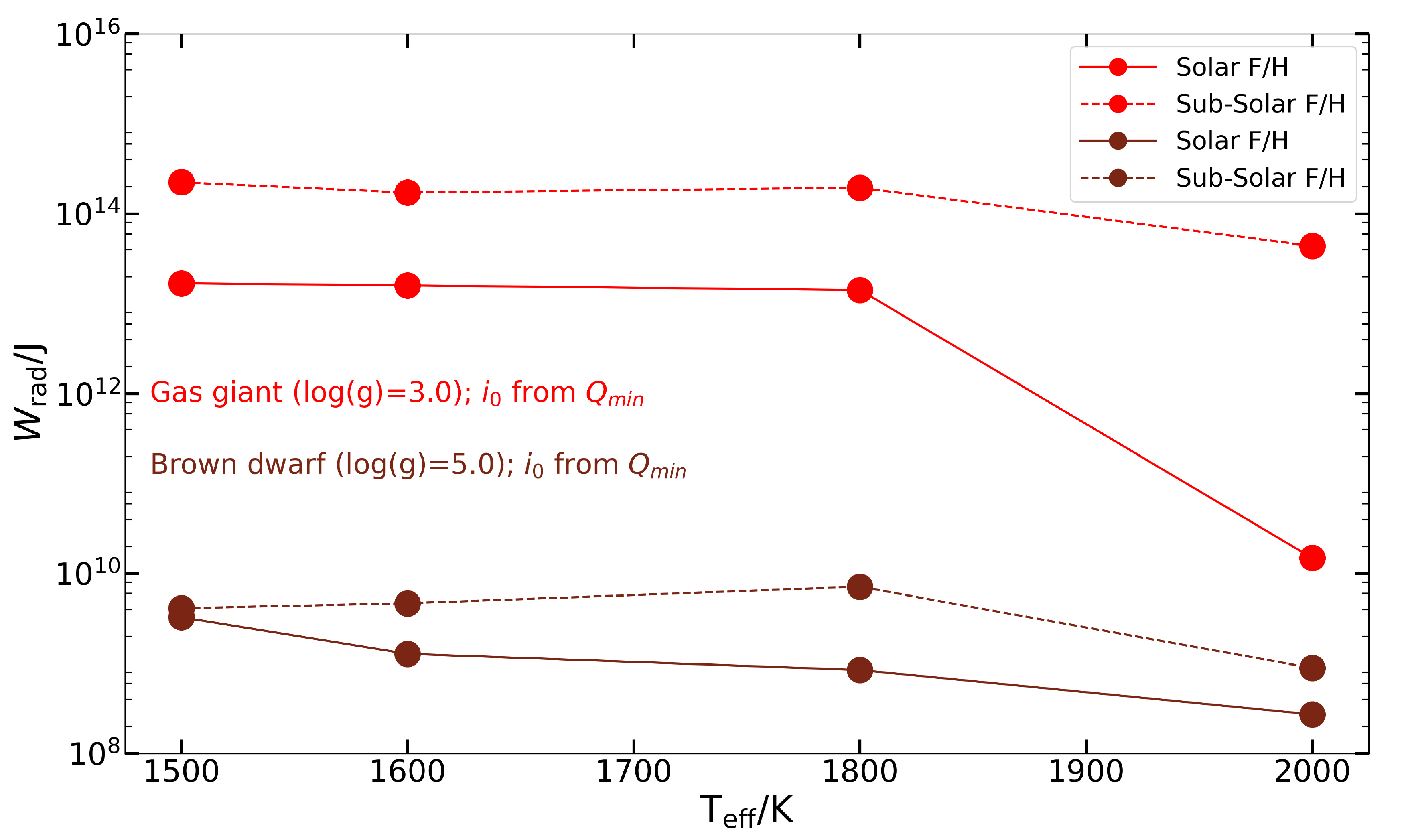}
  \includegraphics[width=\columnwidth, trim=0cm 0cm 0cm 0cm]{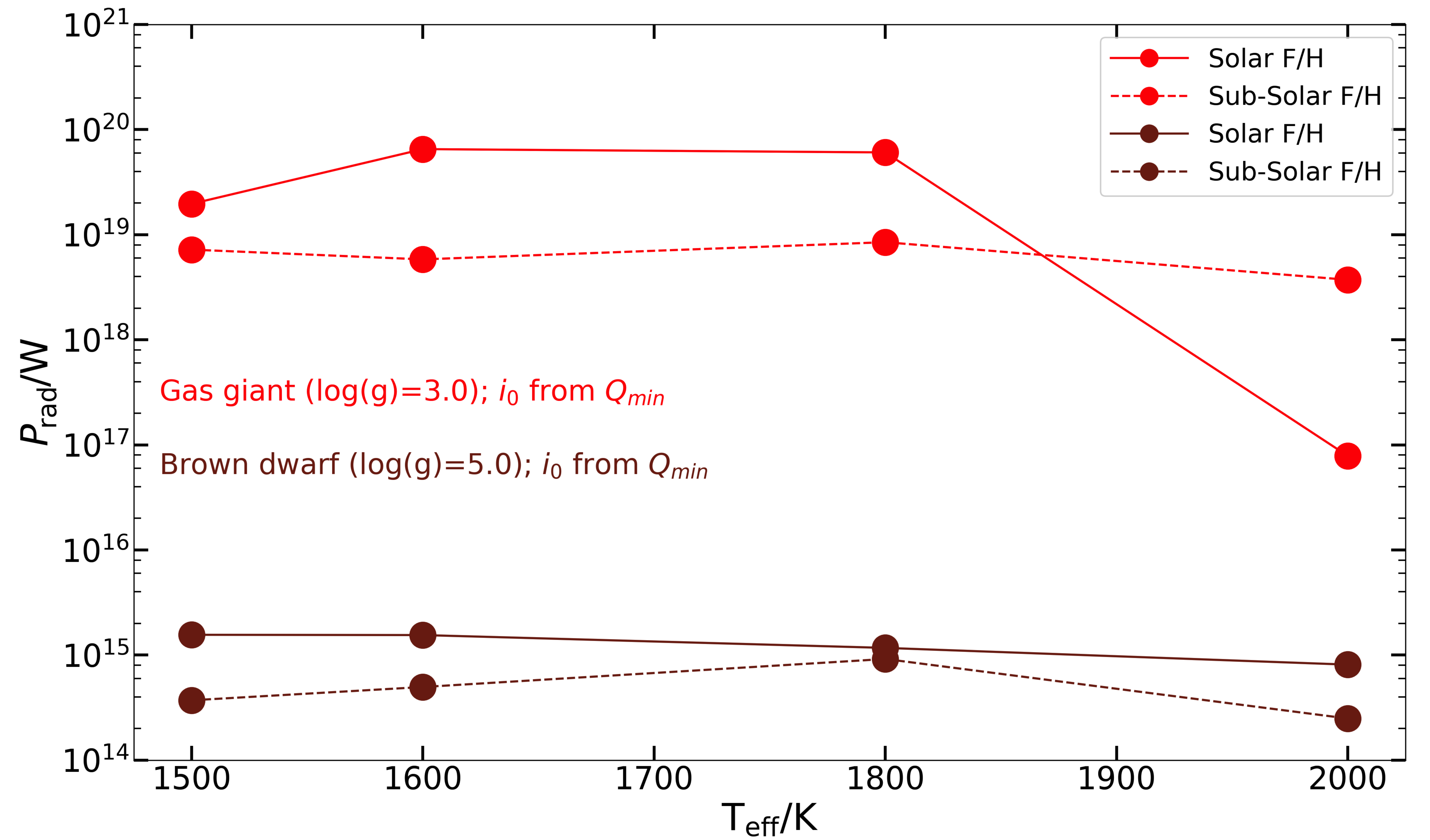}
	}
  \caption{Exo-Case (i): Total radiated energy (left) and total radio power (right) from  extrasolar lightning in giant gas planets and brown dwarfs.}
  \label{fig:exoiq}
\end{figure*}
 \begin{figure}
  \centering
  \includegraphics[width=0.7\columnwidth, trim=0cm 0cm 0cm 0cm]{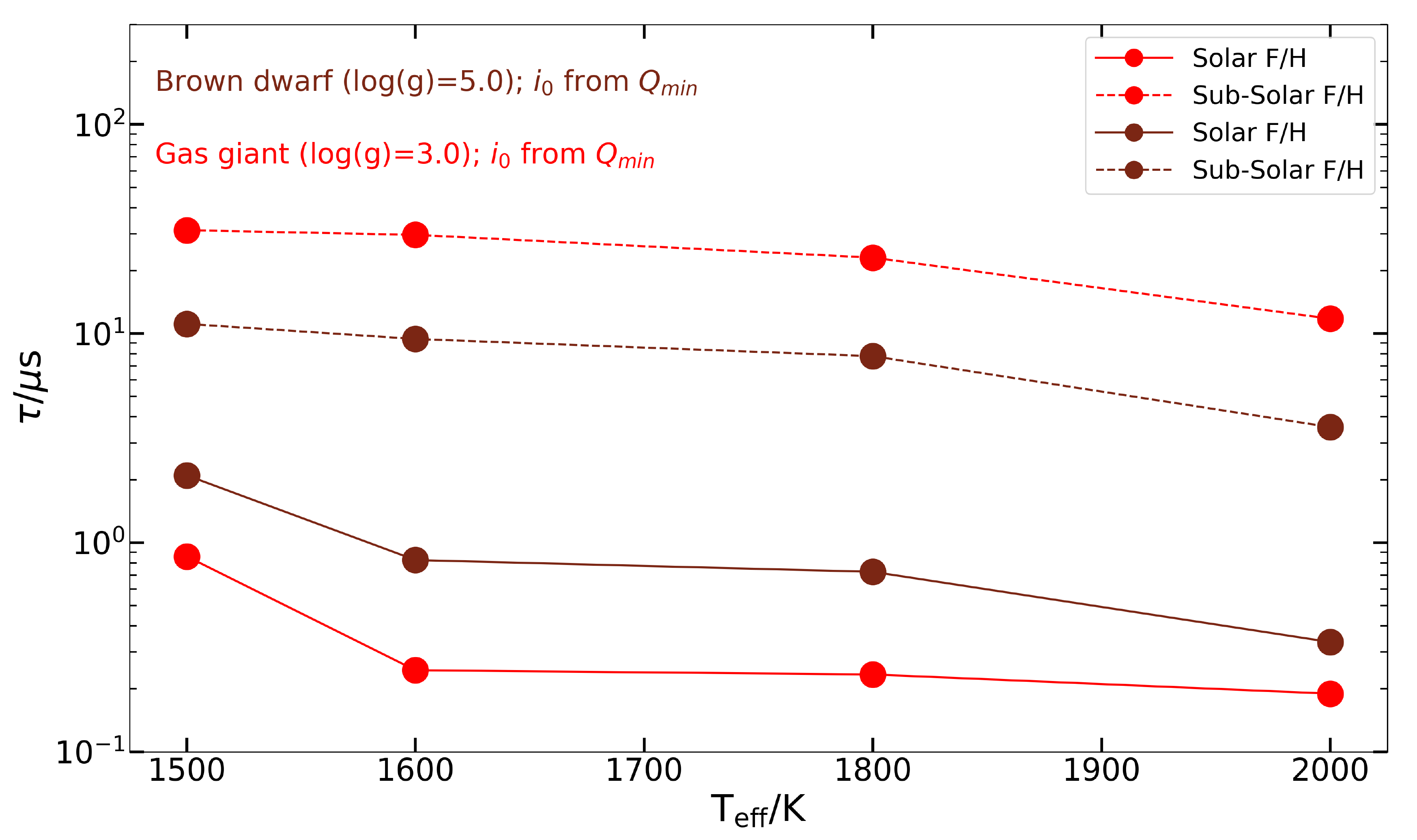}
  \caption{Exo-Case (i): Discharge duration, $\tau$,  for different types of extrasolar bodies based on discharge lengths obtained by \citet{bailey2014} and listed in Table \ref{table:h}. $\tau$ was calculated from Eq. \ref{eq:tau1} using a current velocity of $\v v_0 = 0.3$c. }
  \label{fig:exotau}
\end{figure}

%__________________________________________________________________
\subsection{Energy-release of lightning on exoplanets and brown dwarfs} \label{sec:exoen}

The radio energy radiated by lightning discharges depends on the peak current, $i_0$, and the duration of the discharge, $\tau$. $\tau$ also determines the frequency at which the peak power is released, while the strength of the electric field is determined by $i_0$. In our modelling approach, $\tau$ depends on the extrasolar object's properties through the extension of the discharge, $h$, and $i_0$ can be linked to the minimum number of charges, $Q_{\rm min}$, necessary to initiate a breakdown \footnote{For comparison, the average number of charges in a lightning channel on Earth is 30 C, \citet{bruce1941}.}. We note, however, that it is unclear how many of the breakdown-triggering charges contribute to the  lightning  plasma channel current $i(t)$ eventually. $Q_{\rm min}$ must therefore be considered as another parameter, which however, allows to link to the cloud properties in terms of their numbers and sizes, but more importantly, it allows us to link to global properties of exoplanets and brown dwarfs, hence, moving from single-object studies to studying an ensemble of objects.

\citet{bailey2014} utilised a grid of model atmospheres (\textsc{Drift-Phoenix}) that is determined by the global properties effective temperature, $T_{\rm eff}$, metallicity, [M/H], and surface gravity, log($g$), which determine the local temperature and pressure profiles. They found that the extension of the discharge, $h$, is larger in high-pressure atmospheres where the surface gravity is large or the metallicity is low and that $h$ decreases with increasing effective temperature. High-pressure atmospheres require a larger $Q_{\rm min}$ for lightning to be initiated. However, \citet{bailey2014} showed that in brown dwarfs, where clouds form at higher pressures, the minimum number of charges necessary for breakdown is smaller than in giant gas planets. They reason this with the extension of the cloud deck in brown dwarfs being shorter than in giant gas planets, resulting in a larger electric field throughout the cloud. This means that a lower number of charges is sufficient to initiate the breakdown in those atmospheres.  
 
In the following, we explore various parameter combinations in order to map out a potential parameter space of exo-lightning. To explore lightning energy and power release on exoplanets and brown dwarfs, we carry out three studies: Exo-Case (I): Both $Q_{\rm min}$ and $h$ are taken from \citet{bailey2014}. Exo-Case (II):  $h$ is from \citet{bailey2014}, and $i_0=$1000kA, 100kA, 30kA are guided by  Solar System observations (e.g. \citealt{farrell1999}). This assumes that charge accumulation in the channel on extrasolar objects is similar to their Solar System counterparts. Exo-Case (III): $Q_{\rm min}$ from \citet{bailey2014}, but $h$ is chosen from Solar System values: $h = 259$ km, 7.89 km, 2 km \citep[][]{bruning2015,rakov2003,baba2007}. Such discharges would have an "extrasolar-like" current, but Solar System-like discharge channel length. Equation \ref{eq:tau1} is used to calculate the discharge duration from $h$ and the current velocity in the channel, $\v v_0 = 0.3$c.

\smallskip

\textbf{Exo-Case (I) ($h$ and $Q_{\rm min}$, Table \ref{table:h}):}
Figure~\ref{fig:exoiq} summarises how the lightning power and energy released in exoplanet and brown dwarf atmospheres depends only marginally on the (global) effective temperature, but strongly on the surface gravity (brown vs read lines), and somewhat on the gas-phase metallicity (i.e. the abundance of elements heavier then hydrogen, linking to stellar/planetary evolutionary aspects). Hence, lightning in giant gas planets, or low-gravity, young, brown dwarfs (log($g$)$=3.0$) reaches higher energies than in older brown dwarfs (log($g$)$=5.0$).  This is of interest as older brown dwarfs have been observed to have a considerable magnetic field strength which may create a substantial magnetosphere shielding any radio emission from the atmosphere. Though the radiated radio power of lightning is higher in cold solar composition atmospheres (Fig. \ref{fig:exoiq}, right), the radiated energy is higher for sub-solar compositions (Fig. \ref{fig:exoiq}, left). Figure \ref{fig:exotau} illustrates the changing associated discharge time scales, $\tau$, which links well with the increased $h$ for decreasing metallicity ([M/H]). 

The breakdown field that determines whether a lightning discharge will develop or not, does not strongly depend on the chemical composition (i.e. ionisation energy) of the gas. However, it depends on the local pressure, which is determined by the opacity in an atmosphere \citep{helling2013a}. The very high currents resulting from high $Q_{\rm min}$ (Table \ref{table:h}) produce an electric field that will release very high energy and power in the radio bands: $\sim$$10^{8}$--$10^{10}$ J and $\sim$$10^{14}$--$10^{15}$ W in brown dwarf atmospheres, and $10^{13}$--$10^{14}$ J and $10^{18}$--$10^{20}$ W in giant gas planet atmospheres. Applying a $k = 0.01$ radio efficiency \citep[$\sim$1\% on Earth][]{volland1984}, the total dissipation energy for these objects is $W_d \sim$$10^{15}$--$10^{16}$ J for gas giants and $W_d \sim$$10^{10}$--$10^{12}$ J for brown dwarfs, latter one being comparable to lightning on Jupiter and Saturn. If we assume that the radio efficiency is 0.001  \citep{borucki1987}, the resulting dissipation energy is $W_d \sim$$10^{16}$--$10^{17}$ J for gas giants and $W_d \sim$$10^{11}$--$10^{13}$ J for brown dwarfs. If we further consider the factor of four to ten underestimate of the energy, as is suggested by our tests in Sect. \ref{sec:val}, the resulting energies will further increase. We note that the principle dependencies on global parameters (T$_{\rm eff}$, log(g), [M/H]) found for the Exo-Case (i) will not be affected by uncertainties of the peak current $i_0$ strength due to the use of $Q_{\rm min}$, however, the actual values will need to be revalidated once the link between the field breakdown and the channel current can be modelled.

 \begin{figure*}
  \centering
  \includegraphics[width=0.49\columnwidth, trim=0cm 0cm 0cm 0cm]{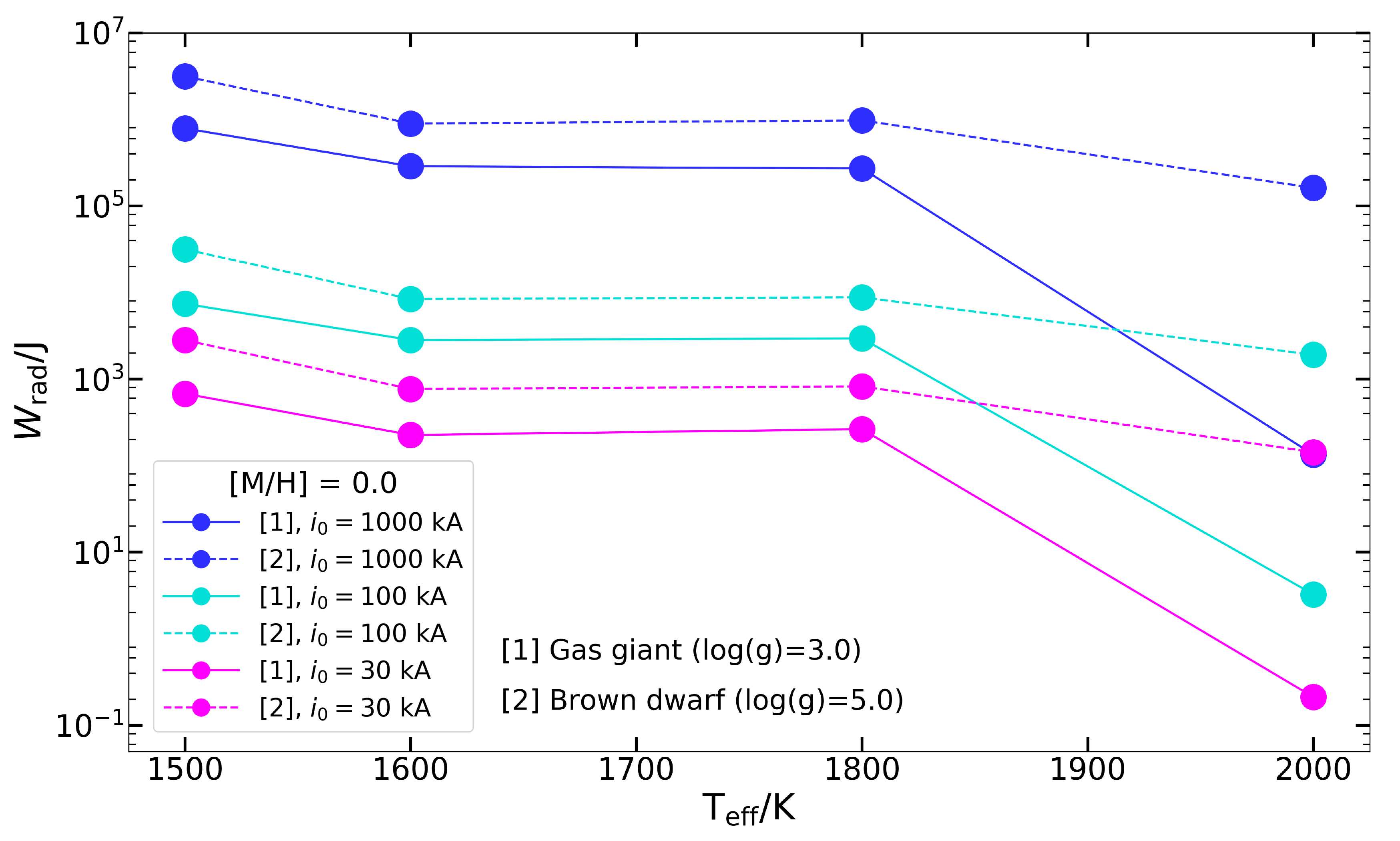}
  \includegraphics[width=0.49\columnwidth, trim=0cm 0cm 0cm 0cm]{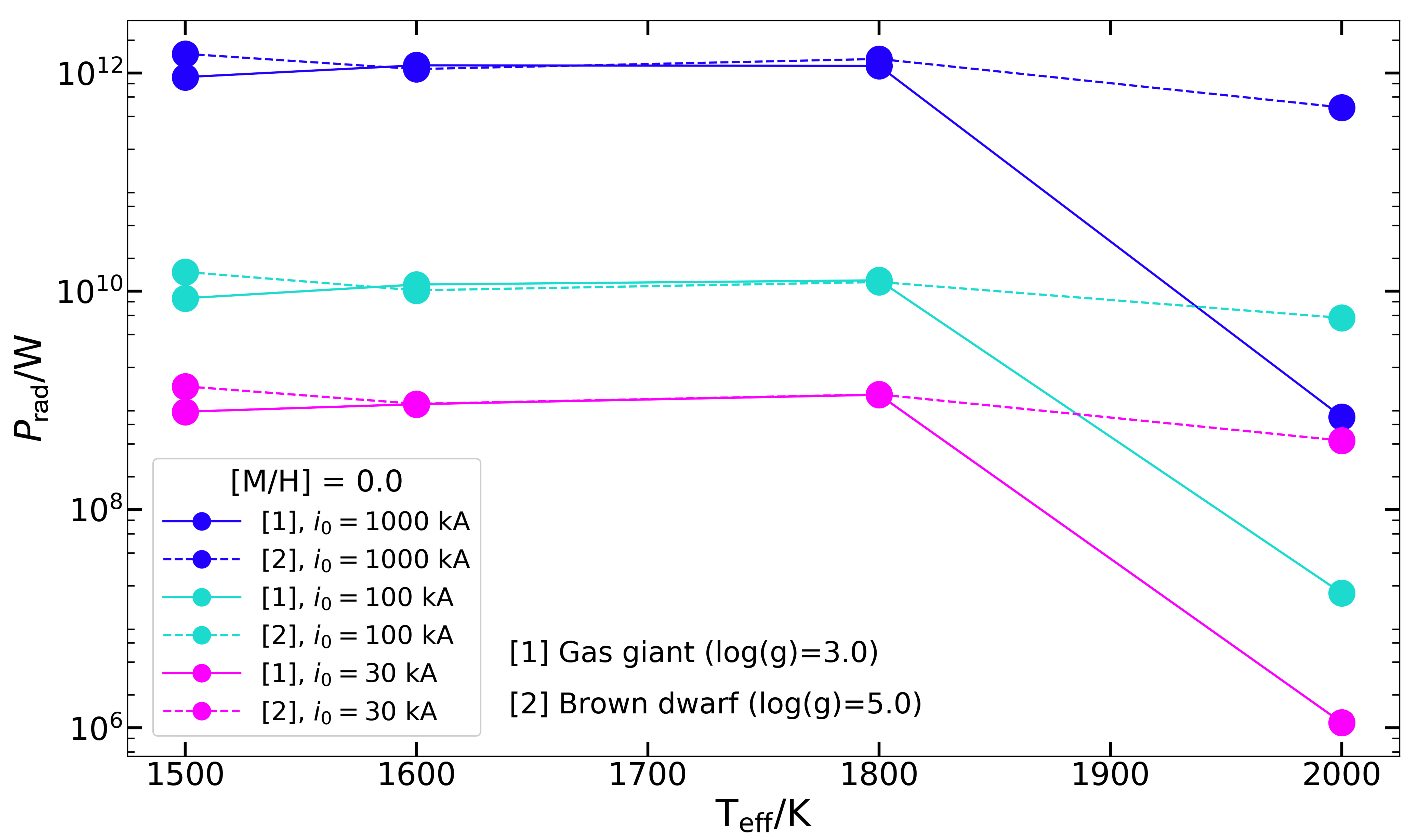} \\
  \includegraphics[width=0.49\columnwidth, trim=0cm 0cm 0cm 0cm]{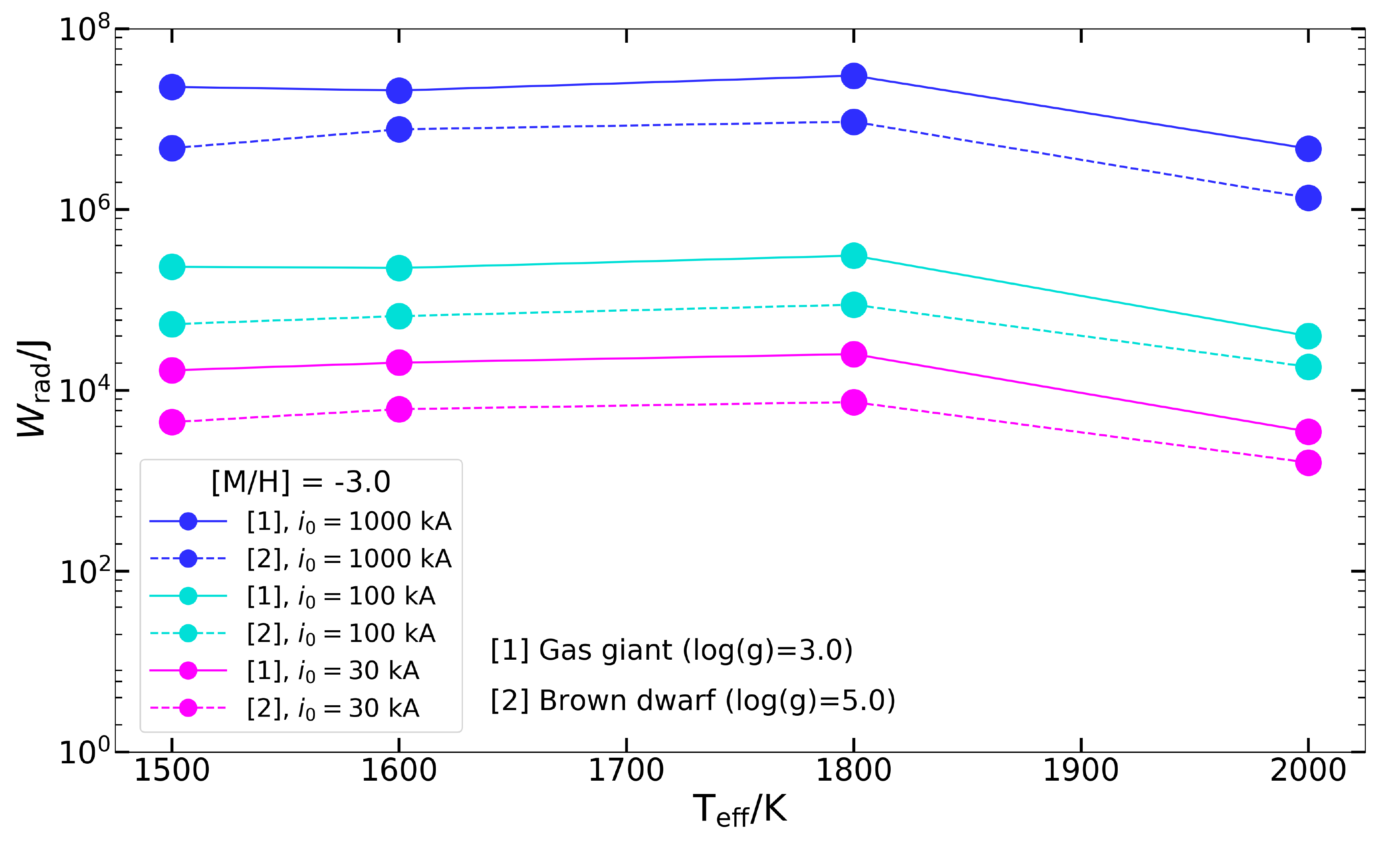}
  \includegraphics[width=0.49\columnwidth, trim=0cm 0cm 0cm 0cm]{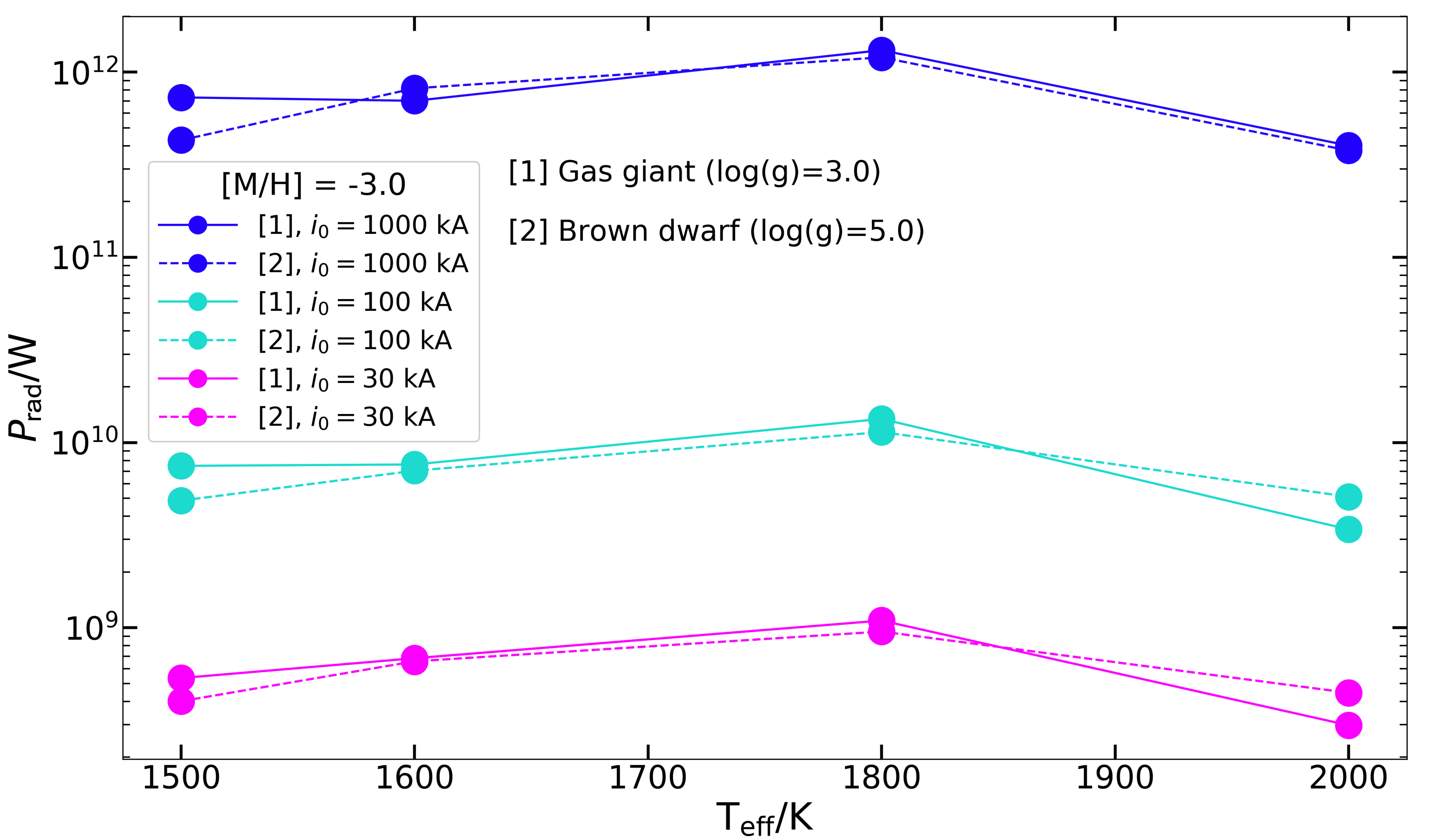} \\
  \caption{Eco-Case (ii): Total radiated energy (left) and total radio power (right) released from extrasolar lightning  in gas giants and brown dwarfs. \textbf{Top panels:}  solar metallicity atmospheres ([M/H]=0.0), \textbf{Bottom panels:} sub-solar metallicity atmospheres ([M/H]=-3.0). The peak currents $i_0 =$ 30 kA, 100 kA, 1000 kA (magenta, cyan, and blue colours, respectively), and the discharge extension $h$ are prescribed, with $h$ linking to the global parameters T$_{\rm eff}$, log(g) and [M/H] (Table~\ref{table:h}). }
  \label{fig:exo1}
\end{figure*}

\begin{figure*}
  \hspace*{-0.5cm}\includegraphics[width=0.75\textwidth, angle=-90]{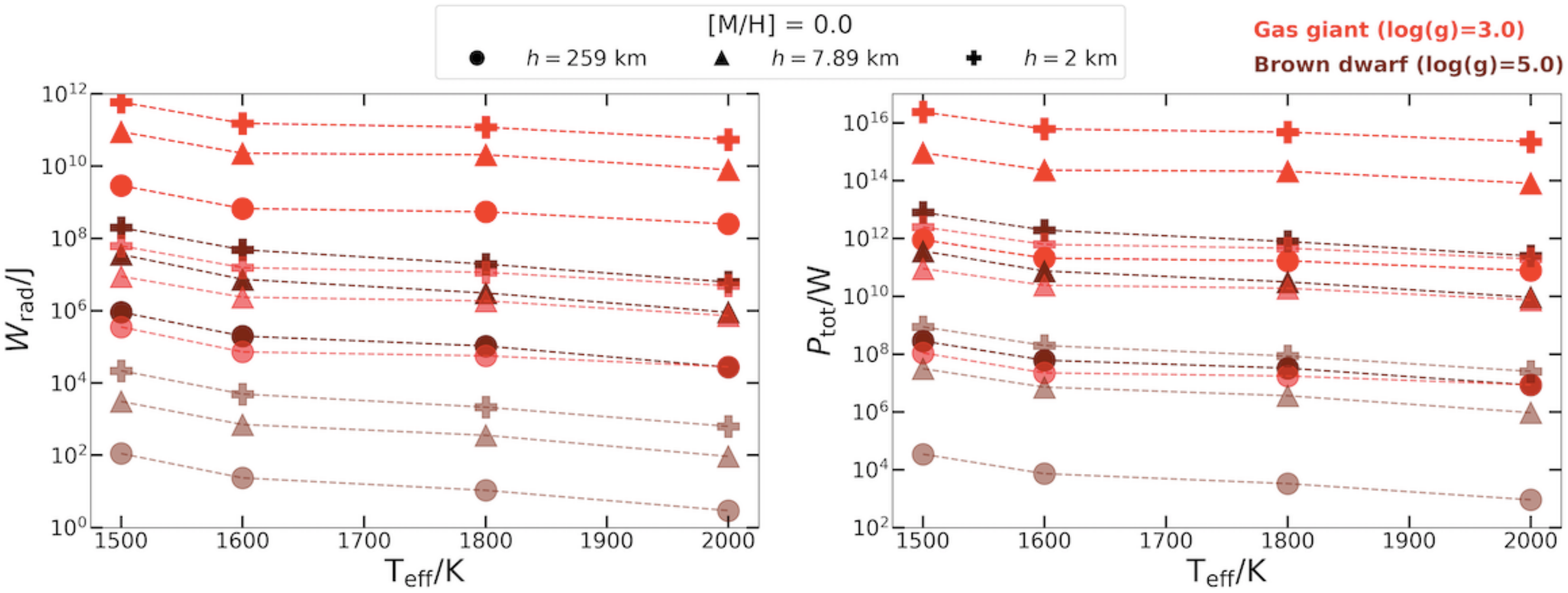}\\[-5cm]
  \includegraphics[width=0.75\textwidth, angle=-90]{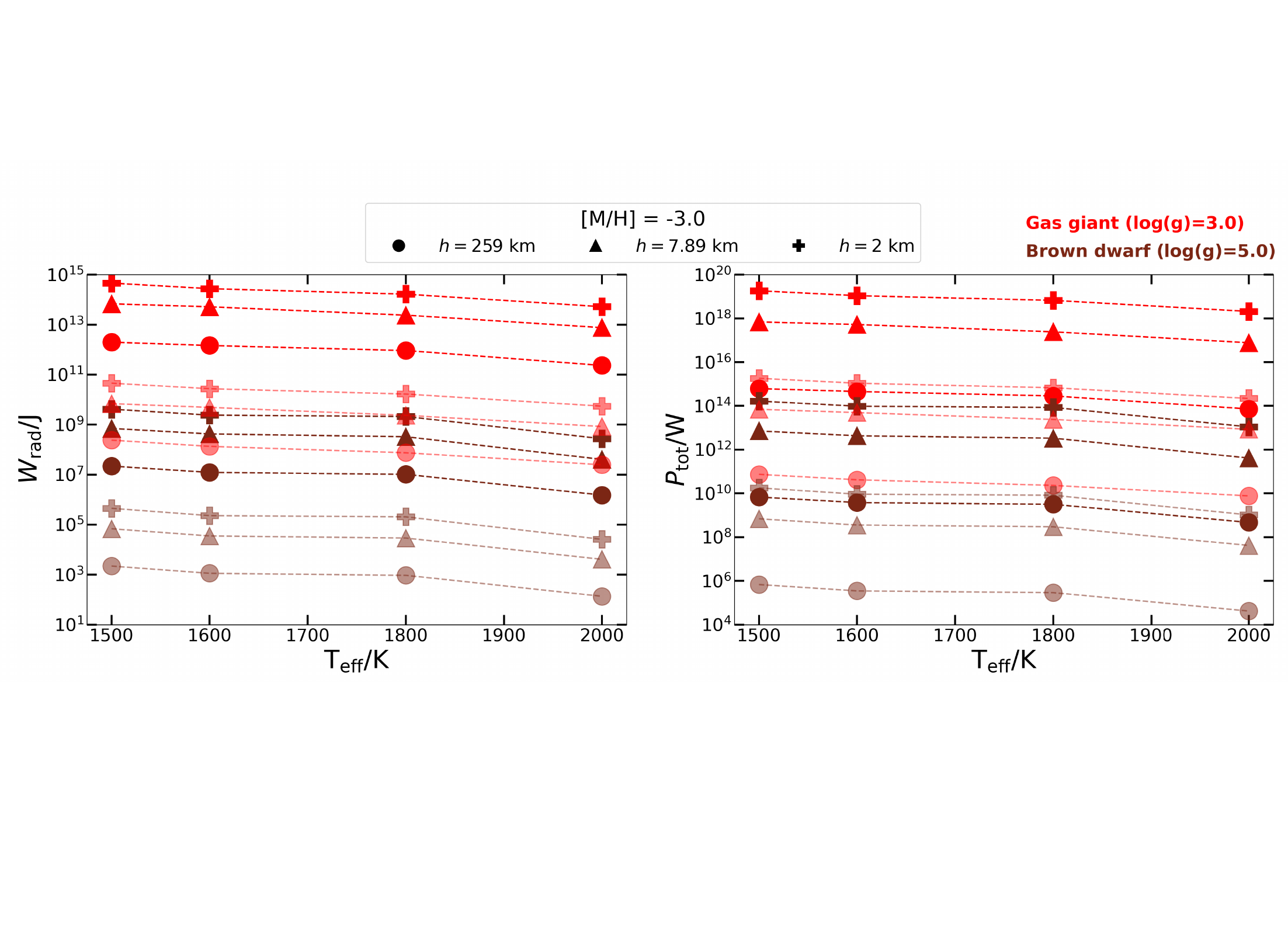}\\[-3cm]
  \caption{Exo-Case (iii): Total radiated energy (left) and total radio power (right) released from exrasolar lightning in gas giant (dark red) and brown dwarf (dark brown) atmospheres. \textbf{Top panels:}  solar metallicity atmospheres ([M/H]=0.0), \textbf{Bottom panels:} sub-solar metallicity atmospheres ([M/H]=-3.0). The discharge extensions $h =$ 2km,  7.89km, 259 km (cross, triangle, circle symbols; resulting in $\tau = 2.5 \times 10^{-5}, 9.9 \times 10^{-5}, 3.2 \times 10^{-3}$ s with $\v v = 0.3$c)  and the minimum charges necessary to initiate a discharge, $Q_{\rm min}$,  are prescribed, with $Q_{\rm min}$ linking to the global parameters T$_{\rm eff}$, log(g) and [M/H] (Table~\ref{table:h}). The light red and brown colours show results for 1\% of $Q_{\rm min}$.}
   \label{fig:exo2}
\end{figure*}

\textbf{Exo-Case (II) ($h$ as in Table \ref{table:h}, $i_0 = 1000, 100, 30$ kA):}
We test how the lightning radio power and energy release behaves assuming a fix peak current for each atmosphere for a physically motivated discharge extension $h$ which links to global object parameters (like log(g)). We do not use $Q_{\rm min}$. Figure \ref{fig:exo1} shows the results for the three example current peaks (purple-blue colours). The larger the peak current, the more power and energy is released from lightning. We note that in solar metallicity atmospheres lightning is more energetic in brown dwarfs than in giant gas exoplanets, in sub-solar compositions lightning releases more energy in lower gravity environments (giant gas planets, young brown dwarfs). With the highest peak current, $i_0 = 1000$ kA (similar to the Saturnian super-bolt), the released radio energy is $W_{\rm rad} \sim$$10^6$--$5 \times 10^7$ J for both gas giants and brown dwarfs, with slightly lower energy release from the latter type of objects.

\textbf{Exo-Case (III) ($Q_{\rm min}$ as in Table \ref{table:h}, $h =$ 2 km, 7.89 km, 259 km:}
We fix $h$ and use $Q_{\rm min}$ which links to the object's global parameters T$_{\rm eff}$, log(g) and [M/H] (Table~\ref{table:h}). We note that the $Q_{\rm min}$ values for the solar metallicity giant gas planets compare well with the values that have been empirically {\bf} derived for Jupiter (Table~\ref{table:val2}), those for solar metallicity {\bf} brown dwarfs are comparable to some of the empirically derived values for Saturn lightning (Table~\ref{table:val3}). Figure \ref{fig:exo2} shows that sub-solar metallicity atmospheres produce lightning with more radio energy and power release than solar compositions. For the same $h$, in higher surface gravity environments (i.e. brown dwarfs) lightning releases less energy than in lower surface gravity objects. The released power and energy slightly decreases with effective temperature. Also, the shorter the discharge channel, $h$, the larger $P_{\rm rad}$ and $W_{\rm rad}$ for a given $Q_{\rm min}$. In this case, giant gas planetary lightning produces higher energy every time. The released radio energy for gas giants is $W_{\rm rad} \sim$$10^8$--$10^{15}$ J, and for brown dwarfs $W_{\rm rad} \sim$$5 \times 10^4$--$10^{10}$ J. Applying various radio efficiencies, the total dissipation energy can be 2 to 3 orders of magnitude larger than $W_{\rm rad}$, and further applying the uncertainty factor from our tests in Sect. \ref{sec:val}, the results can be even higher by an order of magnitude. Figure~\ref{fig:exo2} (light red and light brown) shows that {\bf} lightning power drops by $\approx 4$ orders of magnitude if only 1\% of the required charges for a field breakdown contribute to the lightning current.

Our results suggest that the discharge energy will strongly depend on the process through which cloud particles are charged and the processes that cause the electrostatic potential to build up. A discussion of processes for brown dwarfs and giant gas planets can be found in \citet{helling2016b} and in comparison to the Solar System in \citet{helling2016}. However, no consistent description of large-scale lightning discharges is available for the atmospheres discussed here, nor for the Solar System planets.

%__________________________________________________________________
\subsection{The effect of parameter uncertainties} \label{subs:uncert}

%Table Alpha test
\begin{table}  
 \begin{center}
 \caption{Statistical analysis of the changes in the total discharge energy, $W_d$, and the total radio power of lightning, $P_{\rm rad}$ due to changes in the $\alpha$ parameter. $\alpha$ is randomly picked for 100 times, for each extrasolar case study in Table \ref{table:h} (in total 16 cases, as in Approach (I) in Sect. \ref{sec:exoen}). The statistical values here are the minimum, maximum, average and median of the results of the 16 cases. The maximum value suggest that there is a 200\% change in the results, however, that is only valid for one data point, which corresponds to the outlier data points in Figs \ref{fig:exoiq} and \ref{fig:exo1} ([M/H]=0.0, log($g$)=3.0, $T_{\rm eff} = 2000$ K). After removing this outlier, we get a more informative result.}
	\vspace{0.3cm}
  \begin{tabular}{@{}lllll@{}}	
	\hline
	 & minimum & maximum & average & median \\
	\hline
	$\alpha$ & 8.7\% & 9.5\% & 9.3\% & 9.3\% \\
	$W_{\rm d}$ & 12.5\% & 200.0\% & 32.9\% & 21.4\%  \\
	$P_{\rm rad}$ & 12.5\% & 200.0\% & 32.9\% & 21.4\% \\
	\hdashline
	\multicolumn{5}{c}{After removing the 200\% outlier} \\
	$\alpha$ & 8.7\% & 9.5\% & 9.3\% & 9.3\% \\
	$W_{\rm d}$ & 12.5\% & 36.2\% & 21.8\% & 19.8\%  \\
	$P_{\rm rad}$ & 12.5\% & 36.2\% & 21.8\% & 19.8\% \\
	\hline
  \label{table:test}
  \end{tabular}
 \end{center}
\end{table}
%Table 

%Table tau, i0 test
\begin{table}  
 \begin{center}
 \caption{Effects of changing discharge duration, $\tau$ (left), and current peak, $i_0$ (right), compared to a base value, on the total radio power, $P_{\rm rad}$, and the radiated discharge energy, $W_{\rm rad}$. All other parameters remain unchanged. The base value for $\tau$ was 100 $\mu$s, while for $i_0$ it was 30 kA. $\Delta$ sign represents the change or uncertainty in the value.}
  \vspace{0.3cm}
  \begin{tabular}{@{}llllllll@{}}	
	\cline{1-3}\cline{6-8}
  $\Delta \tau$ [s] & $\Delta P_{\rm rad}$ & $\Delta W_{\rm rad}$ & & & $\Delta i_0$ [A] & $\Delta P_{\rm rad}$ & $\Delta W_{\rm rad}$ \\ \cline{1-3}\cline{6-8}
	$10^{-6}$ & 0.13 \% & 1.13 \% & & & $10^3$ & 6.8\% & 6.8\% \\
	$10^{-5}$ & 1.2 \% & 11.3 \% & & & $10^4$ & 77.8\% & 77.8\% \\ \cline{6-8}
	$10^{-4}$ & 6.6 \% & 113.3 \% \\ \cline{1-3}
  \label{table:test2}
  \end{tabular}
 \end{center}
\end{table}
%Table 

\begin{enumerate}

\item[a)] \textbf{$\alpha$ $[{\rm s}^{-1}]$:}
We tested our results against the uncertainty in the only (semi-)randomly chosen parameter, $\alpha$, a frequency type constant introduced in the current function (Eq. \ref{eq:1}). We randomly choose $\alpha$ from a normal distribution with mean and standard deviation that ensures that $\alpha$ is $\sim$1 order of magnitude lower than $\beta$, the other frequency type parameter of the bi-exponential current function (Eq. \ref{eq:1}). Our choice is based on the commonly used $\alpha$ and $\beta$ pairs in the literature (Table \ref{table:3}). In addition, the mean and standard deviation of the normal distribution are proportional to $\alpha \beta$ as expressed from Eq. \ref{eq:tau}. We carried out the numerical experiment for a hundred runs for each exoplanet and brown dwarf types (Table \ref{table:h}). The results are listed in Table \ref{table:test}, and show that a roughly 9.3\% change in $\alpha$ as a result of the random pick throughout the 100 runs for each case, results in a 21.8\% average and 19.8\% median variation in both the total power ($P_{\rm rad}$) and the total discharge energy ($W_d$). The variation in both $W_d$ and $P_{\rm rad}$ is between 12.5\% and 36.2\% depending on the object, and with that on the extension of the discharge, $h$, and the number of charges in the channel, $Q_{\rm min}$. However, the tests did not include the testing of the effects of $h$ and $Q_{\rm min}$. These values are calculated after removing the outlier of the data set, appearing in both Figs \ref{fig:exoiq} and \ref{fig:exo1} ([M/H]=0.0, log($g$)=3.0, $T_{\rm eff} = 2000$ K). This atmosphere alone suggests that the variations caused by $\alpha$ can be up to 200\%. 

\vspace{0.2cm}

\item[b)] \textbf{$\tau$ [s] and $i_0$ [A]:}
The discharge duration, $\tau$, and the peak current, $i_0$, are the two values that will affect the results the most (Sect. \ref{sec:resdis}). Therefore, we tested how much changing these values compared to a base value will affect the resulting radiated discharge energy, $W_{\rm rad}$, and total radio power, $P_{\rm rad}$. We gradually increased $\tau$ and $i_0$ separately, starting from a base or comparison value, while all the rest of the input parameters were fixed (i.e $\alpha$, $\beta$, see Sect. \ref{sec:current}). The base value for $\tau$ was 100 $\mu$s, while for $i_0$ it was 30 kA. Our results are shown in Table \ref{table:test2}. The tests showed, that the energy is fairly sensitive to the changes in the discharge duration, while the power seems to be less sensitive. A 1 $\mu$s change in $\tau$ causes only about a percent change in the energy, and 0.1\% change in the power. Doubling the base value of $\tau$ changes the power only by 6.6\%, but changes the energy by more than a 100\%. We also point out that increasing $\tau$ results in {\bf} increasing $P_{\rm rad}$ and $W_{\rm rad}$. Similarly, we tested the sensitivity of the end results to $i_0$. Our results show that a 10 kA change in the peak current results in a $\sim$77.8\% change in both the total radio power, and the radiation energy. This is twice as much as the uncertainty caused by the $\alpha$ parameter.
\end{enumerate}

%__________________________________________________________________
\subsection{Notes on exo-lightning observability}

We have noted in the introductory paragraphs of Sect. \ref{sec:resdis} that attempts to observe lignting on extrasolar objects are yet to be successful \citep[e.g.][]{zarka2012, 2019A&A...624A..40T, 2020MNRAS.495.3881H}. However, it is important to point out that these studies focus on a single signature of lightning, either in the radio or in the optical range. \citet{bailey2014} included a summary table of lightning signatures observed on Earth. Looking at this table, one could conclude that observing exo-lightning should follow a multi-signature, multi-wavelength, multi-technique approach. For example, simultaneous optical and radio observations may reveal unexplained flux increase, e.g. in the presense of cyclotron radio emission \citep{vorgul2016}, from an extrasolar object. Follow-up observations focusing on the chemical changes caused by lightning can confirm that such increase is the result of large lightning activity on the observed explanet or brown dwarf. \citep{hodosan2017}.

Observations with existing and future, more sensitive instruments will allow us to give further constraints on the findings of this paper summarized in Sect. \ref{sec:conc}. Such observations, while they could be unsuccessful in actually detecting lightning, will be valuable in providing lower limits of lightning properties, be that the power released from it \citep{hodosan2017, 2020MNRAS.495.3881H}, or the spatial and time distribution of large-scale discharges \citep[e.g.][]{hodosan2016a}, and hence they will further increase our understanding of cloud-phenomena on exoplanets and brown dwarfs.

%__________________________________________________________________
%__________________________________________________________________
\section{Conclusions} \label{sec:conc}

Our current knowledge of clouds in exoplanets and brown dwarfs \citep[e.g.][]{sing2015,2019NatAs...3..813B,2019AJ....157..101M}, and lightning activity in the Solar System \citep[e.g][]{rakov2003,yair2008,yair2012,helling2016} suggests that lightning may occur in extrasolar atmospheres. The electrostatic field breakdown that is associated with lightning is relatively independent of the chemical composition of the atmospheric gas, but the local gas pressure affects the location of the breakdown if a sufficient ambient electric field exists \citep{helling2013a}. Cosmic rays will enhance the local population of mostly thermal seed electrons in particular in the upper part of the atmosphere \citep{rimmer2013}, and cosmic rays may trigger the network of plasma channels known as lightning (\citealt{1933Natur.132..712T,2020JGRD..12531433T}). \citet{trinh2017} measured that cosmic ray air showers during thunderstorms have a much larger fraction of strong circularly polarized radio emission than when measured during fair-weather conditions. They directly link this observation to the changes in electric filed in thunderclouds, and suggest that such observations may help characterize the electric filed of thunderstorms. Most, if not all, of our knowledge about lightning is derived from detailed observations on Earth for which, hence, well adopted parameterisations exist. Lesser is known of lightning on other Solar System planets. On Jupiter and Saturn, only the highest energy tail of the lightning distribution has be detected (\citealt{hodosan2016a}). Due to the lack of exo-lightning detection, we do not know how similar or different lightning is in extrasolar planetary atmospheres compared to what is known from Earth and the Solar System. We therefore utilise a modelling approach based on Earth lightning parameterisations, part of which has been applied to Saturn and Jupiter, in order to explore possible parameters that may describe lightning in exoplanet and brown dwarf atmospheres. Tests demonstrate that this approach works best for combinations of Earth parameters, but that observations for Jupiter and Saturn do not constrain parameters like discharge extension or channel peak current well.

The radiated power and emitted energy of a lightning discharge depends on two properties: the discharge duration, $\tau$, and the peak current, $i_0$. The longer in time the lightning discharge, the larger the energy released for a constant peak current. Short lightning channels can be the result of high local gas pressure in the atmosphere, which is the case for objects of large surface gravity (like brown dwarfs) and low metallicity (like planets and brown dwarfs being formed with population III stars). The larger the lightning  channel peak current, the larger the power and energy released from the lightning discharge, hence, atmospheres that enable high-efficient cloud charging may produce a large current flow, and therefore a large radio signal. Possible exoplanet candidates maybe those who receive a high flux of cloud-ionising (but no evaporating) irradiation in combination with a large-scale charge separation.

We suggest that lightning on extrasolar, planetary objects can be expected to be very different compared to the Solar System. We relate our modelling  approach to extrasolar atmospheres through the extension of the discharge, $h$, and the charges in the current channel, $Q_{\rm min}$. We note that the actual number of charges producing the necessary electric field breakdown and converts into the lightning current, as well as the extension of the discharge channel are unknown. However, we can draw some general conclusions about extrasolar lightning:\\
-- The emitted lightning power changes only marginally with the effective temperature of cloud-forming giant gas planets or brown dwarfs. Therefore, young brown dwarfs and giant gas planets can be expected to show similar lightning power.\\
-- The emitted lightning power changes with  the global metallicity of the objects because of its effect in the atmospheric temperature - pressure structure.\\
-- The emitted lightning power depends strongly on the surface gravity of the object because this affects the atmospheric pressure stratification, which in turn determines the location and extension of the lightning discharge.\\[0.1cm]
-- Low-gravity atmospheres (giant gas planets, young brown dwarfs) reach a higher total lightning power ($W_{\rm rad}$ [J]) and higher radio power ($P_{\rm rad}$ [W]) than compact, high-gravity atmospheres (e.g. brown dwarfs with log(g)=5). However, for a given peak current, solar element abundance, low gravity atmospheres of giant gas planets produce less energetic lightning than compact, high-gravity atmospheres (population II objects in terms of stellar generations).\\
 -- For the same discharge extension $h$, higher surface gravity objects host less powerful and energetic flashes. The shorter the discharge channel, the higher the released energy and power are. $W_{\rm rad} \sim$$10^8$--$10^{15}$ J for gas giant planets, and $W_{\rm rad} \sim$$5 \times 10^4$--$10^{10}$ J for brown dwarfs.\\
-- In the case when lightning discharges form very short channels with a large number of charges within the channel, {\bf} very quick discharges occur with very large peak currents in giant gas planets. The total dissipated energies would reach values of $10^{11}$--$10^{13}$ J in brown dwarfs (log($g$)=5.0) and $10^{16}$--$10^{17}$ J in giant gas planets (log($g$)=3.0).

\smallskip

\noindent We note that our results may underestimate the actual energy-release by a factor of four to ten, as it is suggested by our tests with Solar System lightning (Sect. \ref{sec:val}). Uncertainty in the results is also introduced by the random pick of the $\alpha$ parameter (Sect. \ref{subs:uncert}), which can be on average 20\%.

Our results suggest that lightning may release $10^5\times$ more energy and radio power in extrasolar gas giant and young brown dwarf atmospheres than in Solar System planetary atmospheres. Extrasolar gas giants and brown dwarfs are different from Earth, Jupiter, and Saturn in terms of, for example, atmospheric extension, cloud particle population and size distributions, external radiation and atmospheric dynamics, 
therefore such energy release may not be unreasonable.

% \end{linenumbers}
%__________________________________________________________________
%__________________________________________________________________
\section*{Acknowledgments}

We thank William M. Farrell and Yoav Yair for useful discussions. 
We highlight financial support of the European Community under the FP7 by an ERC starting grant number 257431. Ch. H. acknowledges funding from the European Union H2020-MSCA-ITN-2019 under Grant Agreement no. 860470 (CHAMELEON).

%\end{linenumbers}

%%%%%%%%%%%%%%%%%%%%%%%%%%%%%%%%%%%%%%%%%%%%%%%%%%

%%%%%%%%%%%%%%%%% APPENDICES %%%%%%%%%%%%%%%%%%%%%

\appendix

%__________________________________________________________________
%__________________________________________________________________
\section{List of symbols and units}

%__________________________________________________________________
\subsection{Current functions}

%Table 
\begin{table*}  
 \centering
 \small
 \caption{Symbols; Section \ref{sec:current}}
  \begin{tabular}{@{}llll@{}}	
	\hline
	Symbol & Definition & Units & Reference \\ 
	\hline \hline
	$i(t)$ & current in the lightning channel & A & \\
	$i_0$ & current peak & A & \\
	$\eta$ & correction factor for the current peak & - & \\
	$m \in \mathbb{N}$ & & - & \\
	$\alpha$ & frequency-type constant & s$^{-1}$ & \citet{bruce1941} \\
	$\frac{1}{\alpha}$ & overall duration of the return stroke & s & \citet{dubrovin2014} \\
	$\beta$ & frequency-type constant & s$^{-1}$ & \citet{bruce1941} \\
	$\frac{1}{\beta}$ & rise time of the current wave & s & \citet{dubrovin2014} \\
	$\tau_1$ & time constant determining the current-rise time & s & \citet{diendorfer1990} \\
	$\tau_2$ & time constant determining the current-decay time & s & \citet{diendorfer1990} \\
	\hline
  \label{table:isym}
  \end{tabular}
\end{table*}
%Table 

%__________________________________________________________________
\subsection{Electric field, frequency and power spectrum}

%Table Parameters
\begin{table*}  
 \centering
 \small
 \caption{Symbols; Section \ref{sec:efield} and \ref{sec:freqsp}}
  \begin{tabular}{@{}lll@{}}	
	\hline
	Symbol & Definition & Units \\ 
	\hline \hline
	$E(t)$ & electric field & V m$^{-1}$ \\ 
	$c$ & speed of light & m s$^{-1}$  \\
	$r$ & distance of the source and observer & m \\
	$M(t)$ & electric dipole moment formed between two charged regions & C m \\
	$Q(t)$ & electric charge & C \\
	$h$ & separation of the charged regions & m \\
	$\v v(t)$ & velocity of the return stroke \citep{bruce1941} & m s$^{-1}$ \\
	$\gamma$ & frequency-type constant \citep{bruce1941} & s$^{-1}$ \\
	$\epsilon_0$ & permittivity of the vacuum & F m$^{-1}$ \\
	$E(f)$ & electric field frequency spectrum & - \\
	$f$ & frequency & Hz \\
	$P(f)$ & power spectrum & - \\
	$P'(f)$ & radiated power spectral density & W Hz$^{-1}$ \\
	\hline
  \label{table:esym}
  \end{tabular}
\end{table*}
%Table 

%__________________________________________________________________
\subsection{Discharge dissipation energy}

%Table Parameters
\begin{table*} 
 \centering
 \small
 \caption{Symbols; Section \ref{sec:disen}.}
  \begin{tabular}{@{}lll@{}}	
	\hline
	Symbol & Definition & Units \\ 
	\hline \hline
	$W_{\rm rad}$ & discharge energy radiated into radio frequencies & J \\
	$W_d$ & discharge dissipation energy & J \\
	$\frac{P_0}{\Delta f}$ & peak spectral power density & W Hz$^{-1}$ \\
	$f_0$ & frequency of the peak of the power spectrum & Hz \\
	$n$ & negative spectral roll-off at high frequencies & - \\ 
	$P_{\rm rad}$ & Total radiated radio power & W \\
	\hline
  \label{table:wsym}
  \end{tabular}
\end{table*}

\bibliographystyle{elsarticle-harv}
\bibliography{bib.bib}




%% \bibitem must have one of the following forms:
%%   \bibitem[Jones et al.(1990)]{key}...
%%   \bibitem[Jones et al.(1990)Jones, Baker, and Williams]{key}...
%%   \bibitem[Jones et al., 1990]{key}...
%%   \bibitem[\protect\citeauthoryear{Jones, Baker, and Williams}{Jones
%%       et al.}{1990}]{key}...
%%   \bibitem[\protect\citeauthoryear{Jones et al.}{1990}]{key}...
%%   \bibitem[\protect\astroncite{Jones et al.}{1990}]{key}...
%%   \bibitem[\protect\citename{Jones et al., }1990]{key}...
%%   \harvarditem[Jones et al.]{Jones, Baker, and Williams}{1990}{key}...
%%


% \end{thebibliography}

\end{document}